\documentclass[10pt,twocolumn,tightenlines,floats,aps,amsmath,amssymb,nofootinbib,prd,floatfix,longbibliography,superscriptaddress]{revtex4-2}

\usepackage[utf8]{inputenc}
\usepackage{wasysym}
\usepackage{amsmath,amssymb,amsfonts,amsthm,mathrsfs}
\usepackage{graphicx,wrapfig}
\usepackage{enumerate}
\usepackage{color}

\usepackage{physics}
\usepackage{mathtools}
\usepackage{dcolumn}
\usepackage{orcidlink}

\usepackage{hyperref}
\hypersetup{
	bookmarks=true,         % show bookmarks bar?
	unicode=false,          % non-Latin characters in Acrobat’s bookmarks
	pdftoolbar=true,        % show Acroba,t’s toolbar?
	pdfmenubar=true,        % show Acrobat’s menu?
	pdffitwindow=false,     % window fit to page when opened
	pdfstartview={FitH},    % fits the width of the page to the window
	pdftitle={Primordial power spectrum at N3LO in effective field theories of inflation},    % title
	pdfauthor={Mauricio Gamonal, Eugenio Bianchi},     % author
	pdfkeywords={inflation, cosmology, quantum gravity}, % list of keywords
	pdfnewwindow=true,      % links in new PDF window
	colorlinks=true,       % false: boxed links; true: colored links
	linkcolor=red,          % color of internal links (change box color with linkbordercolor)
	citecolor=blue,        % color of links to bibliography
	filecolor=cyan,         % color of file links
	urlcolor=blue        % color of external links
}

\newcommand{\ii}{\mathrm{i}}
\newcommand{\ee}{\mathrm{e}}

\begin{document}

\title{Primordial power spectrum at N3LO in effective theories of inflation}

\author{Eugenio Bianchi\,\orcidlink{0000-0001-7847-9929}}
\email[]{ebianchi@psu.edu}
\author{Mauricio Gamonal\,\orcidlink{0000-0002-0677-4926}}
\email[]{mgamonal@psu.edu}

\affiliation{Institute for Gravitation and the Cosmos, The Pennsylvania State University, University Park, Pennsylvania 16802, USA\\[.2em]}
\affiliation{Department of Physics, The Pennsylvania State University, University Park, Pennsylvania 16802, USA\\[-0.5em]
${}$}

\date{\today}

\begin{abstract}
We develop a systematic framework to compute the primordial power spectrum up to next-to-next-to-next to leading order (N3LO) in the Hubble-flow parameters for a large class of effective theories of inflation. We assume that the quadratic action for perturbations is characterized by two time-dependent functions---the kinetic amplitude and the speed of sound---that are independent of the Fourier mode $k$. Using the Green's function method introduced by Stewart and Gong and extended by Auclair and Ringeval, we determine the primordial power spectrum fully expanded around a pivot scale up to N3LO, starting from a given generic action for perturbations. As a check, we reproduce the state-of-the-art results for scalar and tensor power spectra of the simplest ``vanilla'' models of single-field inflation. The framework applies to Weinberg's effective field theory of inflation (with the condition of no parity violation) and to the effective theory of spontaneous de Sitter symmetry breaking. As a concrete application, we provide the expression for the N3LO power spectrum of $R+R^2$ Starobinsky inflation in metric variables, without a field redefinition. All expressions are provided in terms of an expansion in one single parameter, the number of inflationary e-foldings $N_\ast$. Surprisingly, we find that, compared to previous leading-order calculations, for $N_\ast = 55$ the N3LO correction results in a $7\%$ decrease of the predicted tensor-to-scalar ratio, in addition to a deviation from the consistency relation and a prediction of a negative running $\alpha_\mathrm{s}=-\frac{1}{2}(n_\mathrm{s}-1)^2+\ldots$ of the scalar tilt. These results provide precise theoretical predictions for the next generation of CMB observations. 
\end{abstract}

\maketitle

\section{Introduction}
\label{sec:Introduction}

Cosmic inflation \cite{Brout1978, Starobinsky1979, Starobinsky1980, Guth1981, Mukhanov1981,Linde1982,Albrecht1982,Guth1982,Hawking1982,Linde1983} provides a mechanism for the production of primordial perturbations that is successful in explaining a wide range of cosmological observations, including the nature of anisotropies in the temperature fluctuations of the cosmic microwave background (CMB), and the quantum origin of the large scale structure of the Universe \cite{Akrami2020,Martin:2024qnn}. Within this theoretical framework, there is a plethora of inflationary models which range from quantum-gravity motivated models to phenomenological parametrizations of potentials \cite{Martin2014,Odintsov:2023weg},
and additional observations are required to distinguish between different models. Thus, as upcoming cosmological observations are expected to improve the constraints on many of the primordial observables \cite{CORECollaboration2016, S4Collaboration2020, LiteBIRDCollaboration2022}, precise theoretical predictions for these future observations are also required. We address
this issue directly for the large class of inflationary models summarized in Table \ref{Tab:Models}.

The primordial power spectrum is one of the most relevant cosmological observables. The Green's function method introduced by Stewart and Gong in \cite{Stewart2001} was recently extended by Auclair and Ringeval \cite{Auclair2022} to provide a detailed computation of the power spectrum at next-to-next-to-next to leading order (N3LO) in the framework of single-field slow-roll inflation---a phase driven by a minimally coupled scalar field slowly rolling down its potential---together with an extension to non minimal kinetic terms obtained via a mapping method \cite{BeltranJimenez:2013ikr}. These models belong to a broader class of effective theories of inflation: a prototypical example is the action for the free propagation of scalar curvature perturbations $\mathcal{R}$,
\begin{equation}
	\label{eq:Action-EFT-Example}
S_{2}[\mathcal{R}] =\frac{1}{2}  \int\! \dd[4]{x} \, a(t)^3\,Z_\textrm{s}(t)\Big( \dot{\mathcal{R}}^2- \frac{c_\textrm{s}(t)^2}{a(t)^2} (\partial_i \mathcal{R})^2 \Big)\,,
\end{equation}
where the nontrivial time-dependence of the background fields and geometry, assumed to be homogeneous and isotropic,
is encoded in three functions of time---the scale factor $a(t)$, the kinetic amplitude $Z_s(t)$ and the speed of sound $c_s(t)$ \cite{Cheung2007}. While in a given model these functions take a specific form, it is useful to treat them as independent to obtain general formulas which apply to all perturbations in a scalar-vector-tensor decomposition. We work in a spatially flat quasi-de Sitter background and assume that there is no gravitational parity violation, which implies that the functions $Z_s(t)$ and $c_s(t)$ admit a Hubble-flow expansion and are independent of the mode $k$ in Fourier space. The goal of this paper is to provide a N3LO computation of the primordial power spectrum and its associated power-law quantities for the broad family of effective theories of inflation  (Table~\ref{Tab:Models}) which have a quadratic action of this form. 
\begin{table}[t]
	\caption{\label{Tab:Models}
Effective models of inflation. The functions $Z(t)$ and $c(t)$ are defined by the quadratic action for scalar ($\mathrm{s}$) and tensor ($\mathrm{t}$) perturbations of the form \eqref{eq:Quadratic-Action-psi}. The symbols indicate that the functions reported are of the form considered here (\checked) or not ($\times$). The symbol ``$\dagger$" indicates that the effective theory requires an additional assumption of no parity violation to exclude a dependence on $k$ in the functions $Z(t)$ and $c(t)$.}
	\begin{ruledtabular}
		\begin{tabular}{l c c c c}
			Theory  &$Z_{\mathrm{s}}(t)$ &$c_{\textrm{s}}(t) $ &$Z_{\mathrm{t}}(t)$ &$c_{\textrm{t}} (t)$     \\
			\hline 
			Single-field \cite{Mukhanov1992} &   Eq. \eqref{eq:Z-s-SingleField}  & $1$ & Eq. \eqref{eq:Z-t-SingleField} & $1$ \\
			$R+\alpha R^2$ \cite{DeFelice2010,Nojiri:2010wj,Nojiri:2017ncd} & Eq. \eqref{eq:Z-s-Starobinsky} 
			& $1$ &  Eq. \eqref{eq:Z-t-Starobinsky} & $1$ \\
			$K$-inflation \cite{Garriga1999} & \checked   &\checked  & \checked&$1$ \\
			LQC+inflaton \cite{Agullo2013,Fernandez-Mendez:2012poe} & \checked  &\checked  &\checked  &\checked  \\
			$f(\varphi)$-Gauss Bonnet \cite{Satoh:2008ck}  & \checked & \checked & \checked& \checked \\
				$f(\varphi)$-Chern Simons \cite{Lue:1998mq,Alexander:2009tp} &  \checked&\checked &  $\times$  &  $\times$  \\
			General scalar-tensor \cite{DeFelice2010} & \checked &\checked  &\checked & \checked \\
			Goldston mode EFT \cite{Cheung2007} &  \checked   & \checked& \checked   & \checked  \\
			Multifield EFT \cite{Achucarro2012} &  \checked   & \checked& \checked & \checked  \\
			Minimally broken CFT \cite{Baumann2015}  & \checked  & \checked & $\checked\dagger$   & $\checked\dagger$ \\
			Weinberg's EFT \cite{Weinberg2008} &  \checked & \checked&  $\checked\dagger$ &  $\checked\dagger$ \\
		\end{tabular}
	\end{ruledtabular}
\end{table}

In an exactly de Sitter background, the Mukhanov-Sasaki equation admits an exact solution for the mode functions, which corresponds to the choice of the Bunch-Davies vacuum for the quantum perturbations. The constant Hubble rate $H(t)=H_\ast$ results in a scale-invariant power spectrum $P\sim H_\ast^2$. This leading order (LO) prediction is corrected by a next-to-leading order contribution (NLO) that takes into account the fact that the Hubble rate $H(t)$ decreases slowly, imprinting more power in red than in blue modes---a red tilt $n_s\lesssim 1$. At this order, the approximate equation admits again an exact solution which defines a quasi-Bunch-Davies vacuum with mode functions given by a combination of Bessel functions \cite{Stewart1993}. However, the solution in terms of Bessel functions is not easily extended to higher orders, and various approximation schemes have been developed, including the uniform approximation \cite{Habib2002,Martin2003,Casadio2005,Easther2005,Kinney2007,Lorenz2008,DeFelice2011a}. The next-to-next-to-leading order (N2LO) corrections for scalar perturbations were derived in \cite{Schwarz:2001vv} using the constant-horizon approximation, and in \cite{Stewart2001} as a systematic expansion using the Green's function method, while the N2LO corrections to tensor modes were obtained in \cite{Leach2002}. The fully expanded N3LO corrections for slow-roll inflation were derived in \cite{Auclair2022}, which is the method we adopt and extend here. Motivated by these recent results, we address the problem of finding the contributions to the power spectrum up to N3LO for the broad class of models with perturbations described by the quadratic action  \eqref{eq:Quadratic-Action-psi}. In particular we work out the N3LO predictions of the model of inflation introduced by Starobinsky, motivated by quantum gravity considerations \cite{Starobinsky1979,Starobinsky1980}. Remarkably, this model provides the best account of current observations in terms of a single free parameter, the number of inflationary e-foldings $N_\ast$. While its analysis is generally done via a field redefinition that maps it into an inflaton potential, here we work in the geometric framework where inflation is driven by higher curvature terms. Our explicit N3LO computations show a tensor-to-scalar ratio that is $7\%$ smaller compared to its standard expression.

The manuscript is structured as follows: In Sec.~\ref{sec:generic-setting}, we describe the assumptions, the general framework and the Hubble-flow expansion of the background variables. In Sec.~\ref{sec:Quantization}, we discuss the quantization of perturbations, a choice of variables analogous to Mukhanov-Sasaki variables, and a logarithmic expansion of the Hubble-flow parameters. In Sec.~\ref{sec:Mode-Green}, we find the mode equation satisfied by our dynamical variables and describe the Green's function method introduced in \cite{Stewart2001,Auclair2022}. In Sec.~\ref{sec:Final-Expressions}, we report the final expressions for the power spectrum, which takes the schematic form,
\begin{equation}
\mathcal{P}_0^{(\psi)}(k) =  \frac{\hbar H_\ast^2}{4\pi^2 c_\ast^3 Z_\ast} \left[ 1 + p_1(k) + p_2 (k) + p_3(k) \right],
\end{equation}
with $H$, $Z$, and $c$ evaluated at a pivot scale $k_\ast$, as described in \eqref{eq:PowerSpectrum_Full}. The functions $p_{n}(k)$ have a logarithmic dependence, i.e., include powers of $\ln(k/k_\ast)$, and represent the N$n$LO correction to the power spectrum. The explicit form of these functions is reported in Tables~\ref{Tab:p0s}-\ref{Tab:p3s}. From the full power, we can also extract the predictions for the amplitude, tilt, running of the tilt, and running-of-the-running of the tilt, which are reported in Tables~\ref{Tab:GenericTheory-Features} and ~\ref{Tab:GenericTheory-Amplitude}. In Sec.~\ref{sec:SingleField}, we discuss our results in the context of single-field inflation. In Sec.~\ref{sec:Starobinsky} , we analyze Starobinsky inflation and compute the N3LO corrections as an expansion in the single parameter $N_\ast$, reported in Table~\ref{Tab:Starobinsky-Results}. Finally, in Sec.~\ref{sec:discussion}, we discuss the results obtained in this work and outline possible extensions. 

Throughout the paper, we adopt units with the speed of light $\mathrm{c}=1$, while we keep track of Planck's constant $\hbar$ and Newton's gravitational constant $G$. The metric signature is $(-+++)$, a derivative with respect to cosmic time is denoted by  $(\dot{\phantom{a}})$, and a derivative with respect to other variables by $(\phantom{1})^\prime$. Complex conjugation is denoted by $(\phantom{1})^\ast$, and evaluation at a pivot scale by $(\phantom{1})_\ast$. 

\section{Action and perturbations}
\label{sec:generic-setting}

\subsection{Quadratic action}
We consider an inflationary background geometry
given by the spatially-flat Friedman-Lema\^itre-Robertson-Walker (FLRW) metric
\begin{equation}
	\label{eq:FLRW-Metric}
	\bar{g}_{\mu\nu} \dd{x}^\mu \dd{x}^\nu =- \dd{t}^2 + a(t)^2\, \delta_{ij} \dd{x}^i \dd{x}^j\,,
\end{equation}
which, together with other homogeneous and isotropic fields, satisfies the background equations of motion of an inflationary theory of gravity and matter. Because of the symmetry of the background, perturbations of the geometry and of matter fields decompose in scalar, vector, and tensor (SVT) modes. Once one fixes a gauge and solves the Hamiltonian and diffeomorphism constraints perturbatively, the action of perturbations decouples at quadratic order and takes the general form
\begin{widetext}
	\begin{equation}
		\label{eq:Quadratic-Action-psi}
		S_{2}[\psi] =  \int \dd{t} \int \frac{ \dd[3]{\vb{k}} }{(2\pi)^3}\, a(t)^3 Z_\psi(t) \left[ \frac{1}{2} \big|\dot{\psi}(\vb{k},t)\big|^2 - \frac{1}{2} c_\psi(t)^2\,  \frac{k^2}{a(t)^2} \abs{\psi(\vb{k},t)}^2 \right],
	\end{equation}
	\end{widetext}
where $k=\abs{\vb{k}}$, and we used the generic name $\psi( \vb{k},t )$ for the Fourier transform of each of the SVT modes $\Psi(\vb{x},t)$,
\begin{equation}
\Psi(\vb{x},t) = \int \frac{\dd[3]{\vb{k}} }{(2\pi)^3}\,  \psi( \vb{k},t )\, \ee^{\ii  \vb{k}\cdot \vb{x}}.
\label{eq:Fourier}
\end{equation}
Together with the scale factor $a(t)$, the quadratic action encodes the coupling of the SVT modes to the background via two functions of time $t$: the kinetic amplitude, $Z_\psi(t)$, and the speed of sound, $c_\psi(t)$. In a given model, their time dependence can be expressed in terms of time-derivatives of the Hubble rate $H(t) = \dot{a}(t)/a(t)$, but here we treat them as independent as it allows us to derive general results. We make the assumption that they do not depend on the mode $k$ (which excludes some models of inflation), and require both a no-ghost condition, $Z_\psi(t)>0$, and a no-Laplacian-instability condition, $c_\psi(t)^2 >0$. 

The well-studied case of single-field inflation corresponds to the two functions being constant for tensor modes, while scalar curvature perturbations have a constant speed of sound and a time-dependent kinetic amplitude $Z_s(t)$ proportional to the slow-roll parameter $-\dot{H}(t)/H(t)^2$. In the case of Starobinsky inflation treated in the geometric framework, $Z_{\mathrm{s}}(t)$ and $Z_{\mathrm{t}}(t)$ depend nontrivially on higher time-derivatives of the Hubble rate both for scalar and for tensor modes, but the speed of sound is still constant for both. 

The effective field theory of single-field \cite{Cheung2007} and multifield \cite{Achucarro2012} inflation has a quadratic action of this form, with a nontrivial speed of sound $c_\psi(t)$ that needs to be determined via observations, and a nontrivial kinetic amplitude $Z_\psi(t)$ that depends on the slow-roll parameter and on the speed of sound.

In loop quantum cosmology, quantum geometry effects modify the Mukhanov-Sasaki equation which, in a self-consistent approximation, can also be cast in the form \eqref{eq:Quadratic-Action-psi} \cite{Agullo2013,Fernandez-Mendez:2012poe}.

In models of $K$-inflation \cite{Garriga1999}, the action of the inflaton field includes higher derivative terms of the form
\begin{equation}
S_{K\mathrm{infl}}=\int\! \dd[4]{x} \sqrt{-g}\, f_K(\varphi)\,\big(g^{\mu\nu}\partial_\mu\varphi\partial_\nu \varphi\big)^2\,,
\label{Kinfl}
\end{equation}
which result again in an action for scalar perturbations of the form \eqref{eq:Quadratic-Action-psi} with a nontrivial speed of sound. Similarly, models that include a coupling $f_{\textrm{GB}}(\varphi) \mathcal{L}_{\textrm{GB}}$ to the Gauss-Bonnet density \cite{Satoh:2008ck}, 
 \begin{equation}
 \mathcal{L}_{\textrm{GB}}=R_{\mu\nu\rho\sigma}R^{\mu\nu\rho\sigma}\!-\!4R_{\mu\nu}R^{\mu\nu}\!+\!R^2\,,\label{L_GB}
 \end{equation}
 can be cast in the form \eqref{eq:Quadratic-Action-psi} for scalar pertubations. On the other hand, in models with a coupling $f_{\textrm{CS}}(\varphi) \mathcal{L}_{\textrm{CS}}$ to the Chern-Simons density, 
 \begin{equation}
 \mathcal{L}_{\textrm{CS}}=-(\sqrt{-g})^{-1}\,\epsilon^{\mu\nu\rho\sigma}\,R^\alpha{}_{\beta\mu\nu}\,R^\beta{}_{\alpha\rho\sigma}\,,
 \label{L_CS}
 \end{equation}
tensor perturbations with circular polarization $(\pm)$ have a kinetic term $Z_\pm(t,k)$ and speed of sound $c_\pm(t,k)$ which depend linearly on $\pm k$, resulting in gravitational parity violation  \cite{Lue:1998mq,Alexander:2009tp}. Therefore, they cannot be cast in the form \eqref{eq:Quadratic-Action-psi} assumed here because of the dependence on $k$.

Weinberg's formulation of an effective field theory of single-field inflation starts from the most general action $S[g_{\mu\nu},\varphi]$ which includes all diffeomorphism invariant terms, organized in an order-by-order expansion in the number of spacetime derivatives, up to fourth order \cite{Weinberg2008}. In particular, it includes a coupling of the inflaton field $\varphi$ to quadratic terms in the Weyl tensor, $C_{\mu\nu\rho\sigma}C^{\mu\nu\rho\sigma}$ and $(\sqrt{-g})^{-1}\epsilon^{\mu\nu\rho\sigma}\,C^\alpha{}_{\beta\mu\nu}\,C^\beta{}_{\alpha\rho\sigma}$, and a term of the $K$-inflation type \eqref{Kinfl}. Up to field redefinitions, and using reduction of order of the four-derivative terms with respect to the (two-derivative) Einstein equations of motion
\begin{equation}
R_{\mu\nu}=8\pi G \;(T_{\mu\nu}-\tfrac{1}{2}g_{\mu\nu}T)\,,
\end{equation}
one can reabsorb terms quadratic in the Ricci tensor into a $K$-inflation term. Therefore, the effective action takes the form of single-field inflation with an inflaton potential $V(\varphi)$, a $K$-inflation term  \eqref{Kinfl}, a GB-coupling \eqref{L_GB}, and a CS-coupling \eqref{L_CS}. If we further assume that the theory is invariant under all diffeomorphisms, including diffeomorphisms not connected with the identity such as orientation reversals, then the parity-violating coupling to $\mathcal{L}_{\mathrm{CS}}$ has to vanish, and the quadratic action for scalar and tensor perturbation takes once more the form \eqref{eq:Quadratic-Action-psi}, with the kinetic amplitude and the speed of sound expressed in terms of time-derivatives of the Hubble rate $H(t)$, of the background inflaton field $\bar{\varphi}(t)$, and of the couplings $V(\bar{\varphi}(t))$, $f_K(\bar{\varphi}(t))$, and $f_{\mathrm{GB}}(\bar{\varphi}(t))$.

In summary, the formalism that we develop in this work applies to the broad class of effective theories of inflation in which quadratic action can be cast in the form of \eqref{eq:Quadratic-Action-psi} and, among others, include most of the models discussed above, as shown in Table~\ref{Tab:Models}.

\subsection{Hubble-flow expansion}
The variational principle for the quadratic action \eqref{eq:Quadratic-Action-psi} results in equations of motion for the SVT perturbations that depend on the background functions $a(t)$, $H(t)$, $Z_\psi(t)$, $c_\psi(t)$ and their time derivatives. In the simplest models, the mechanism that produces a frozen power spectrum for tensor modes is a transition in the mode equation $\ddot{u}(t)+3 H(t) \,\dot{u}(t)+(k/a(t))^2\, u(t)=0$ from an oscillatory phase with small friction $H(t)\ll k/a(t)$, to an overdamped phase with large friction $H(t)\gg k/a(t)$. For a given mode $k$, the transition requires an increasing comoving scale $a(t) H(t)$, i.e., an accelerating phase
\begin{equation}
0<\ddot{a} = \dv{}{t} \, \big( a(t) H(t)  \big) = \big(1-\epsilon_{1H}(t) \big)\, a(t)H(t)^2\,,
\label{eq:ddot-a}
\end{equation}
or equivalently a Hubble-flow parameter $\epsilon_{1H}(t)<1$ with
\begin{equation}
\epsilon_{1H}(t) \equiv -\frac{\dot{H}(t)}{H(t)^2}\,.
\end{equation}
During a de Sitter phase, $a_0(t)= \ee^{H_0 t}$, the Hubble rate is constant and therefore the Hubble-flow parameter vanishes exactly. In a quasi-de Sitter phase, the Hubble rate is assumed to change slowly and it is useful to introduce a systematic expansion, called \textit{Hubble-flow expansion}, defined recursively in terms of the dimensionless parameters $\epsilon_{nH}$, with
\begin{equation}
 \epsilon_{(n+1)H}(t) \equiv - \frac{ \dot{ \epsilon}_{nH} (t) }{ H(t)  \,\epsilon_{nH}(t)}\,.
\end{equation}
A phase of quasi-de Sitter inflation is characterized by $\abs{\epsilon_{nH}} \ll 1$. Different conventions have been used in the literature for the Hubble-flow expansion, and to avoid confusion a dictionary is provided in Table \ref{Tab:dictionary}. Generally speaking, our definition has a relative sign of $\epsilon_{nH}$, for $n\ge 2$, with respect to the definition used in \cite{Auclair2022}. Similarly, we introduce new Hubble-flow parameters for the kinetic amplitude $Z_\psi(t)$,
\begin{align}
	\epsilon_{1Z}(t)\, &\equiv-\frac{\dot{Z}_\psi( t)}{H(t) Z_\psi (t)}\,,\\
	\epsilon_{(n+1)Z}(t)\, &\equiv - \frac{ \dot{ \epsilon}_{nZ} (t) }{ H(t)  \,\epsilon_{nZ}(t)}\,,
\end{align}
and for the speed of sound $c_\psi(t)$,
\begin{align}
		\epsilon_{1c}(t) \,& \equiv -\frac{\dot{c}_\psi( t)}{H(t) c_\psi (t)}\,,\\
		\epsilon_{(n+1)c}(t) \,& \equiv - \frac{ \dot{ \epsilon}_{nc} (t) }{ H(t)  \,\epsilon_{nc}(t)}\,.
\end{align}
All the flow parameter are assumed to be of the same order, e.g., $\order{\epsilon_{nH}} = \order{\epsilon_{n'Z}} = \order{\epsilon_{n''c}}$. To simplify the notation, we use a placeholder $\epsilon$ to track the corresponding order of the expansion, i.e., $\order{\epsilon_{1H}} = \order{\epsilon}$, $\order{\epsilon_{1Z}^2} = \order{\epsilon^2}$, and so on. In a specific model of inflation, this assumption is to be checked \textit{a posteriori}. In this work, we consider corrections up to order $\epsilon^3$. 

\begin{table}[t]
	\caption{Dictionary of definitions for the first Hubble-flow parameters. Compare also with the conversion table in \cite{Schwarz:2001vv}.  }
	\label{Tab:dictionary}
	%	\rowcolors{2}{gray!25}{white}
	\begin{ruledtabular}
		\begin{tabular}{c c c}
			This work & Stewart and Gong \cite{Stewart2001} & Auclair and Ringeval \cite{Auclair2022}  \\
			\hline
			$\epsilon_{1H}$ & $\epsilon_{1}$ & $\epsilon_{1}$ \\
			\hline
			$\epsilon_{2H}$ & $-2\epsilon_1 - 2 \delta_1$ & $-\epsilon_{2}$ \\
            \hline
            $\epsilon_{2H}\epsilon_{3H}$ & $\epsilon_1^2 + 6 \epsilon_1 \delta_1 - 2\delta_1^2 + 2\delta_2$ & $\epsilon_{2}\epsilon_{3}$ \\
		\end{tabular}
	\end{ruledtabular}
\end{table}

\section{Quantization of perturbations}
\label{sec:Quantization}

\subsection{Field representation and mode expansion}
Inflation explains the anisotropies in the CMB temperature and the seeds of large-scale structures in terms of vacuum fluctuations that are assumed to be homogeoneous and isotropic. Specifically, the SVT fields $\hat{\Psi} (\vb{x},t)$ are assumed to be initially in a vacuum state $|0 \rangle$ with a two-point correlation function that, at an equal time $t$, is invariant under rotations and translations,

%\newpage

\begin{equation}
\bra{0} \hat{\Psi} (\vb{x},t) \hat{\Psi} (\vb{x}',t) \ket{0}\;=\; G\big(|\vb{x}-\vb{x}'|,t\big)\,.
\label{2-point-G}
\end{equation}
Neglecting interactions and non-Gaussianities, this condition can be encoded in a Gaussian vacuum state in Fock space, defined by the condition $\hat{a}(\vb{k}) \ket{0} = 0\,,\;\forall \vb{k}$, with bosonic creation and annihilation operators, $[\hat{a}(\vb{k}),\hat{a}^\dagger(\vb{k^\prime})] = (2\pi)^3 \delta^{(3)} (\vb{k}-\vb{k}^\prime)$, together with a mode expansion of the quantum field
\begin{equation}
\hat{\Psi} (\vb{x},t) =\! \int\!\frac{\dd[3]{\vb{k}}}{(2\pi)^3} \left( u(k,t) \hat{a}(\vb{k}) + u^\ast (k,t) \hat{a}^\dagger (-\vb{k}) \right) \ee^{i \vb{k}\cdot \vb{x}}.
\label{eq:psi-modes}
\end{equation}
The assumption of homogeneity and isotropy \eqref{2-point-G} implies that the modes $u(k,t)$ depend only on $k=|\vb{k}|$. Moreover, at the classical level, the conjugate momentum derived from the action  \eqref{eq:Quadratic-Action-psi} is defined by
\begin{equation}
	\label{eq:Conjugate-Momentum-pi}
	\pi(\vb{k},t) \equiv \fdv{S_{2}[\psi]}{\dot{\psi}(\vb{k},t)} = \frac{a(t)^3 Z_\psi(t)}{(2\pi)^3} \dot{\psi}(-\vb{k},t)\,,
\end{equation}
and denoted $\Pi(\vb{x},t)$ in position space. Canonical quantization of the Poisson brackets results in the canonical commutation relation (CCR),  $[\hat{\Psi}(\vb{x},t),\hat{\Pi} (\vb{x}', t)] = \ii\, \hbar \,\delta^{(3)} (\vb{x}-\vb{x}')$, which, together with \eqref{eq:psi-modes}, implies the canonical Wronskian condition for the mode functions
\begin{equation}
	\label{eq:CCR-u(t)}
u(k,t) \dot{u}^\ast (k,t) - \dot{u}(k,t) u^\ast(k,t) = \frac{\ii \,\hbar}{a(t)^3\, Z_\psi (t)},
\end{equation}
where $u^\ast$ is the complex conjugate of $u$. Finally, the equations of motion (EoM) derived from the action \eqref{eq:Quadratic-Action-psi} for each of the SVT modes results in the mode equation
\begin{equation}
	\label{eq:EoM-u(t)}
	\ddot{u}(k,t) + (3-\epsilon_{Z1}(t)) H(t) \dot{u}(k,t) + c_\psi(t)^2\frac{k^2}{a(t)^2}  \, u(k,t) = 0\,.
\end{equation} 
A choice of initial conditions $u(k,t_0), \dot{u}(k,t_0)$ for the EoM \eqref{eq:EoM-u(t)}, satisfying the CCR constraint \eqref{eq:CCR-u(t)}, or equivalently a choice of solution of the two equations, defines a Gaussian state $|0 \rangle$ that is homogeneous and isotropic, \eqref{2-point-G}. In the next section, we discuss how to select an adiabatic solution of \eqref{eq:CCR-u(t)} and \eqref{eq:EoM-u(t)} that generalizes the Bunch-Davies vacuum to quasi-de Sitter space, order-by-order in the Hubble-flow expansion.

\subsection{Mukhanov-Sasaki variables}
In de Sitter space, the Hubble rate is constant, $a_0(t)= \ee^{H_0 t}$, and in the simplest models of a test quantum field, the EoM reduces to $\ddot{u}+3 H_0 \,\dot{u}+(k/a_0(t))^2\, u=0$ and the CCR equation to $u\, \dot{u}^\ast  - \dot{u}\, u^\ast = \ii \,\hbar/a_0(t)^3$. Using a reparametrization of time $t\to \eta$ and a change of variables $u\to w$, 
\begin{equation}
\eta= -\frac{1}{a_0(t)\, H_0}\,,\quad w(\eta) =\sqrt{a_0(t)^2/\hbar\,}\: u(t)\,,
\label{eq:eta-conformal}
\end{equation}
the EOM and CCR take the simpler form
\begin{align}
w''(\eta)+\bigg(k^2-\frac{2}{\eta^2}\bigg)\,w(\eta)=&\;0\,,\label{eq:w-EoM}\\[.5em]
w(\eta)\,w'{}^\ast(\eta)-w'(\eta)\,w^\ast(\eta)=&\;\ii\,.
\end{align}
These two equations admit a basis of linearly independent solutions $\big(w(\eta),\,w^\ast(\eta)\big)$, with
\begin{equation}
w(\eta)=\frac{1}{\sqrt{2\, k}}\bigg(1-\frac{\ii}{k\,\eta}\bigg)\:\ee^{-\ii k\,\eta}\,,
\label{eq:w-BD}
\end{equation}
which defines the Bunch-Davies vacuum, i.e., the de Sitter invariant state $|0 \rangle$ for the test quantum field \cite{Bunch:1978yq}. In quasi-de Sitter space, one can follow a similar strategy as first done by Mukhanov and Sasaki \cite{Mukhanov:1985rz,Sasaki:1986hm}, with the conformal time $\eta$ and a choice of pivot scale $k_\ast$ defined by the horizon-crossing condition $k_\ast\equiv a(t_\ast) H(t_\ast)$ or, equivalently, $k_\ast \eta_\ast=-1$, which characterizes the transition from the oscillatory to the frozen overdamped phase discussed earlier. Mathematically, the first step of this strategy relies on recasting a second-order linear differential equation with time-dependent coefficients, $v''(x) +  f_1(x) v'(x) + f_2(x) v(x) = 0$, in the standard form $w''(\eta)+Q(\eta)\,w(\eta)=0$,
where $Q(\eta)$ is made as nearly free from poles and branch points as is conveniently possible,
by changing both independent and dependent variables \cite{Dingle74,White2005}. 

In this section, we address this problem for the Eqs. \eqref{eq:CCR-u(t)} and \eqref{eq:EoM-u(t)}, extending the analysis of \cite{Stewart2001,Auclair2022}. We start by noticing that the EoM for $u$ includes a term proportional to $\dot{u}$ that we can remove via a change to a new variable $v(x)$, which also includes a time reparametrization $t\to x(t)$ to be determined. Starting from the ansatz
\begin{equation}
u(t) = \frac{v(x(t))}{\sqrt{\mu(x(t))}},
\end{equation}
the EoM takes the form
\begin{equation}
v''(x) +  f_1(x) v'(x) + f_2(x) v(x) = 0\,.
\end{equation}
The function $f_1(x)$ is given by
\begin{equation}
	f_1(x) =\frac{\ddot{x} }{\dot{x}^2}  + \Big(3 H+ \frac{\dot{Z}_\psi}{Z_\psi} - \frac{\dot{\mu}}{\mu}\Big) \frac{\dot{x} }{\dot{x}^2}\,,
\end{equation}
and $f_2(x)$ reads
\begin{equation}
	f_2(x)= \frac{k^2 c_\psi^2}{a^2 \dot{x}^2} + \frac{  3 \dot{\mu}^2 - 2 \mu \ddot{\mu} + 2 H \mu \dot{\mu} (\epsilon_{Z1}-3) }{4 \mu^2 \dot{x}^2}				.
\end{equation}
By imposing the condition $f_1(x) =0$, we find $\mu(x(t)) =  \mu_0  \, a(t)^3 \, Z_\psi(t)\, \dot{x}/\hbar$, where $\mu_0$ is an integration constant that can be determined to be $\mu_0 = 1$ by imposing the canonical normalization of the CCR,
\begin{align}
\label{eq:CCR-v(x)}
v(x) \,v'{}^\ast(x) - v'(x)\, v^\ast(x) = \ii .
\end{align}
The next step is to impose that a term of the form $ \big((k/k_\ast)^2 - \ldots \big)v(x)$, where $k_\ast$ is a pivot scale later defined in terms of a horizon-crossing time. Imposing this condition on $f_2(x)$, we find the equation 
\begin{equation}
	\label{eq:xdot}
	\dot{x} = - k_\ast \frac{c_\psi(t)}{a(t)},
\end{equation}
which defines the change of time variables $t\to x$. Note that the time variable $x$ is not the conformal time $\eta$, i.e., \eqref{eq:eta-conformal}, but a generalized version that we denote by $\tau$, as a consequence of the time-dependent speed of sound $c_\psi(t)$. With this identification, the map between $u(t)$ and $v(x)$ is fully characterized,
\begin{equation}
	\label{eq:Map-u(t)-v(x)}
	u(t) \to  \frac{v(x)}{\sqrt{\mu(x)}} = \frac{v(x)}{\sqrt{k_\ast\, a^2\, c_\psi\, Z_\psi/\hbar }}\,.
\end{equation}
Now, notice that the EoM \eqref{eq:EoM-u(t)} becomes
\begin{equation}
	\label{eq:EoM-v(x)-aH}
v''(x) +  \left[   \frac{k^2}{k_\ast^2} + \frac{a(t)^2 H(t)^2}{k_\ast^2 c_\psi^2(t)} q(t) \right] v(x)  = 0,
\end{equation}
where
\begin{align}
q(t) & = -2 + \epsilon_{H1}(t)  + \frac{3}{2} \epsilon_{Z1}(t) + \frac{\epsilon_{1c}(t)}{2}  \nonumber \\
&\quad -\frac{\epsilon_{1H} (t) \epsilon_{1Z} (t)}{2}  - \frac{\epsilon_{1Z} (t)^2 }{4} - \frac{ \epsilon_{1Z} (t) \epsilon_{2Z} (t)   }{2} \nonumber \\
&\quad - \frac{ \epsilon_{1c} (t) \epsilon_{1H} (t)   }{2}-
\frac{ \epsilon_{1c}(t) \epsilon_{2c}(t)}{2}      + \frac{\epsilon_{1c}^2(t)}{4}  .
\end{align}
This equation is exact in the flow parameters $\epsilon_{1H}(t)$, $\epsilon_{1Z}(t)$, $\epsilon_{1c}(t)$, etc. The last step is to write $t$ in terms of the time $x$ by solving \eqref{eq:xdot}. This can be done self-consistently in a Hubble-flow expansion as discussed in Appendix~\ref{app:conformaltime}, where we find
\begin{equation}
\label{eq:x-in-terms-of-aH}
x (t) = - k_\ast \tau(t) =  \frac{k_\ast \tilde{c}_\psi(t)}{a(t)H(t)} \, ,
\end{equation}
with $\tilde{c}_\psi(t)$ defined as
\begin{widetext}
\begin{align}
	&\tilde{c}_\psi(t) \equiv  c_\psi(t) \Big[  1 + \epsilon_{1H}(t)  - \epsilon_{1c}(t)  + \epsilon_{1H}(t)^2 - \epsilon_{1H}(t) \epsilon_{2H}(t)  - 2 \epsilon_{1c}(t) \epsilon_{1H}(t)  + \epsilon_{1c}(t)\epsilon_{2c}(t) + \epsilon_{1c}(t)^2   \nonumber \\
	&\;\; 
+\epsilon_{1H}(t)^3 + \epsilon_{1H}(t) \epsilon_{2H}(t) \epsilon_{3H}(t)  - 
3 \epsilon_{1H}(t)^2 \epsilon_{2H}(t) + \epsilon_{1H}(t) \epsilon_{2H}(t)^2 - \epsilon_{1c}(t) \epsilon_{2c}(t)^2 + 
3 \epsilon_{1c}(t) \epsilon_{1H}(t) \epsilon_{2H}(t)  \nonumber \\
	&\;\; - 
	3 \epsilon_{1c}(t)^2 \epsilon_{2c}(t)  +  
	3 \epsilon_{1c}(t) \epsilon_{1H}(t) \epsilon_{2c}(t)  - 
	\epsilon_{1c}(t) \epsilon_{2c}(t) \epsilon_{3c}(t) 	-\epsilon_{1c}(t)^3 + 3 \epsilon_{1c}(t)^2 \epsilon_{1H}(t) - 
	3 \epsilon_{1c}(t) \epsilon_{1H}(t)^2 \;+\mathcal{O}(\epsilon^4)  \Big]\,.\label{eq:c-tilde-expansion}
\end{align}
	\end{widetext}
One can use \eqref{eq:x-in-terms-of-aH} whenever we need to write $a(t)$ exclusively in terms of $x$, order-by-order. The next step is to write each flow parameter in terms of the new time variable $x$, which can be done via a logarithmic expansion around the value $x_\ast\equiv 1$. The time $t_\ast$ with $x_\ast=x(t_\ast)=1$, encodes a ``generalized horizon-crossing'' condition in the expansion \eqref{eq:x-in-terms-of-aH}, i.e., $\tilde{c}_\ast\, k_\ast = a_\ast \,H_\ast$. Note that the standard ``horizon-crossing'' condition is only recovered when $\tilde{c}_\ast = 1$, i.e., in exact de Sitter and with $c_\psi = 1$. In quasi-de Sitter, where the Hubble-flow parameters are generically non-zero, even if $c_\psi=1$ there will be contributions to $\tilde{c}_{\psi}$. However, in that scenario the contributions come only from $\epsilon_{1H}, \epsilon_{2H}$ and $\epsilon_{3H}$, i.e., the background geometry. Then, the expansion around a particular pivot scale $k_\ast$ will be the same for scalar and tensor modes. A different situation occurs if $c_{\psi}\neq 1$, where the additional contributions make the comparison of two different SVT modes at the same pivot scale, as in the case of the tensor-to-scalar ratio $r$, not immediate. In Appendix ~\ref{app:comparing-pivot-scales}, we describe a procedure to expand the power spectra of two different SVT modes, with speeds of sound $c^{(A)}\neq c^{(B)}$, around the same pivot scale in such a way that both spectra can be consistently compared with each other.

\subsection{Logarithmic expansion}
To illustrate the procedure, let us consider an arbitrary function $\rho(t)$, such as $H(t)$, $Z_\psi(t)$, or $c_\psi(t)$. For definiteness, suppose that this function is smooth and only depends on time. The goal is to write it explicitly in terms of the new time variable $x$, i.e., $\rho(x)$, as a perturbative expansion around a pivot value $x_\ast = x(t_\ast) = 1$, i.e., $k_\ast \tau_\ast = -1$. In a quasi-de Sitter phase of cosmic inflation, these geometric functions are assumed to be slowly varying with respect to time $x$. Thus, a logarithmic expansion is appropriate in this context. In other words, we look for an expression of the form
\begin{align}
\label{eq:general-log-expansion}
		\rho(x) &= \rho_\ast + \rho_{1\ast} \ln(x) + \rho_{2\ast} \ln(x)^2 \nonumber \\
		&\quad\quad \;+ \rho_{3\ast} \ln(x )^3 + \order{\epsilon^4},
\end{align}
where the expansion is always understood to be around $x_\ast=1$, the first term $\rho_\ast = \rho(x_\ast)$, and the rest of coefficients are given by $	\rho_{n\ast}= (n!)^{-1} \eval{\dv*[n]{\rho(x)}{\ln(x)}}_{x\to x_\ast}$, $n=1,2,3$. 
These coefficients are given by expressions that can be recursively expanded by using \eqref{eq:xdot} and its derivatives and then replacing \eqref{eq:x-in-terms-of-aH} in combination with the flow expansion. In this way, the \textit{logarithmic expansion} is extended to all the relevant variables, including the flow parameters. The generic expressions for $\rho = H, Z_\psi, c_\psi$ and the associated flow parameters are given by \eqref{eq:LogExpansion},
\begin{widetext}
	\begin{align}
			\label{eq:LogExpansion}
		\rho (x)
			&= \rho_\ast \left[ 1 + \Big(\epsilon_{1\rho\ast}  + \epsilon_{1\rho\ast} ( \epsilon_{1H\ast}-\epsilon_{1c\ast} )  + \epsilon_{1\rho\ast} [\epsilon_{1H\ast} (\epsilon_{1H\ast} - \epsilon_{2H\ast}) + \epsilon_{1c\ast} (-2 \epsilon_{1H\ast} + \epsilon_{2c\ast}) +\epsilon_{1c\ast}^2 ] \Big) \ln\left(x\right) \right.\nonumber  \\
			&\quad + \frac{1}{2}\,\Big( \epsilon_{1\rho\ast} (\epsilon_{1\rho\ast} + \epsilon_{2\rho\ast})  + \epsilon_{1\rho\ast} [-\epsilon_{1c\ast} (2 \epsilon_{1\rho\ast} + \epsilon_{2c\ast} + 2 \epsilon_{2\rho\ast}) + \epsilon_{1H\ast} (2 \epsilon_{1\rho\ast} + \epsilon_{2H\ast} + 2 \epsilon_{2\rho\ast})] \Big) \ln\left(x\right)^2 \nonumber \\
			&\quad \left. + \frac{1}{6} \, \epsilon_{1\rho\ast} \Big(\epsilon_{1\rho\ast}^2 + 3 \epsilon_{1\rho\ast} \epsilon_{2\rho\ast} + \epsilon_{2\rho\ast} (\epsilon_{2\rho\ast} + \epsilon_{3\rho\ast})\Big) \ln\left(x\right)^3 \right] + \order{\epsilon^4}\,, \nonumber 
		\end{align}
		\begin{align}	
			\epsilon_{1\rho}(x) 
				&=\epsilon_{1\rho\ast}  + \Big(\epsilon_{1\rho\ast} \epsilon_{2\rho\ast} + (-\epsilon_{1c\ast} + \epsilon_{1H\ast}) \epsilon_{1\rho\ast} \epsilon_{2\rho\ast} \Big) \ln\left(x\right) + \Big( \frac{1}{2} \epsilon_{1\rho\ast} \epsilon_{2\rho\ast} (\epsilon_{2\rho\ast} + \epsilon_{3\rho\ast})  \Big) \ln\left(x\right)^2 + \order{\epsilon^4}\,, \nonumber \\
				\epsilon_{2\rho}(x) &= \epsilon_{2\rho\ast} + \epsilon_{2\rho\ast} \epsilon_{3\rho\ast} \ln\left(x\right) + \order{\epsilon^3}\,, \nonumber \\
				\epsilon_{3\rho}(x) &= \epsilon_{3\rho\ast} + \order{\epsilon^2}\,.
	\end{align}
	\end{widetext}
In this way, for instance, the Hubble rate $H(x)$ in a neighborhood of $x_\ast$ can be  completely written in terms of $\ln(x)^n$ and other coefficients evaluated at the pivot time, e.g., $\epsilon_{1H\ast} = \epsilon_{1H}(t_\ast)$,  $\epsilon_{1c\ast} = \epsilon_{1c}(t_\ast)$, etc.

\section{Green's function method}
\label{sec:Mode-Green}

The logarithmic expansion is a crucial tool to write the final expression of the equation of motion that describes the dynamics of the mode functions $v(x)$ and, therefore $u(k,t)$, during the inflationary epoch. Using the expansion \eqref{eq:x-in-terms-of-aH} in Eq. \eqref{eq:EoM-v(x)-aH}, the EoM for the modes becomes
\begin{align}
	\label{eq:EoM-v(x)-Final}
	v''(x) +  \left[ \left(\frac{k}{k_\ast}\right)^2 - \frac{2}{x^2} \right] v(x) =  \frac{g(x)}{x^2} v(x) ,
\end{align}
where $g(x) \equiv g_{1\ast}  +  g_{2\ast} \ln(x) + g_{3\ast} \ln(x)^2$, and the coefficients $g_{1\ast}$, $g_{2\ast}$, and $g_{3\ast}$ are given by
\begin{widetext}
	\begin{align}
 \label{eq:g's-full-expression}
		g_{1\ast} &=3 \epsilon_{H1\ast} - \frac{3  \epsilon_{Z1\ast} }{2} - \frac{9  \epsilon_{c1\ast} }{2}  \nonumber \\
		&\qquad +  4 \epsilon_{1H\ast}^2 - \frac{5 \epsilon_{1H\ast}   \epsilon_{1Z\ast}}{2} + \frac{ \epsilon_{1Z\ast}^2}{4} 
		- 4  \epsilon_{1H\ast}  \epsilon_{2H\ast} + \frac{ \epsilon_{1Z\ast}  \epsilon_{2Z\ast}}{2} - \frac{21  \epsilon_{1H\ast}  \epsilon_{1c\ast}}{2} + 3  \epsilon_{1Z\ast}  \epsilon_{1c\ast} + \frac{27  \epsilon_{1c\ast}^2}{4} + \frac{9}{2}  \epsilon_{1c\ast}  \epsilon_{2c\ast}		\nonumber \\
		& \qquad  \quad +5 \epsilon_{1H\ast}^3 + 4 \epsilon_{1H\ast} \epsilon_{2H\ast} \epsilon_{3H\ast}	- 14 \epsilon_{1H\ast}^2 \epsilon_{2H\ast}  + 4 \epsilon_{1H\ast} \epsilon_{2H\ast}^2	 - \frac{7}{2} \epsilon_{1H\ast}^2 \epsilon_{1Z\ast} + \frac{1}{2} \epsilon_{1H\ast} \epsilon_{1Z\ast}^2 + 3 \epsilon_{1H\ast} \epsilon_{1Z\ast} \epsilon_{2H\ast} \nonumber \\
		& \qquad \quad + \epsilon_{1H\ast} \epsilon_{1Z\ast} \epsilon_{2Z\ast}- \frac{1}{2} \epsilon_{1c\ast} \epsilon_{1Z\ast}^2 - 18 \epsilon_{1c\ast}^2 \epsilon_{2c\ast} + 15 \epsilon_{1c\ast} \epsilon_{1H\ast} \epsilon_{2c\ast} - 3 \epsilon_{1c\ast} \epsilon_{1Z\ast} \epsilon_{2c\ast} - 4 \epsilon_{1c\ast} \epsilon_{2c\ast}^2 + 17 \epsilon_{1c\ast} \epsilon_{1H\ast} \epsilon_{2H\ast}  \nonumber \\
		& \qquad   \quad  - \epsilon_{1c\ast} \epsilon_{1Z\ast} \epsilon_{2Z\ast}  - 4 \epsilon_{1c\ast} \epsilon_{2c\ast} \epsilon_{3c\ast}  -9 \epsilon_{1c\ast}^3 + \frac{45}{2} \epsilon_{1c\ast}^2 \epsilon_{1H\ast} - \frac{37}{2} \epsilon_{1c\ast} \epsilon_{1H\ast}^2  - \frac{9}{2} \epsilon_{1c\ast}^2 \epsilon_{1Z\ast} + 8 \epsilon_{1c\ast} \epsilon_{1H\ast} \epsilon_{1Z\ast}\,, \nonumber \\
		g_{2\ast} &= 3  \epsilon_{1H\ast}  \epsilon_{2H\ast} - \frac{3}{2}  \epsilon_{1Z\ast}  \epsilon_{2Z\ast} - \frac{9}{2}  \epsilon_{1c\ast}  \epsilon_{2c\ast} \nonumber \\
		& \qquad 
			- 4 \epsilon_{1H\ast} \epsilon_{2H\ast} \epsilon_{3H\ast}+ 11 \epsilon_{1H\ast}^2 \epsilon_{2H\ast}  - 4 \epsilon_{1H\ast} \epsilon_{2H\ast}^2 + \frac{1}{2} \epsilon_{1Z\ast} \epsilon_{2Z\ast} \epsilon_{3Z\ast} - \frac{5}{2}  \epsilon_{1H\ast} \epsilon_{1Z\ast} \epsilon_{2H\ast}  + \frac{9}{2} \epsilon_{1Z\ast} \epsilon_{2Z\ast} \epsilon_{1c\ast} \nonumber \\
			&\qquad - 4\epsilon_{1Z\ast} \epsilon_{1H\ast}  \epsilon_{2Z\ast} + \frac{1}{2} \epsilon_{1Z\ast}^2\epsilon_{2Z\ast}  + \frac{1}{2} \epsilon_{1Z\ast} \epsilon_{2Z\ast}^2  + 3 \epsilon_{1c\ast} \epsilon_{1Z\ast} \epsilon_{2c\ast} + \frac{9}{2} \epsilon_{1c\ast}  \epsilon_{2c\ast}^2- \frac{27}{2} \epsilon_{1H\ast} \epsilon_{2H\ast} \epsilon_{1c\ast}  \nonumber \\
			& \qquad + \frac{9}{2} \epsilon_{1c\ast} \epsilon_{2c\ast} \epsilon_{3c\ast}  +18 \epsilon_{1c\ast}^2\epsilon_{2c\ast}  - 15 \epsilon_{1c\ast} \epsilon_{1H\ast} \epsilon_{2c\ast} \,, \nonumber \\
		g_{3\ast} &=  \frac{3 \epsilon_{1H\ast} \epsilon_{2H\ast} \epsilon_{3H\ast}}{2} + \frac{3 \epsilon_{1H\ast} \epsilon_{2H\ast}^2}{2}  - \frac{3 \epsilon_{1Z\ast} \epsilon_{2Z\ast} \epsilon_{3Z\ast}}{4} - \frac{3 \epsilon_{1Z\ast} \epsilon_{2Z\ast}^2}{4} - \frac{9 \epsilon_{1c\ast} \epsilon_{2c\ast} \epsilon_{3c\ast}}{4} -\frac{9 \epsilon_{1c\ast} \epsilon_{2c\ast}^2}{4} \,,
 	\end{align}
	\end{widetext}
	with all quantities truncated at order $\mathcal{O}(\epsilon^3)$, i.e., N3LO. Notice that the lowest order flow parameter contained in $g_{n\ast}$ is of order $\order{\epsilon^n_\ast}$. Given the functional form of $g(x)$, it is clear that Eq.~\eqref{eq:EoM-v(x)-Final} does not admit a closed-form analytical solution, but in this form it can now be solved in an order-by-order expansion as we discuss below.

%\subsection*{Green's function method}
Following and extending \cite{Stewart2001,Auclair2022}, we use the Green's function method to correct the Bunch-Davies vacuum order by order in a systematic expansion. To simplify the intermediate formulas, we rescale our variables $x\to y$  and $v\to w$ to remove the dependence on the pivot scale $k_\ast$:
\begin{equation}
	v(x) = \sqrt{\frac{k_\ast}{2k}} \; w(y) \quad \textrm{and}\quad 
y = \frac{k}{k_\ast} x \,.
\label{eq:v-fact2}
\end{equation}
Notice that under this rescaling we have $y=-k\tau$ and $x=-k_\ast \tau$, where the generalization of conformal time is given by $\tau(t)$, which solves
\begin{equation}
\dot{\tau}(t)=\frac{c_\psi(t)}{a(t)}\,,
\label{eq:eta-x-def}
\end{equation} 
as it can be seen from \eqref{eq:xdot}. Then, the EoM \eqref{eq:EoM-v(x)-Final} reads as
\begin{equation}
	\label{eq:EoM-tilde v(y)}
 w''(y) + \left[  1 - \frac{2}{y^2}   \right]  w(y) = \frac{g(y)}{y^2}     w(y),
\end{equation}
where $g(y) = g_{1\circledast} + g_{2\circledast} \ln(y) + g_{3\circledast} \ln(y)^2$, i.e., the coefficients $g_{n\circledast}$ are of the same form as \eqref{eq:g's-full-expression} but are implicitly assumed to be evaluated at $y_\circledast = 1$, and the CCR relation of \eqref{eq:CCR-v(x)} becomes 
\begin{equation}
	\label{eq:CCR-tilde v(y)}
 w(y) w'{}^\ast (y) - w'(y)  w^\ast (y)  = -2 \,\ii.
\end{equation}
Note the factor of $2$ in the mode \eqref{eq:v-fact2} and in the CCR normalization \eqref{eq:CCR-tilde v(y)}, fixed to match the normalization used in \cite{Auclair2022}.
Now, the goal is to solve \eqref{eq:EoM-tilde v(y)} in a systematic expansion around the Bunch-Davies solution \eqref{eq:w-BD},
\begin{equation}
	 w_0 (y) = \bigg(1+\frac{\ii}{y}\bigg)\, \ee^{\ii y}.
\end{equation}
The LHS of \eqref{eq:EoM-tilde v(y)} takes the same form as \eqref{eq:w-EoM}, and we can capture the correction from the RHS introducing the advanced Green's function in the variable $y$ (which is the causal Green's function for cosmic time $t$):
\begin{equation}
G(y,s) = \tfrac{\ii}{2} \big(   w_0(y)  w^\ast_0(s) -  w_0(s)   w_0^\ast(y)  \big) \,\Theta(s-y),
\end{equation}
where $\Theta(s-x)$ is the Heaviside step function. The solution of the EoM \eqref{eq:EoM-tilde v(y)} can be found recursively as
\begin{equation}
 w(y) =  w_0 (y) +  \int_{y}^{\infty} \frac{g(y)}{s^2}  w(s) \, G(y,s) \dd{s}.
\end{equation}
We note that the structure of the function $q(y)$ allows us to write an expansion of the form $ w(y) =  w_0(y) +  w_1(y) +  w_2(y) +  w_3(y) + \order{\epsilon^4}$, where
\begin{align}
 w_1(y) &= g_{1\circledast} \int_{y}^{\infty} \frac{G(y,s)}{s^2}  w_0(s) \dd{s}\,, \\
 w_2(y) & = g_{2\circledast} \int_{y}^{\infty} \frac{G(y,s)}{s^2} \ln(s)  w_0(s) \dd{s} \nonumber\\
&\quad +g_{1\circledast}  \int_{y}^{\infty} \frac{G(y,s)}{s^2}  w_1(s) \dd{s}\,, \\
 w_3(y) &= g_{3\circledast}  \int_{y}^{\infty} \frac{G(y,s)}{s^2} \ln(s)^2  w_0(s) \dd{s} \nonumber \\
&\quad + g_{2\circledast} \int_{y}^{\infty} \frac{G(y,s)}{s^2} \ln(s)  w_1(s) \dd{s} \nonumber \\
&\qquad + g_{1\circledast} \int_{y}^{\infty} \frac{G(y,s)}{s^2}   w_2(s) \dd{s}\,.
\end{align}   
It can be checked that $ w_n(y) = \order{\epsilon^n}$ in the flow expansion. These integrals are  difficult to work with, especially at the third order. However, as we discuss in the next section on the late-time power spectrum, to extract the physically relevant information from the vacuum state described by the mode functions, we only need to know the asymptotic behavior of $\abs{ w(y)}^2$ in the limit $y\to 0^{+}$. The freezing in the power spectrum corresponds to a finite value of the ratio $\abs{ w(y)}^2/\mu(y)$ as $y\to 0^{+}$, despite the divergent behavior of $ w(y)$ and $\mu(y)$ by themselves. Fortunately, it was shown in \cite{Auclair2022} that this asymptotic behavior is fully captured by a family of one-dimensional integrals, starting with:
\begin{align}
		F_0(y)  &= \int_{y}^{\infty} \frac{e^{2is}}{s} \dd{s} = - \ln(y)  - B + \order{y} \,,\nonumber \\
		F_1(y)  &= \int_{y}^{\infty} \frac{e^{2is}}{s} \ln(s) \dd{s} \nonumber \\
		&= -\frac{1}{2} \ln(y)^2 + \frac{B^2}{2} + \frac{\pi^2}{12} + \order{y} \,, \\
		F_2(y)  &= \int_{y}^{\infty} \frac{e^{2is}}{s} \ln(s)^2 \dd{s}  \nonumber \\
		&= - \frac{1}{3} \ln(y)^3 - \frac{B^3}{3} - \frac{\pi^2}{6} B - \frac{2}{3} \zeta(3) + \order{y}\,,
\end{align}
where $B = \gamma_E + \ln(2) - \ii\pi/2$, with $\gamma_E \simeq  0.577$ being the Euler-Mascheroni constant and $\zeta(3) \simeq 1.202 $ being the Riemann zeta function evaluated at $3$. Next, one considers the two-dimensional integral,
\begin{align}
	  F_{00}(y) &= \int_y^\infty \dfrac{e^{-2is}}{s} F_{0}(s) \dd{s}, \nonumber \\ 
	  &= \frac{\pi^2}{4} + \frac{B^2}{2} + B \ln(y) +\frac{1}{2} \ln(y)^2 + \order{y} ,
	  \end{align}
and the three-dimensional integral,
\begin{align}
	F_{000}(y) &= \int_y^\infty \frac{e^{+2is}}{s} F_{00}(s)\dd{s} \nonumber \\
	&= - \frac{7}{3}  \zeta(3) -\frac{\pi^2}{4} B - \frac{1}{6}
	B^3 - \left(\frac{\pi^2}{4} +  \frac{B^2}{2}\right) \ln(y)\nonumber  \\ &\quad  - \frac{B}{2} \ln^2(y) - \frac{1}{6} \ln^3(y) + \order{y}.
\end{align}
The details of the asymptotic expansions and the explicit expressions of the solution $ w(y)$ in terms of these integrals can be found in \cite{Auclair2022}. As in the limit $y\to 0^+$ we have $\mu \sim a^2 \sim x^{-2}\sim y^{-2}$, we have that the relevant quantity to compute in detail in the limit is 
 $\abs{ y \,  w(y)}^2$. In terms of the integrals given above, it is given by
\begin{widetext}
\begin{align}
		\big|y\,  w(y) \big|^2 &= 1
	+ \frac{2}{3} g_{1\circledast} \qty[2 + \Re(F_{0})]
	\nonumber \\
	&\qquad + \frac{2}{27} g_{1\circledast}^2 \qty[4 + 3 \abs{F_{0}}^2 + 11 \Re(F_{0})]
	+ \frac{2}{9} g_{2\circledast} \qty[8 + 7 \Re(F_{0}) + 3 \Re(F_{1}) + 6 \ln(y)] \nonumber
	\\ & \qquad \quad+ \frac{2}{243} g_{1\circledast}^3 \qty[-8 + 14 \Re(F_{0}) + 30 \abs{F_{0}}^2 + 9 \Re(\overline{F_{00}} F_{0}) + 9 \Re(F_{000})] \nonumber 
	\\ & \qquad \quad+ \frac{4}{81} g_{1\circledast} g_{2\circledast} \qty[- 4 + 21 \abs{F_{0}}^2 + 9 \Re(\overline{F_{1}} F_{0}) + 40 \Re(F_{0}) + 15 \Re(F_{1}) + 12 \ln(y) + 18 \Re(F_{0}) \ln(y)] \nonumber
	\\ &\qquad \quad  + \frac{2}{27} g_{3\circledast} \qty[52 + 50 \Re(F_{0}) + 42 \Re(F_{1}) + 9 \Re(F_{2}) + 48 \ln x + 18 \ln(y)^2] + \order{\epsilon^4, x} \,,
\end{align}
and the asymptotic evaluation of the multidimensional integrals gives
\begin{align}
		\big| y\,  w(y) \big|^2  &= 1 - \frac{2C g_{1\circledast}}{3} - \frac{4g_{1\circledast}^2}{9} + \frac{2C g_{1\circledast}^2}{27} + \frac{2C^2 g_{1\circledast}^2}{9} + \frac{8g_{1\circledast}^3}{27} + \frac{68C g_{1\circledast}^3}{243} - \frac{4C^2 g_{1\circledast}^3}{81}  \nonumber \\
&\quad - \frac{4C^3 g_{1\circledast}^3}{81} - \frac{2C g_{2\circledast}}{9} + \frac{C^2 g_{2\circledast}}{3} - \frac{8g_1 g_2}{27} + \frac{80C g_{1\circledast} g_{2\circledast}}{81} + \frac{2}{27}C^2 g_{1\circledast} g_{2\circledast} - \frac{2}{9}C^3 g_{1\circledast} g_{2\circledast}  \nonumber\\
&\quad + \frac{g_{1\circledast}^2 \pi^2}{18} - \frac{g_{1\circledast}^3 \pi^2}{81} - \frac{1}{27}C g_{1\circledast}^3 \pi^2 - \frac{g_{2\circledast} \pi^2}{36} + \frac{7}{162}g_{1\circledast} g_{2\circledast} \pi^2 - \frac{5}{54}C g_{1\circledast} g_{2\circledast} \pi^2 - \frac{g_{3\circledast} \pi^2}{54} + \frac{1}{18}C g_{3\circledast} \pi^2  \nonumber \\
&\quad + \frac{8g_{3\circledast}}{9} - \frac{4C g_{3\circledast}}{27} + \frac{2C^2 g_{3\circledast}}{9} - \frac{2C^3 g_{3\circledast}}{9} \nonumber  \\
&\qquad + \ln(y) \left(-\frac{2g_{1\circledast}}{3} + \frac{2g_{1\circledast}^2}{27} + \frac{4C g_{1\circledast}^2}{9} + \frac{68g_{1\circledast}^3}{243} - \frac{8C g_{1\circledast}^3}{81} - \frac{4C^2 g_{1\circledast}^3}{27} - \frac{2g_{2\circledast}}{9} + \frac{8g_{1\circledast} g_{2\circledast}}{81}\right.  \nonumber \\
&\qquad + \left.\frac{8C g_{1\circledast} g_{2\circledast}}{27} - \frac{2}{9}C^2 g_{1\circledast} g_{2\circledast} - \frac{4g_{3\circledast}}{27} - \frac{g_{1\circledast}^3 \pi^2}{27} + \frac{1}{54}g_{1\circledast} g_{2\circledast} \pi^2\right) - \frac{14}{81}g_{1\circledast}^3 \zeta(3) - \frac{4}{9}g_{3\circledast} \zeta(3) \nonumber \\
& \quad\qquad+ \left(\frac{2g_{1\circledast}^2}{9} - \frac{4g_{1\circledast}^3}{81} - \frac{4C g_{1\circledast}^3}{27} - \frac{g_{2\circledast}}{3} + \frac{2g_{1\circledast} g_{2\circledast}}{9}  + \frac{2C g_{1\circledast} g_{2\circledast}}{9} - \frac{2g_{3\circledast}}{9}\right) \ln(y)^2 \nonumber \\
& \quad\qquad - \left(\frac{4g_{1\circledast}^3}{81} - \frac{2g_{1\circledast} g_{2\circledast}}{9} + \frac{2g_{3\circledast}}{9}\right) \ln(y)^3
\,, \label{eq:asymptotic}
\end{align}
\end{widetext}
where $C = \gamma_E + \ln(2) - 2 \simeq -0.730$.  We report the numerical value of mathematical constants with only few figures but, as the N3LO power spectrum is at order $\mathcal{O}(\epsilon^3)$, with $\epsilon\sim 10^{-2}$, the appropriate number of significant figures of each exact mathematical constant should be used.

\section{Primordial observables}
\label{sec:Final-Expressions}

\subsection{Power spectrum}
We proceed to briefly review the definition of power spectrum. The quantum field $\hat{\Psi}(\vb{x},t)$ is an operator-valued distribution. What we measure with a finite-resolution detector at $\vb{x}_0$ is the smearing of the quantum field against a test function $f(|\vb{x}-\vb{x}_0|)$ that characterizes its response, i.e.,
\begin{align}
	\hat{\Psi}_f(t) &\equiv \int \dd[3]{\vb{x}} f(|\vb{x}-\vb{x}_0|) \hat{\Psi}(\vb{x},t)\nonumber\\
 &= \int \frac{\dd[3]{\vb{k}}}{(2\pi)^3} \tilde{f}(k) \hat{\psi}(\vb{k},t)\,,
\end{align}
where $\hat{\psi}(\vb{k},t) =u(k,t) \hat{a}(\vb{k}) + u^\ast (k,t) \hat{a}^\dagger (-\vb{k}) $. The Fourier transform $\tilde{f}(k)$ of the test function is assumed to be smooth and with a compact support in $ [k_{\textrm{min}},k_{\textrm{max}}]$ which captures the band or range of wavelengths that our observations probe. 
The quasi-Bunch-Davies vacuum $|0\rangle$, discussed earlier in terms of the mode functions $u(k,t)$,  is defined as the state that approaches the Bunch-Davies vacuum in the far past where quasi-de Sitter space approaches de Sitter space. As
$\hat{a}(\vb{k})|0 \rangle=0$, the expectation value of measurements of the smeared field is zero, $ \langle \Psi_f\rangle\equiv\bra{0} \hat{\Psi}_f(t) \ket{0} = 0$, and the variance $(\Delta \Psi_f)^2\equiv  \langle \Psi_f{}^2 \rangle-\langle \Psi_f\rangle^2$ is given by the equal-time two-point correlation function
\begin{align}
&\bra{0} \hat{\Psi}_f(t) \hat{\Psi}_f (t) \ket{0} = \int \frac{\dd[3]{\vb{k}}}{(2\pi)^3} \, |u(k,t)|^2\,\,\big|\tilde{f}(k)\big|^2 \nonumber \\
&\qquad= \int_0^\infty \frac{\dd{k}}{k} \;\frac{k^3}{2\pi^2}  \,|u(k,t)|^2\, \big|\tilde{f}(k)\big|^2\nonumber \\
&\qquad = \int_0^\infty \!\!\dd (\log k)\, \mathcal{P}^{(\psi)} (k,t)\, \big|\tilde{f}(k)\big|^2\,,
\end{align}
where we integrated away the angular variables (using the assumed invariance of the test function under rotations and the homogeneity and isotropy of the state) and defined the power spectrum in a band $\dd(\log k)$ at the time $t$ as usual
\begin{equation}
\mathcal{P}^{(\psi)} (k,t) \equiv \frac{k^3}{2\pi^2} |u(k,t)|^2.
\end{equation}
Therefore, by specifying the vacuum state $|0 \rangle$ in terms of the mode function $u(k,t)$, one has an immediate way to predict the size of quantum fluctuations in terms of $|u(k,t)|^2$. Moreover,  the mode $k_\ast$ that transitions from an oscillatory phase to an overdamped phase around a time $t_\ast$ before the end of inflation has a power $\mathcal{P}^{(\psi)} (k_\ast,t)$ that freezes to a finite value for $t\gg t_\ast$. The \emph{late-time} power spectrum, defined formally as the limit $\mathcal{P}_0 (k) = \lim_{t\to \infty} \mathcal{P}(t,k)$, is then given by
\begin{align}
\mathcal{P}_0^{(\psi)} (k) &= \lim_{t\to \infty } \frac{k^3}{2\pi^2}  |u(k,t)|^2= \lim_{x\to 0^+ } \frac{k^3}{2\pi^2} \frac{\abs{v(x)}^2}{\mu(x)}  \nonumber\\
&\!\!\!=  \lim_{y\to 0^+ }  \frac{\hbar\, H(y)^2  }{4\pi^2 c_\psi(y) \tilde{c}_\psi^2(y) Z_\psi(y)}\abs{ y \,  w(y)}^2 \nonumber\\
&\!\!\!=  \lim_{y\to 0^+ }  \frac{\hbar\, H(y)^2  }{4\pi^2 c_\psi^3(y) Z_\psi(y)}\abs{ y \,  w(y)}^2 \,(1+\ldots\,)\,,
\end{align}
where the ellipses indicate the Hubble-flow expansion of $c_\psi(y)^2/\tilde{c}_\psi(y)^2=(1+\ldots\,)$ as defined in 
\eqref{eq:c-tilde-expansion}. Furthermore, all the $y$ dependent functions will admit a logarithmic expansion as done in \eqref{eq:LogExpansion}. The caveat resides in the new pivot scale associated to this new logarithmic expansion in terms of $\ln(y)$. Notice that, using \eqref{eq:eta-x-def},
\begin{equation}
\ln(y) = \ln(-k\tau) = \ln(\frac{\tau}{\tau_\circledast}) ,
\end{equation} 
where the new reference time is $\tau_\circledast = -1/k$, such that $y_\circledast = -k \tau_\circledast = 1$. Hence, a function $\rho(y)$ could be expanded as 
\begin{align}
\rho(y) &= \rho_\circledast + \rho_{1\circledast} \ln(\frac{\tau}{\tau_\circledast}) + \rho_{2\circledast}\ln(\frac{\tau}{\tau_\circledast}) ^2\nonumber \\  &\quad +\rho_{3\circledast} \ln(\frac{\tau}{\tau_\circledast}) ^3 + \cdots
\end{align}
In this way, the leading order term of the late-time power spectrum becomes 
\begin{align}
\mathcal{P}_0^{(\psi)} (k)  &= \lim_{y\to 0^+ }  \frac{\hbar H_\circledast^2}{4\pi^2 c_\circledast^3 Z_\circledast}\abs{ y \,  w(y)}^2 	(1 +  \cdots ) \nonumber \\
&= \frac{\hbar H_\circledast^2}{4\pi^2 c_\circledast^3 Z_\circledast}\; p_{\circledast}\,.\label{eq:P0-circle}
\end{align}
Note that, both the terms $|y \,w(y)|^2$ and $(1+\ldots)$ contain logarithmically divergent terms in the limit $y\to 0^+$. Remarkably, these logarithms exactly cancel out, and in \eqref{eq:P0-circle} there are no divergences in the limit. The coefficients in $p_\circledast$ can be found in Appendix~\ref{app:finite}, and are a combination of $\epsilon$'s evaluated at the reference time $\tau_\circledast$ without any direct scale dependence. To compare with the pivot scale\footnote{To consider a comparison between the spectra of two SVT modes with different speeds of sound, see Appendix \ref{app:comparing-pivot-scales}.}  associated to the generalized horizon-crossing $k_\ast=a_\ast H_\ast/\tilde{c}_\ast$, equivalent to $x_\ast=1$, we need the $y$--dependent functions to be expanded around $\tau_\ast$ and not $\tau_\circledast$, which can be obtained by noticing that
\begin{equation}
\ln(\frac{\tau_\circledast}{\tau_\ast}) = - \ln(\frac{k}{k_\ast}).
 \end{equation}
On the other hand, the usual expansion in terms of the variable $x$ satisfies $\ln(x) = \ln(-k_\ast \tau) = \ln(\tau/\tau_\ast)$. As a consequence, we can write 
 \begin{align}
 \rho_\circledast \equiv  \rho\qty(\frac{\tau_\circledast}{\tau_\ast}) &= \rho_\ast - \rho_{1\ast} \ln(\frac{k}{k_\ast})\nonumber \\  
 & \quad +\rho_{2\ast} \ln(\frac{k}{k_\ast})^2 -\rho_{3\ast} \ln(\frac{k}{k_\ast})^3.
 \end{align}
Thus, one can check that all the quantities $\rho_{\circledast}$, $\epsilon_{1\rho\circledast}$ that we would obtain in the logarithmic expansion can be expressed in terms of $\rho_{\ast}$, $\epsilon_{1\rho\ast}$, and so on, by only applying the map 
 \begin{align}
 \rho_{\circledast} &\to \rho(x)\big|_{\ln(x)\leftrightarrow -\ln(k/k_\ast)} \nonumber \\[.5em]
  \epsilon_{n\rho\circledast} &\to \epsilon_{n\rho}(x)\big|_{\ln(x)\leftrightarrow -\ln(k/k_\ast)} \, , 
 \end{align}
where $\rho(x)$ and $\epsilon_{n\rho}(x)$ are the functional expressions of \eqref{eq:LogExpansion}.  Now we have all the pieces to compute the third-order corrections to the late-time power spectrum, which can be parametrized as follows:
\begin{align}
	\label{eq:PowerSpectrum_Full}
		\mathcal{P}_0^{(\psi)}(k) &=  \frac{\hbar H_\ast^2}{4\pi^2 c_\ast^3 Z_\ast} \Bigg[ 1 + p_{0\ast} + p_{1\ast} \ln(\frac{k}{k_\ast}) \nonumber\\ 
		&\qquad +\, p_{2\ast} \ln(\frac{k}{k_\ast})^2 + p_{3\ast} \ln(\frac{k}{k_\ast})^3 \Bigg],
\end{align}
where the coefficients $p_{0\ast}$, $p_{1\ast}$, $p_{2\ast}$ and $p_{3\ast}$ are reported in Tables~\ref{Tab:p0s}, \ref{Tab:p1s}, \ref{Tab:p2s} and \ref{Tab:p3s}, respectively. The above expression together with the reported coefficients are directly useful in analyzing data from cosmological observables, and represent the main result of this work. A \textit{Mathematica} notebook with the explicit expressions can be found in \cite{GitHubCode2024}. Additionally, we report in Table~\ref{Tab:GenericTheory-Features} the power-law quantities, and in Appendix~\ref{App:Power-law-generic} the corresponding expression for the amplitude of the primordial power spectrum, i.e., $\mathcal{A}^{(\psi)}_\ast = \mathcal{P}^{(\psi)}_0(k_\ast)$.

\begin{table*}[h]
	\caption{Full expression of $p_{0\ast}$ for a theory with generic $Z_\psi$ and $c_{\psi}$, up to N3LO corrections.}
	\label{Tab:p0s}
	\begin{ruledtabular}
		\begin{tabular}{p{0.2in} p{6.2in}}
			 Order & Expression \\
			\hline
			  $\begin{aligned} ~\\[-3ex]
				\textrm{NLO}:
				\hspace{2.6em}  & \hspace{-1.7em} p_{0\ast}\;=\; - 2 (1 + C) \epsilon_{1H\ast} + C \epsilon_{1Z\ast} +(2 + 3 C) \epsilon_{1c\ast}   \\ 
				\textrm{N2LO}: \hspace{2.6em} &+\frac{1}{24} \Big\{  12 \Big(-6 + 4 C + 4 C^2 + \pi^2 \Big) \epsilon_{1H\ast}^2+ \epsilon_{1Z\ast} \Big[ 
3 \Big(-8 + 4 C^2 + \pi^2 \Big) \epsilon_{1Z\ast} 
+ \Big(-12 C^2 + \pi^2 \Big) \epsilon_{2Z\ast} 
\Big] 
 \\
&- 2 \epsilon_{1H\ast} \Big[ 
6 \Big(-8 + 2 C + 4 C^2 + \pi^2 \Big) \epsilon_{1Z\ast} 
+ \Big(-24 - 24 C - 12 C^2 + \pi^2 \Big) \epsilon_{2H\ast} 
\Big] +3 \Big(-64 + 24 C + 36 C^2 + 9 \pi^2 \Big) \epsilon_{1c\ast}^2  \\
&- 3 \epsilon_{1c\ast} \Big[ 
4 \Big(-20 + 10 C + 12 C^2 + 3 \pi^2 \Big) \epsilon_{1H\ast} 
- 2 \Big(-24 + 4 C + 12 C^2 + 3 \pi^2 \Big) \epsilon_{1Z\ast} 
- \Big(-16 - 16 C - 12 C^2 + \pi^2 \Big) \epsilon_{2c\ast} 
\Big]\Big\}  \\
\textrm{N3LO}: \hspace{2.6em}  & 
	+\frac{1}{24} \Bigg\{
- 8 \epsilon_{1H\ast}^3 \Big(-16 - 24 C + 4 C^3 + 3 C \pi^2 + 14 \zeta(3)\Big)
+ 2 \epsilon_{1H\ast}^2 \Big( (96 - 36 C^2 - 24 C^3 - 13 \pi^2 - 2 C(-36 + 5 \pi^2)) \epsilon_{2H\ast} \\
&+ 6 \epsilon_{1Z\ast} \Big[-16 - 24 C + 4 C^3 + 3 C \pi^2 + 14 \zeta(3)\Big] \Big)
+ \epsilon_{1H\ast} \Big[
  \epsilon_{1Z\ast} \Big((-96 - 72 C + 36 C^2 + 24 C^3 + 13 \pi^2 + 10 C \pi^2) \epsilon_{2H\ast}\\
  &+ 2 C(-48 + 12 C^2 + 5 \pi^2) \epsilon_{2Z\ast} \Big)
  + 2 \epsilon_{2H\ast} (\epsilon_{2H\ast} + \epsilon_{3H\ast}) \Big(-8 - 12 C^2 - 4 C^3 + \pi^2 + C(-24 + \pi^2) - 8 \zeta(3) \Big)\\
 & - 6 \epsilon_{1Z\ast}^2 \Big(-16 - 24 C + 4 C^3 + 3 C \pi^2 + 14 \zeta(3)\Big)
\Big]
+ \epsilon_{1Z\ast} \Big[
- \epsilon_{2Z\ast} (\epsilon_{2Z\ast} + \epsilon_{3Z\ast}) \Big(16- 4 C^3 + C \pi^2 - 8 \zeta(3)\Big) \\
& +C (48 - 12 C^2 - 5 \pi^2) \epsilon_{1Z\ast} \epsilon_{2Z\ast}
+ \epsilon_{1Z\ast}^2 \Big(-16 - 24 C + 4 C^3 + 3 C \pi^2 + 14 \zeta(3)\Big)
\Big]
- 3 \epsilon_{1c\ast}^2 \Big[
  \Big(-96 + 36 C^2 + 36 C^3 \\
  &+ 13 \pi^2 + 15 C(-8 + \pi^2)\Big) \epsilon_{2c\ast}
  + 18 \epsilon_{1H\ast} \Big(-16 - 24 C + 4 C^3 + 3 C \pi^2 + 14 \zeta(3)\Big)\\
 & - 9 \epsilon_{1Z\ast} \Big(-16 - 24 C + 4 C^3 + 3 C \pi^2 + 14 \zeta(3)\Big)
\Big]
+ \epsilon_{1c\ast} \Bigg[
  \epsilon_{1Z\ast} \Big(-\Big( -96 + 36 C^2 + 36 C^3 + 13 \pi^2 + 15 C(-8 + \pi^2) \Big) \epsilon_{2c\ast} \\
 & - 3 C(-48 + 12 C^2 + 5 \pi^2) \epsilon_{2Z\ast}\Big)
  + \epsilon_{2c\ast} (\epsilon_{2c\ast} + \epsilon_{3c\ast}) \Big(-48 C - 24 C^2 - 12 C^3 + 2 \pi^2 + 3 C \pi^2 - 24 \zeta(3)\Big)\\
  &+ 36 \epsilon_{1H\ast}^2 \Big(-16 - 24 C + 4 C^3 + 3 C \pi^2 + 14 \zeta(3)\Big)
  + 9 \epsilon_{1Z\ast}^2 \Big(-16 - 24 C + 4 C^3 + 3 C \pi^2 + 14 \zeta(3)\Big)\\
  &+ \epsilon_{1H\ast} \Big( 2(-96 + 36 C^2 + 36 C^3 + 13 \pi^2 + 15 C(-8 + \pi^2)) \epsilon_{2c\ast}
  + 3(-96 - 72 C + 36 C^2 + 24 C^3 + 13 \pi^2 + 10 C \pi^2) \epsilon_{2H\ast}\\
  &- 36 \epsilon_{1Z\ast} \Big(-16 - 24 C + 4 C^3 + 3 C \pi^2 + 14 \zeta(3)\Big)
  \Big)
\Bigg]
+ 27 \epsilon_{1c\ast}^3 \Big(-16 - 24 C + 4 C^3 + 3 C \pi^2 + 14 \zeta(3)\Big)
\Bigg\}
			\end{aligned}$
		\end{tabular}
	\end{ruledtabular}
\end{table*}

\begin{table*}[h]
	\caption{Full expression of $p_{1\ast}$ for a theory with generic $Z_\psi$ and $c_{\psi}$, up to N3LO corrections.}
	\label{Tab:p1s}
	\begin{ruledtabular}
		\begin{tabular}{p{0.2in} p{6.2in}}
			 Order & Expression \\
			\hline
			  $\begin{aligned} ~\\[-3ex]
				\textrm{NLO}:
				\hspace{2.6em}  & \hspace{-1.7em} p_{1\ast}\;=\; - 2 \epsilon_{1H\ast} + \epsilon_{1Z\ast} + 3 \epsilon_{1c\ast}   \\ 
				\textrm{N2LO}: \hspace{2.6em} &
    +(2 + 4 C) \epsilon_{1H\ast}^2 
+ \epsilon_{1H\ast} \Big[ 
- \epsilon_{1Z\ast} - 4 C \epsilon_{1Z\ast} + 2(1 + C) \epsilon_{2H\ast} 
\Big] 
+ C \epsilon_{1Z\ast} (\epsilon_{1Z\ast} - \epsilon_{2Z\ast}) \\
&+ \epsilon_{1c\ast} \Big[ 
(3 + 9 C) \epsilon_{1c\ast} 
+ (-5 - 12 C) \epsilon_{1H\ast} + \epsilon_{1Z\ast} + 6 C \epsilon_{1Z\ast} 
- 2 \epsilon_{2c\ast} - 3 C \epsilon_{2c\ast} 
\Big]  \\
\textrm{N3LO}: \hspace{2.6em}  & 
+\frac{1}{24} \Bigg\{
-24 \big(-8 + 4C^2 + \pi^2\big) \epsilon_{1H\ast}^3 
+ 4 \epsilon_{1H\ast}^2 \big(9 \big(-8 + 4C^2 + \pi^2\big) \epsilon_{1Z\ast} 
+ \big(36 - 36C - 36C^2 - 5\pi^2\big) \epsilon_{2H\ast}\big)\\
&- 2 \epsilon_{1H\ast} \Big[9 \big(-8 + 4C^2 + \pi^2\big) \epsilon_{1Z\ast}^2 
+ \epsilon_{1Z\ast} \big(- \big(-36 + 36C + 36C^2 + 5\pi^2\big) \epsilon_{2H\ast} 
- \big(-48 + 36C^2 + 5\pi^2\big) \epsilon_{2Z\ast}\big) \\
&- \big(-24 - 24C - 12C^2 + \pi^2\big) \epsilon_{2H\ast} \big(\epsilon_{2H\ast} + \epsilon_{3H\ast}\big)\Big]
+ \epsilon_{1Z\ast} \Big[3 \big(-8 + 4C^2 + \pi^2\big) \epsilon_{1Z\ast}^2 
+ \big(48 - 36C^2 - 5\pi^2\big) \epsilon_{1Z\ast} \epsilon_{2Z\ast} \\
&- \big(-12C^2 + \pi^2\big) \epsilon_{2Z\ast} \big(\epsilon_{2Z\ast} + \epsilon_{3Z\ast}\big)\Big]
- 9 \epsilon_{1c\ast}^2 \big(18 \big(-8 + 4C^2 + \pi^2\big) \epsilon_{1H\ast} 
- 9 \big(-8 + 4C^2 + \pi^2\big) \epsilon_{1Z\ast} \\
&+ \big(-40 + 24C + 36C^2 + 5\pi^2\big) \epsilon_{2c\ast}\big)
+ 3 \epsilon_{1c\ast} \Bigg[36 \big(-8 + 4C^2 + \pi^2\big) \epsilon_{1H\ast}^2 
+ 9 \big(-8 + 4C^2 + \pi^2\big) \epsilon_{1Z\ast}^2 \\
&+ 2 \epsilon_{1H\ast} \big(-18 \big(-8 + 4C^2 + \pi^2\big) \epsilon_{1Z\ast} 
+ \big(-40 + 24C + 36C^2 + 5\pi^2\big) \epsilon_{2c\ast} 
+ \big(-36 + 36C + 36C^2 + 5\pi^2\big) \epsilon_{2H\ast}\big)\\
&+ \epsilon_{1Z\ast} \big(- \big(-40 + 24C + 36C^2 + 5\pi^2\big) \epsilon_{2c\ast} 
- \big(-48 + 36C^2 + 5\pi^2\big) \epsilon_{2Z\ast}\big)\\
&- \big(-16 - 16C - 12C^2 + \pi^2\big) \epsilon_{2c\ast} \big(\epsilon_{2c\ast} + \epsilon_{3c\ast}\big)\Bigg]
+ 81 \big(-8 + 4C^2 + \pi^2\big) \epsilon_{1c\ast}^3
\Bigg\}
			\end{aligned}$
		\end{tabular}
	\end{ruledtabular}
\end{table*}

\begin{table*}[th]
	\caption{Full expression of $p_{2\ast}$ for a theory with generic $Z_\psi$ and $c_{\psi}$, up to N3LO corrections.}
	\label{Tab:p2s}
	\begin{ruledtabular}
		\begin{tabular}{p{0.2in} p{6.2in}}
			 Order & Expression \\
			\hline
			  $\begin{aligned} ~\\[-3ex]
				\textrm{N2LO}: \hspace{2.6em} &\hspace{-1.7em} p_{2\ast}\;=\;
  \frac{1}{2} \Big\{ 
4 \epsilon_{1H\ast}^2 
- 4 \epsilon_{1H\ast} \epsilon_{1Z\ast} 
+ \epsilon_{1Z\ast}^2 
+ 2 \epsilon_{1H\ast} \epsilon_{2H\ast} 
- \epsilon_{1Z\ast} \epsilon_{2Z\ast} 
+ \epsilon_{1c\ast} \Big[ 9 \epsilon_{1c\ast} 
- 3 \Big( 4 \epsilon_{1H\ast} 
- 2 \epsilon_{1Z\ast} 
+ \epsilon_{2c\ast} \Big) 
\Big]
\Big\} \\
\textrm{N3LO}: \hspace{2.6em}  & 
+\frac{1}{2} \Bigg\{
\epsilon_{1H\ast} \epsilon_{2H\ast} \big(-6 \epsilon_{1H\ast} + 3 \epsilon_{1Z\ast} - 2 (\epsilon_{2H\ast} + \epsilon_{3H\ast})\big)
+ C \Bigg( -8 \epsilon_{1H\ast}^3 
+ 12 \epsilon_{1H\ast}^2 \big( \epsilon_{1Z\ast} - \epsilon_{2H\ast} \big)\\
&- 2 \epsilon_{1H\ast} \Big[ 3 \epsilon_{1Z\ast}^2 
- 3 \epsilon_{1Z\ast} (\epsilon_{2H\ast} + \epsilon_{2Z\ast}) 
+ \epsilon_{2H\ast} \big( \epsilon_{2H\ast} + \epsilon_{3H\ast} \big)\Big]
+ \epsilon_{1Z\ast} \Big[ \epsilon_{1Z\ast}^2 
- 3 \epsilon_{1Z\ast} \epsilon_{2Z\ast} 
+ \epsilon_{2Z\ast} (\epsilon_{2Z\ast} + \epsilon_{3Z\ast})\Big]\Bigg)\\
&- 9 \epsilon_{1c\ast}^2 \epsilon_{2c\ast}
+ \epsilon_{1c\ast} \Big[ -3 \epsilon_{1Z\ast} \epsilon_{2c\ast} 
+ \epsilon_{1H\ast} \big( 6 \epsilon_{2c\ast} + 9 \epsilon_{2H\ast} \big) 
+ 2 \epsilon_{2c\ast} (\epsilon_{2c\ast} + \epsilon_{3c\ast})\Big]\\
&+ C \Bigg( 27 \epsilon_{1c\ast}^3 
- 27 \epsilon_{1c\ast}^2 \big( 2 \epsilon_{1H\ast} - \epsilon_{1Z\ast} + \epsilon_{2c\ast}\big)
+ 3 \epsilon_{1c\ast} \Big[12 \epsilon_{1H\ast}^2 
+ 3 \epsilon_{1Z\ast}^2 
+ 6 \epsilon_{1H\ast} (-2 \epsilon_{1Z\ast} + \epsilon_{2c\ast} + \epsilon_{2H\ast}) \\
&- 3 \epsilon_{1Z\ast} (\epsilon_{2c\ast} + \epsilon_{2Z\ast}) 
+ \epsilon_{2c\ast} (\epsilon_{2c\ast} + \epsilon_{3c\ast})\Big]\Bigg)
\Bigg\}
			\end{aligned}$
		\end{tabular}
	\end{ruledtabular}
\end{table*}

\begin{table*}[th]
	\caption{Full expression of $p_{3\ast}$ for a theory with generic $Z_\psi$ and $c_{\psi}$, only containing N3LO corrections.}
	\label{Tab:p3s}
	\begin{ruledtabular}
		\begin{tabular}{p{0.2in} p{6.2in}}
			 Order & Expression \\
			\hline
			  $\begin{aligned} ~\\[-3ex]
\textrm{N3LO}: \hspace{2.6em}  & \hspace{-1.7em} p_{3\ast}\;=\;
\frac{1}{6} \Bigg\{
-8 \epsilon_{1H\ast}^3 
+ 12 \epsilon_{1H\ast}^2 \big(\epsilon_{1Z\ast} - \epsilon_{2H\ast}\big)
- 2 \epsilon_{1H\ast} \Big[3 \epsilon_{1Z\ast}^2 
- 3 \epsilon_{1Z\ast} \big(\epsilon_{2H\ast} + \epsilon_{2Z\ast}\big)
+ \epsilon_{2H\ast} \big(\epsilon_{2H\ast} + \epsilon_{3H\ast}\big)\Big]\\
&+ \epsilon_{1Z\ast} \Big[\epsilon_{1Z\ast}^2 
- 3 \epsilon_{1Z\ast} \epsilon_{2Z\ast}
+ \epsilon_{2Z\ast} \big(\epsilon_{2Z\ast} + \epsilon_{3Z\ast}\big)\Big]
- 27 \epsilon_{1c\ast}^2 \big(2 \epsilon_{1H\ast} - \epsilon_{1Z\ast} + \epsilon_{2c\ast}\big)\\
&+ 3 \epsilon_{1c\ast} \Big[12 \epsilon_{1H\ast}^2 
+ 3 \epsilon_{1Z\ast}^2 
+ 6 \epsilon_{1H\ast} \big(-2 \epsilon_{1Z\ast} + \epsilon_{2c\ast} + \epsilon_{2H\ast}\big)
- 3 \epsilon_{1Z\ast} \big(\epsilon_{2c\ast} + \epsilon_{2Z\ast}\big)
+ \epsilon_{2c\ast} \big(\epsilon_{2c\ast} + \epsilon_{3c\ast}\big)\Big]
+ 27 \epsilon_{1c\ast}^3
\Bigg\}
			\end{aligned}$
		\end{tabular}
	\end{ruledtabular}
\end{table*}
	
\begin{table*}[th]
	\caption{Quantities characterizing deviations from an exact power-law, as defined in \eqref{eq:power-law-quantities-amplitude}-\eqref{eq:power-law-quantities-runrun}, for a theory with generic $Z_\psi$ and $c_{\psi}$, up to N3LO.}
	\label{Tab:GenericTheory-Features}
	\begin{ruledtabular}
		\begin{tabular}{p{0.7in} p{0.67in} p{6in}}
			Quantity &  Order & Expression \\
			\hline
			$\hspace{0.5cm} \theta_\ast^{(\psi)}$  & $\vspace{0.2em}\begin{aligned} ~\\[-3ex] \textrm{NLO}:
				\hspace{0.2cm}  &
				-2 \epsilon_{1H\ast} + \epsilon_{1Z\ast} + 3\epsilon_{1c\ast}  \\ 
				\textrm{N2LO}: \hspace{0.2cm}  &
				- 2 \epsilon_{1H\ast}^2 + 2(1 + C) \epsilon_{1H\ast} \epsilon_{2H\ast} + \epsilon_{1Z\ast} ( \epsilon_{1H\ast} - C  \epsilon_{2Z\ast})  +  \epsilon_{1c\ast}( 5 \epsilon_{1H\ast}-3 \epsilon_{1c\ast} -  \epsilon_{1Z\ast} ) - (2 + 3C) \epsilon_{1c\ast} \epsilon_{2c\ast}   \\
				\textrm{N3LO}: \hspace{0.2cm}&  -
				2 \epsilon_{1H\ast}^3 + (14 + 6C - \pi^2) \epsilon_{1H\ast}^2 \epsilon_{2H\ast}  + \frac{1}{12} (-24 - 24C - 12C^2 + \pi^2) \epsilon_{1H\ast} \epsilon_{2H\ast}^2  \\
				& +   \frac{1}{12} (-24 - 24C - 12C^2 + \pi^2) \epsilon_{1H\ast} \epsilon_{2H\ast} \epsilon_{3H\ast} + \epsilon_{1H\ast}^2 \epsilon_{1Z\ast} + \frac{1}{2} (-10 - 2C + \pi^2) \epsilon_{1H\ast} \epsilon_{1Z\ast} \epsilon_{2H\ast}  \nonumber  \\
				&  + \frac{1}{2} (-8 - 4C + \pi^2) \epsilon_{1H\ast} \epsilon_{1Z\ast} \epsilon_{2Z\ast} + \frac{1}{4} (8 - \pi^2) \epsilon_{1Z\ast}^2 \epsilon_{2Z\ast} + \frac{1}{24} (12C^2 - \pi^2) \epsilon_{1Z\ast} \epsilon_{2Z\ast}^2 \nonumber \\
				& + \frac{1}{24} (12C^2 - \pi^2) \epsilon_{1Z\ast} \epsilon_{2Z\ast} \epsilon_{3Z\ast} +  3 \epsilon_{1c\ast}^3 - 8 \epsilon_{1c\ast}^2 \epsilon_{1H\ast} + 7 \epsilon_{1c\ast} \epsilon_{1H\ast}^2 +  \epsilon_{1c\ast}^2 \epsilon_{1Z\ast}  - 2 \epsilon_{1c\ast} \epsilon_{1H\ast} \epsilon_{1Z\ast}   \\
				&+ \frac{1}{4} (100 + 36C - 9\pi^2) \epsilon_{1c\ast}^2 \epsilon_{2c\ast}  + \frac{1}{2} (-36 - 16C + 3\pi^2) \epsilon_{1c\ast} \epsilon_{1H\ast} \epsilon_{2c\ast} + \frac{1}{4} (28 + 4C - 3\pi^2) \epsilon_{1c\ast} \epsilon_{1Z\ast} \epsilon_{2c\ast} \nonumber  \\
				&  + \frac{1}{8} (16 + 16C + 12C^2 - \pi^2) \epsilon_{1c\ast} \epsilon_{2c\ast}^2 + \frac{1}{2} (-38 - 14C + 3\pi^2) \epsilon_{1c\ast} \epsilon_{1H\ast} \epsilon_{2H\ast} \nonumber \\
				&+ \frac{1}{4} (24 + 8C - 3\pi^2) \epsilon_{1c\ast} \epsilon_{1Z\ast} \epsilon_{2Z\ast}  + \frac{1}{8} (16 + 16C + 12C^2 - \pi^2) \epsilon_{1c\ast} \epsilon_{2c\ast} \epsilon_{3c\ast}  
			\end{aligned}$ \\
			\hline
			$\hspace{0.5cm} \alpha_\ast^{(\psi)}$ & $\vspace{0.2em}\begin{aligned}  ~\\[-3ex]
				\hspace{0.0cm} 	\textrm{N2LO}: & \hspace{0.3cm} 2 \epsilon_{1H\ast} \epsilon_{2H\ast} - \epsilon_{1Z\ast} \epsilon_{2Z\ast}-3 \epsilon_{1c\ast} \epsilon_{2c\ast} 
				\\   \textrm{N3LO}: & \hspace{0.2cm} 
			 	+6 \epsilon_{1H\ast}^2 \epsilon_{2H\ast}  - 2(1 + C) \epsilon_{1H\ast} \epsilon_{2H\ast}^2  - 2(1 + C) \epsilon_{1H\ast} \epsilon_{2H\ast} \epsilon_{3H\ast}     - \epsilon_{1H\ast}  \epsilon_{2H\ast} \epsilon_{1Z\ast} - 2 \epsilon_{1H\ast} \epsilon_{1Z\ast} \epsilon_{2Z\ast}    \\
				&\;\;+ C \epsilon_{1Z\ast} \epsilon_{2Z\ast}^2  + C \epsilon_{1Z\ast} \epsilon_{2Z\ast} \epsilon_{3Z\ast} +9 \epsilon_{1c\ast}^2 \epsilon_{2c\ast} - 8 \epsilon_{1c\ast} \epsilon_{1H\ast} \epsilon_{2c\ast} + \epsilon_{1c\ast} \epsilon_{1Z\ast} \epsilon_{2c\ast} + (2 + 3C) \epsilon_{1c\ast} \epsilon_{2c\ast}^2 \\
				&\; \;- 7 \epsilon_{1c\ast} \epsilon_{1H\ast} \epsilon_{2H\ast}   + 2 \epsilon_{1c\ast} \epsilon_{1Z\ast} \epsilon_{2Z\ast}     + (2 + 3C) \epsilon_{1c\ast} \epsilon_{2c\ast} \epsilon_{3c\ast} \end{aligned}$ \\
			\hline 
			$\hspace{0.5cm} \beta_{\ast}^{(\psi)}$ & $\vspace{0.2em}\begin{aligned}~\\[-3ex] 
				\hspace{0.0cm}  \textrm{N3LO}: & \hspace{0.5em} - 
				2 \epsilon_{1H\ast} \epsilon_{2H\ast} (\epsilon_{2H\ast} + \epsilon_{3H\ast}) + 
				\epsilon_{1Z\ast} \epsilon_{2Z\ast} (\epsilon_{2Z\ast} + \epsilon_{3Z\ast}) + 3 \epsilon_{1c\ast} \epsilon_{2c\ast} (\epsilon_{2c\ast} + \epsilon_{3c\ast}) \end{aligned}$
		\end{tabular}
	\end{ruledtabular}
\end{table*}

\section{Single-field inflation}
\label{sec:SingleField}

As a consistency check of our general formulas, let us consider the well-studied case of single-field inflation: a scalar field $\varphi$ with potential $V(\varphi)$, minimally coupled to Einstein gravity, that is, a system with action
\begin{align}
	\label{eq:Action-Scalar-Field}
S[g_{\mu\nu},\varphi] &= \frac{1}{16 \pi G} \int \dd[4]{x} \sqrt{-g} \, R \nonumber  \\
&\quad + \int \dd[4]{x} \sqrt{-g} \qty(- \frac{1}{2} \partial^\mu \varphi \partial_\mu \varphi - V(\varphi))\,. 
\end{align}
Once we choose a homogeneous and isotropic solution $\bar{g}_{\mu\nu}(t)$, $\bar{\varphi}(t)$, the action for the perturbations $\delta g_{\mu\nu}(\vb{x},t)$, $\delta \varphi(\vb{x},t)$ can be expanded to quadratic order, decomposed in SVT modes, and once gauge conditions are imposed and the constraints solved, it takes the form \eqref{eq:Quadratic-Action-psi}. The kinetic amplitude and the speed of sound for scalar and tensor\footnote{This form of $Z_{\textrm{t}}(t)$ already considers the trace over the two polarizations, i.e., an extra factor of 4 to the total power.} modes takes the form \cite{Mukhanov1992}:
\begin{align}
& Z_{\textrm{s}} (t) = \frac{\epsilon_{1H}(t)}{4\pi G}\,, \qquad c_{\textrm{s}} (t)=1\,,\label{eq:Z-s-SingleField}\\[.5em]
& Z_{\textrm{t}}(t) =\;\, \frac{1}{64\pi G}\,, \qquad c_{\textrm{t}} (t)=1\,.\label{eq:Z-t-SingleField}
\end{align}
Using these expressions, one can determine the full power spectrum for both scalar and tensor modes. From the results of Table~\ref{Tab:GenericTheory-Features}, we can compute the quantities characterizing power-law quantities, i.e., the tilt $\theta^{(\psi)}_\ast$,  the running $\alpha^{(\psi)}_\ast$, and the running-of-the-running $\beta^{(\psi)}_\ast$, discussed in Appendix \ref{App:Power-law-single-field}. The quantities for scalar modes are reported in Table~\ref{Tab:SingleField-ScalarFeatures} (with the scalar spectral index defined as $n_\textrm{s} \equiv 1 +  \theta_\ast^{(\textrm{s})}$ as usual). The power-law quantities for tensor modes are reported in Table~\ref{Tab:SingleField-TensorFeatures} (with the tensor spectral index defined  as $n_\textrm{t} = \theta_\ast^{(\textrm{t})}$ as usual). These expressions fully reproduce\footnote{Note that the extra minus signs in $\epsilon_{2H}$, $\epsilon_{3H}$ and $\epsilon_{4H}$ are simply due to the different sign in the definition of Hubble-flow parameters, as shown in Table \ref{Tab:dictionary}.} the state--of--the--art results of Auclair and Ringeval  \cite{Auclair2022} where they first derive the N3LO formula for tensor modes ($Z=\textrm{const}$, $c=\textrm{const}$), and then derive the formula for scalar modes ($Z(t)$, $c=\textrm{const}$) via a mapping method \cite{BeltranJimenez:2013ikr} from the one for tensor modes. Hence, as a consistency check, our formalism completely reproduces previous calculations for single-field inflation. 

In general, one could also consider an extension of the mapping method of \cite{BeltranJimenez:2013ikr} that applies to the effective action \eqref{eq:Quadratic-Action-psi} and, via a suitable redefinition of time and of the scale factor, brings it into a reference action with $Z_\psi=1$, $c_\psi=1$ for which N3LO results are already available \cite{Auclair2022}. Developing this method would provide an additional consistency check of the general formulas in Table~\ref{Tab:p0s}--\ref{Tab:GenericTheory-Features}. The new framework introduced here does not require a mapping and provides directly the N3LO expressions for the effective action \eqref{eq:Quadratic-Action-psi}. Moreover, by keeping the form of $Z_\psi(t)$ and $c_\psi(t)$ general, we obtain a single set of expressions that apply to the broad class of inflationary models of Table~\ref{Tab:Models}. Note that, since in any given model these functions are assumed to be determined by the Hubble rate $H(t)$ and its time-derivatives, the remaining non-trivial step
is to express the Hubble-flow parameters $\epsilon_{1Z}$, $\epsilon_{1c}$, and so on, in terms of the background Hubble-parameters $\epsilon_{1H}$, $\epsilon_{2H}$, etc. A concrete example of this procedure is described in the next section, for the particular case of Starobinsky inflation in the geometric framework, which requires a more sophisticated machinery in comparison with single-field inflation.

\section{Starobinsky inflation in the geometric framework}
\label{sec:Starobinsky}

The Starobinsky model \cite{Starobinsky1979,Starobinsky1980}  is described by the action for gravity with a higher-curvature term,
\begin{equation}
	\label{eq:Starobinsky_Action}
S[g_{\mu\nu}] = \frac{1}{16\pi G} \int \dd[4]{x} \sqrt{-g} \, (R+\alpha R^2)\,.
\end{equation}
It is the oldest proposed model of inflation, originally motivated by quantum-gravity considerations on the renormalization of the energy-momentum tensor. To date, it provides an accurate account of primordial-power-spectrum observations in terms of one single parameter  \cite{Akrami2020}, the coupling constant $\alpha$ of dimensions $[\alpha]= \textrm{length}{}^2$. The theory is purely gravitational, with the inflationary phase driven by the higher-order curvature term, without the need of any additional inflaton field. The technique generally used for computing the predictions of the power spectrum for this model does not directly use the geometric framework (or Jordan frame) described by \eqref{eq:Starobinsky_Action}, but instead involves a mapping to an action of the form \eqref{eq:Action-Scalar-Field} via a field redefinition $g_{\mu\nu}\to (\tilde{g}_{\mu\nu},\varphi)$. The auxiliary metric $\tilde{g}_{\mu\nu}$ (Einstein frame) is conformally related to the metric $g_{\mu\nu}$ and the potential $V(\varphi)$ depends only on the single parameter $\alpha$ \cite{Whitt1984,Vilenkin1985}. While a field redefinition can simplify calculations without affecting physical predictions (once the same observable is identified in the new variables) \cite{Capozziello1997,Faraoni1999,Karam2017,Ketov:2024klm,Toyama:2024ugg}, it is important to remark that observations of the reheating phase can in principle distinguish between the minimal coupling of the metric $g_{\mu\nu}$ to the standard model of particle physics, as opposed to the minimal coupling to the auxiliary metric $\tilde{g}_{\mu\nu}$ \cite{Martin:2024qnn,Dorsch:2024nan}. The goal of this section is to use the formalism introduced in the previous sections to compute the power spectrum of Starobinsky inflation at N3LO, working purely in the geometric framework \cite{DeFelice2010} and expressing all observables in terms of the number of inflationary e-foldings $N_\ast$ measured with respect to the metric $g_{\mu\nu}$.

The variational principle for the action \eqref{eq:Starobinsky_Action} results in the Einstein equation with a higher curvature term,
\begin{equation}
G_{\mu\nu}\,+\,\alpha\,\mathcal{H}_{\mu\nu}\,=\,0\,,
\label{eq:G+H}
\end{equation}
in vacuum ($T_{\mu\nu}=0$) and with the covariantly conserved tensors  ($\nabla_\mu G^{\mu\nu}=0$ and $\nabla_\mu \mathcal{H}^{\mu\nu}=0$) defined by
\begin{align}
G_{\mu\nu}&\;=\frac{1}{\sqrt{-g}}\frac{\delta}{\delta g^{\mu\nu}}\int \dd[4]{x} \sqrt{-g} \, R\,,\\[.5em]
\mathcal{H}_{\mu\nu}&\;=\frac{1}{\sqrt{-g}}\frac{\delta}{\delta g^{\mu\nu}}\int \dd[4]{x} \sqrt{-g} \, R^2\,,
\end{align}
and given by
\begin{align}
G_{\mu\nu}&\;= R_{\mu\nu}-\tfrac{1}{2} R\, g_{\mu\nu}\,,\label{eq:G-def}\\[.5em]
\mathcal{H}_{\mu\nu}&\;= 2\big(R\,G_{\mu\nu}-\nabla_{(\mu} \nabla_{\nu )}R+ (\square R+\tfrac{1}{4} R^2)g_{\mu\nu}\big) \,,\label{eq:H-def}
\end{align}
where $\square=g^{\mu\nu}\nabla_\mu\nabla_\nu$.

\subsection{Background dynamics}

\begin{figure}[t]
	\centering
	\includegraphics[width=\linewidth]{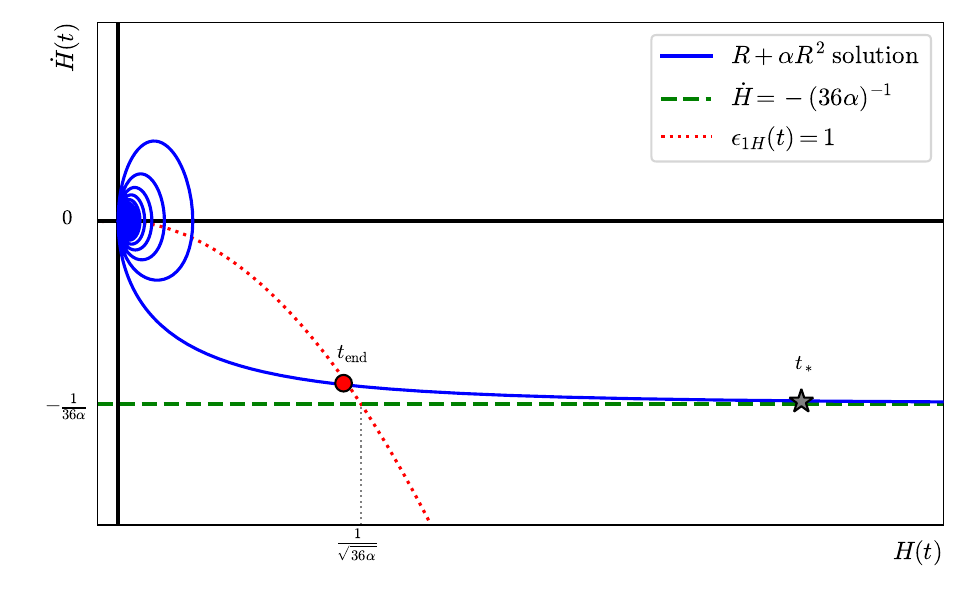}
	\centering
	\caption{Diagram of a typical solution of $R+\alpha R^2$ inflation in the $(H,\dot{H})$ plane. The red dot indicates the end of the inflationary phase, while the gray star illustrates a point associated to a pivot scale $k_\ast$. When $\epsilon_{1H}\to0$, $\dot{H}$ reduces to $-1/36\alpha$, as shown by the green dashed line. The dotted gray line indicates the approximate scale $H\approx 1/\sqrt{36\alpha}$. }
	\label{fig:starobinsky}
\end{figure}

Evaluating the vacuum Einstein equations \eqref{eq:G+H} on the FLRW metric \eqref{eq:FLRW-Metric}, we obtain the Friedmann equation with the Starobinsky higher-curvature term,
\begin{equation}
	\label{eq:Friedmann_Modified}
H(t)^2 + 6\alpha\, H(t)^4\, \epsilon_{1H}(t)\, \big(3 \epsilon_{1H}(t) + 2\epsilon_{2H}(t) - 6\big) = 0\,.
\end{equation}
This theory admits an inflationary phase with approximately constant $\dot{H}\approx -1/36\alpha$  \cite{Ruzmaikina1969}, as shown in the $(H,\dot{H})$ plot in Fig.~\ref{fig:starobinsky}.  From the Friedmann equation \eqref{eq:Friedmann_Modified}, we can find a systematic and self-consistent expansion of $H(t)$ in terms of $\epsilon_{1H}(t)$,
\begin{align}
\label{eq:H-to-eps}
	H(t) &= \frac{1}{6\sqrt{\alpha\, \epsilon_{1H}(t)}} \left[  1 - \frac{1}{12} \epsilon_{1H}(t)+ \frac{19}{288} \epsilon_{1H}(t)^2  \right.  \nonumber \\
	& \;\;\;\left. - \frac{373}{3456} \epsilon_{1H}(t)^3 + \frac{44035}{165888} \epsilon_{1H}(t)^4 + \order{\epsilon^5} \right] .
\end{align}
The derivation of the above expression is discussed in Appendix \ref{app:derivation-H(eps)}. Similarly, for $\epsilon_{2H}(t)$ and $\epsilon_{3H}(t)$, we find
\begin{align}
	\label{eq:Epsilon_Expansion_234}
\epsilon_{2H}(t)&= -2 \epsilon_{1H}(t) + \frac{1}{3} \epsilon_{1H}(t)^2 - \frac{5}{9} \epsilon_{1H}(t)^3 \nonumber \\ &\quad + \frac{38}{27} \epsilon_{1H}(t)^4 + \order{\epsilon^5}\,, \nonumber \\
\epsilon_{3H}(t) &= -2 \epsilon_{1H}(t) + \frac{2}{3} \epsilon_{1H}(t)^2 - \frac{5}{3} \epsilon_{1H}(t)^3 + \order{\epsilon^4} \,,\nonumber \\
\epsilon_{4H}(t) &= -2   \epsilon_{1H}(t)  +  \epsilon_{1H}(t)^2 + \order{\epsilon^3}\,.
\end{align}
It can be directly checked that the expression \eqref{eq:H-to-eps} in combination with \eqref{eq:Epsilon_Expansion_234} solve the Friedmann equation \eqref{eq:Friedmann_Modified}, up to $\epsilon^3$ corrections. Inflation ends at a time $t_{\mathrm{end}}$ defined by $\ddot{a}(t_{\mathrm{end}})=0$ or, in terms of Hubble flow parameters \eqref{eq:ddot-a}, when $\epsilon_{1H}(t_\textrm{end}) = 1$. The expansion from a reference time $t_\ast$ during inflation until the end of inflation or, equivalently, the e-folding number $N_\ast$ in 
\begin{equation}
a(t_{\mathrm{end}})=\ee^{N_\ast}\,a(t_\ast)\,,
\end{equation}
can be computed by noticing that $N_\ast$ can be written as
\begin{equation}
N_\ast \equiv\int_{t_\ast}^{t_{\textrm{end}}} H(t) \dd{t} = - \int_{\epsilon_{1H\ast}}^{1} \frac{\dd{\epsilon_{1H}}}{ \epsilon_{1H}\;\, \epsilon_{2H}\big(\epsilon_{1H}\big)  },
\end{equation}
where the second Hubble-flow parameter is expressed as a function of the first, $\epsilon_{2H}=\epsilon_{2H}(\epsilon_{1H})$, using \eqref{eq:Epsilon_Expansion_234}. Integrating order-by-order we find
\begin{align}
	\label{eq:Ns-to-eps}
N_\ast (\epsilon_{1H\ast}) &= \frac{1}{2\epsilon_{1H\ast}} + \frac{1}{8} \epsilon_{1H\ast}+ \frac{19}{864} \epsilon_{1H\ast}^2- \frac{71}{7776} \epsilon_{1H\ast}^3 \nonumber \\[.5em]
&\quad - \frac{\ln(\epsilon_{1H\ast})}{12} + D_0 + \order{\epsilon_{1H\ast}^4},
\end{align}
with  
\begin{align*}
 D_0& \textstyle \; =-\frac{1}{2}+\frac{\ln (2)}{24}+\frac{\ln(3)}{12}-\frac{\ln(20)}{24}\\[.5em]
&\textstyle\qquad -\frac{19 \tan ^{-1}\left(\frac{1}{\sqrt{39}}\right)}{12 \sqrt{39}}-\frac{19 \cot ^{-1}\left(\frac{3 \sqrt{39}}{7}\right)}{12 \sqrt{39}} \simeq -0.635\,.
\end{align*}
It is clear that, for small values of $\epsilon_{1H\ast}$, the number of e-foldings is determined by the first term but, in our analysis of the N3LO power spectrum, we will need also the higher order terms. The relation \eqref{eq:Ns-to-eps} can be perturbatively inverted, to find the following expression
\begin{align}
	\label{eq:eps-to-Ns}
&\epsilon_{1H\ast} (N_\ast) = \frac{1}{2N_\ast} + \frac{\frac{D_0}{2} - \frac{1}{24} \ln(\frac{1}{2N_\ast})}{N_\ast^2}\; +\nonumber \\[.5em]
&\;\; + \frac{D_1 -D_2 \ln(\frac{1}{2N_\ast}) +  \frac{1}{288}  \ln(\frac{1}{2N_\ast})^2 }{N_\ast^3} 
+ \mathcal{O}(N_\ast^{-4}),
\end{align}
where $D_1 = (3-4D_0 + 48D_0^2)/96 \simeq 0.259$, and $D_2 = (-1+24 D_0)/(288) \simeq -0.056$. Again, for large values of $N_\ast$, the main contribution comes from the first term.  In the range $N_\ast \in [50,60]$, we have the associated range $\epsilon_{1H\ast} \in [0.00995, 0.00830]$. By combining the expansions \eqref{eq:Epsilon_Expansion_234} with the expression for $\epsilon_{1H\ast}$ given in \eqref{eq:eps-to-Ns}, we can express all the features of the power spectra for Starobinsky inflation in terms of $N_\ast$. As our N3LO calculations can be trusted only up to order $\order{\epsilon^3}$, a truncation of the cosmological observables up to order $\order{N_\ast^{-3}}$ will remain consistent for predictive purposes. In this way, the N3LO corrections allow us to check and improve the precision of the predictions in the geometric frame with respect to the known expressions of order $\order{N^{-2}_\ast}$ in the Einstein frame.

\subsection{Perturbations}
We derive the quadratic action for SVT perturbations in Starobinsky inflation, working purely in the geometric framework. The starting point is the tensor $F^{\mu\nu} [g_{\mu\nu}]$ obtained from the variation of the action \eqref{eq:Starobinsky_Action},
\begin{equation}
F^{\mu\nu} [g_{\mu\nu}] \equiv\fdv{S}{g_{\mu\nu}}\,=\,-\frac{1}{16\pi G}\sqrt{-g}\,\big(G^{\mu\nu}+\alpha \mathcal{H^{\mu\nu}}\big)\,,
\end{equation}
where we used \eqref{eq:G-def}--\eqref{eq:H-def} and $\delta g_{\mu\nu}=-g_{\mu\alpha}g_{\nu\beta}\delta g^{\alpha\beta}$. Expanding around the FLRW metric \eqref{eq:FLRW-Metric}, we write 
\begin{equation*}
F^{\mu\nu} [\bar{g}_{\mu\nu}+\delta g_{\mu\nu}] =\bar{F}_0^{\mu\nu}(t)\,+\,\bar{F}_1^{\mu\nu\rho\sigma}(t)\,\delta g_{\rho\sigma}+\mathcal{O}(\delta g^2)\,.
\end{equation*}
As we assume that the background metric satisfies the Friedman equation \eqref{eq:Friedmann_Modified}, the term $\bar{F}_0^{\mu\nu}=0$ vanishes. Therefore, the action \eqref{eq:Starobinsky_Action}, at quadratic order in the perturbation, can be written as
\begin{equation}
S[\bar{g}_{\mu\nu}+\delta g_{\mu\nu}]=\bar{S}(t)+\int\!\dd[4]{x}\tfrac{1}{2}\,\delta g_{\mu\nu}\bar{F}_1^{\mu\nu\rho\sigma}(t)\,\delta g_{\rho\sigma}+\mathcal{O}(\delta g^3)\,.
\label{eq:S-pert}
\end{equation}
We can then use the homogeneity and isotropy of the background to organize perturbations into SVT representations of the $3d$ Euclidean group, $\delta g_{\mu\nu}=\delta g^{(\mathrm{s})}_{\mu\nu}+\delta g^{(\mathrm{v})}_{\mu\nu}+\delta g^{(\mathrm{t})}_{\mu\nu}$ which in the quadratic action decouple, as most easily shown by working with the Fourier transforms.

\medskip

Let us consider scalar perturbations first. In the Fourier domain \eqref{eq:Fourier},  using the Arnowitt–Deser–Misner (ADM) variables $g_{\mu\nu}\dd x^\mu \dd{x}^\nu = - N^2 \dd{t}^2 + h_{ij} (N^i \dd{t} + \dd{y}^i) (N^j \dd{t} + \dd{y}^j) $ (with $i,j=1,2,3$), the perturbation $\delta g^{(\mathrm{s})}_{\mu\nu}$ can be written in terms of the lapse $N=1+\delta N$ and of the shift $N^i=0+\delta N^i$, with scalar perturbations $\delta N$ and $S$,
\begin{align}
&\delta N (\vb{k},t),\\
&\delta N^i (\vb{k},t) = \ii \,k^i S(\vb{k},t) \,
\end{align}
together with the $3d$ metric $h_{ij}=a(t)^2 \delta_{ij}+ \delta h_{ij}^{(\mathrm{s})}$, with scalar perturbations $\mathcal{R}$ and $\mathcal{C}$
\begin{equation}
 \delta h_{ij}^{(\mathrm{s})} (\vb{k},t) = 
-2\, \mathcal{R}(\vb{k},t) a(t)^2 \delta_{ij} - k_i k_j \,\mathcal{C}(\vb{k},t) \,.
\end{equation}
We work in the comoving gauge $\mathcal{H}^{0}{}_i=0$, which generalizes the comoving gauge $T^{0}{}_i=0$ for the energy-momentum in general relativity with matter. Solving perturbatively the Hamitonian constraint $F^0{}_0\approx 0$ and the diffeomorphism constraint $F^0{}_i\approx 0$, we can express the scalar perturbations $\delta N (\vb{k},t)$, $\delta N^i$, and $\mathcal{C}$ in terms of the curvature perturbation $ \mathcal{R}$.  At first order in the perturbation, constant-$t$ spatial sections have scalar curvature given by the $3d$ Ricci scalar ${}^{(3)\!}R=4a(t)^{-2}\,\delta^{ij}\partial_i \partial_j \mathcal{R}$. Substituting these expressions into \eqref{eq:S-pert} and introducing the useful definitions
\begin{align}
\chi(t) &\equiv 1+2\alpha \bar{R}(t)= 1+24\alpha H(t)^2 (1-\tfrac{1}{2}\epsilon_{1H}(t))\,,\\[.5em]
\epsilon_\chi(t) &= - \frac{\dot{\chi}(t)}{H(t)\,\chi(t)}\,,
\end{align}
we find that the quadratic action for the single scalar mode, the curvature perturbation $\mathcal{R}(\vb{k},t$), takes the form \eqref{eq:Quadratic-Action-psi} with kinetic amplitude and speed of sound:
\begin{align}
Z_{\textrm{s}}(t) \, &= \frac{3\chi(t)}{16\pi G}  \qty(\frac{\epsilon_\chi(t)}{1+\frac{1}{2}\epsilon_\chi(t)})^2\,, 	\label{eq:Z-s-Starobinsky}\\[.5em]
c_{\textrm{s}}(t) \, &=\,1\,.
\end{align}
\begin{figure*}[t]
	\centering
	\hspace{-1em}
	\includegraphics[width=0.95\linewidth]{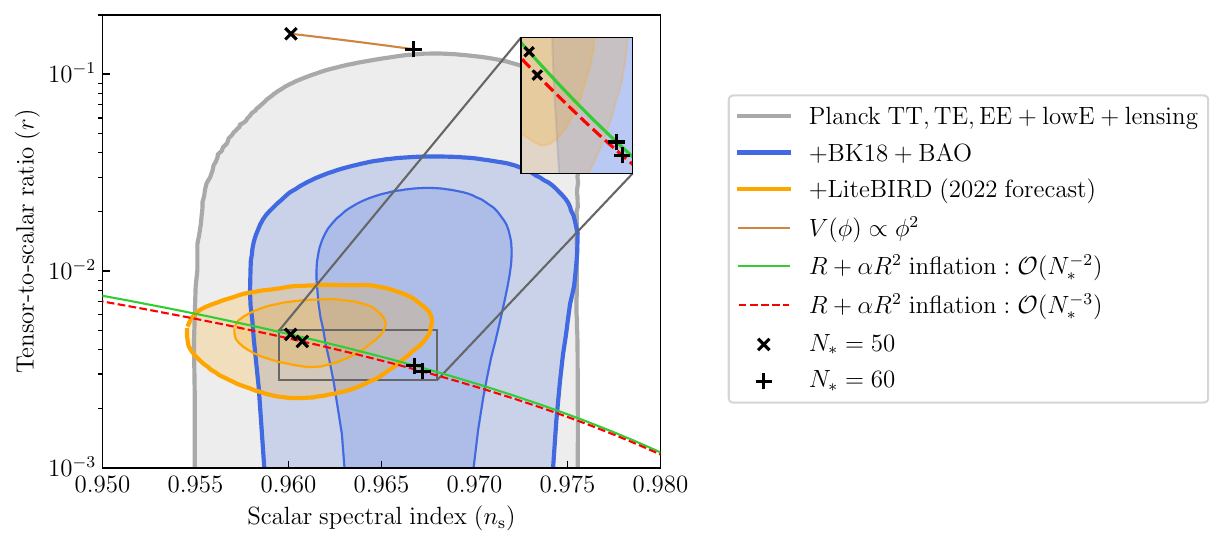}
	\caption{Marginalized joint 68\% and 95\% C.L. regions for $n_{\mathrm{s}}$ and $r$ at $k_\ast = 0.002$ Mpc$^{-1}$ as reported by the Planck Collaboration \cite{Akrami2020}, and the BICEP2/Keck Collaboration \cite{KeckCollaboration2021}. The orange region represents the forecast of the upcoming LiteBIRD experiment for a fiducial model with $r=0.005$ \cite{LiteBIRDCollaboration2022}. Note that the C.L. regions are obtained assuming a power spectrum of the form $ \mathcal{A}_{\mathrm{s}}(k/k_\ast)^{n_{\mathrm{s}}-1}$. Our results for Starobinsky inflation up to N3LO, in the $r$-$n_{\mathrm{s}}$ plane, are shown in the dashed red line. We note the $7\%$ decrease for $r$ and $0.05\%$ increase for $n_\mathrm{s}$ with respect to the standard NLO expressions, for $N_\ast = 55$. }
	\label{fig:rnsplane}
\end{figure*}
For vector perturbations, working in the same comoving gauge, introducing the transverse vector fields $\delta N^a_T$  for the shift and $B_a(\vb{x},t)$ for the ADM metric $ \delta h_{ab}^{(\mathrm{v})} (\vb{k},t)  = \ii\, a(t) (   
 k_a B_b(\vb{x},t) + k_b B_a(\vb{k},t) )$, and solving the transverse part of the diffeomorphism constraint, one finds as usual that there is no propagating vectorial perturbation. Finally, for transverse-traceless tensor perturbations
 \begin{equation}
  \delta h_{ab}^{(\mathrm{t})} (\vb{k},t)  =  e_{ab}^{(+)} (\vb{k}) \gamma_{(+)} (\vb{k},t) + e_{ab}^{(-)} (\vb{k}) \gamma_{(-)} (\vb{k},t)\,,
 \end{equation}
 one finds again that the action takes the form \eqref{eq:Quadratic-Action-psi} with kinetic amplitude and speed of sound:
\begin{align}
Z_{\textrm{t}} (t)\,&= \frac{\chi(t)}{64 \pi G}\,, \label{eq:Z-t-Starobinsky}\\[.5em]
c_{\textrm{t}}(t) \, &=\,1\,.
\end{align}
As a check of this expression, note that in the geometric framework for Starobinsky inflation discussed here, the kinetic amplitude of tensor modes \eqref{eq:Z-t-Starobinsky} reduces to the familiar one in general relativity \eqref{eq:Z-t-SingleField} in the limit $\alpha\to 0$.
Note also that these expressions for the kinetic amplitude and speed of sound are exact as we did not use up to this point any Hubble-flow expansion. Since $c_{\mathrm{s}} = 1$, $c_{\mathrm{t}} = 1$, we have $\tau = \eta$, and the pivot scale considered in this case is the same for both scalar and tensor modes, as previously discussed, so no further shift is needed to compare the predictions to data. More explicitly, the pivot time $t_\ast$ considered in this section is such that the associated pivot scale $k_\ast$ is defined by $k_\ast \, \eta (t_\ast) = -1$, where the conformal time $\eta$ is fully expanded in \eqref{eq:eta-qdS-full}.

\subsection{Power spectrum}

Let us now use the Hubble-flow expansion to express the kinetic amplitude in terms of a series in the single parameter $\epsilon_{1H\ast}$, the first Hubble flow parameter evaluated at the pivot time $t_\ast$. For scalar perturbations, we find
\begin{align}
	Z_{\textrm{s}\ast} &= \frac{\epsilon_{1H\ast}}{2 \pi G} \left[ 1-\frac{19}{6} \epsilon_{1H\ast} + \frac{74}{9}\epsilon_{1H\ast}^2  \right]  + \order{\epsilon_{1H\ast}^4}\,,
\end{align}
with Hubble-flow parameters for $Z_{\textrm{s}}$ given by
\begin{align}
	\epsilon_{1Z\ast}^{(\textrm{s})}&=-2\epsilon_{1H\ast} + \frac{20}{3}\epsilon_{1H\ast}^2 - \frac{130}{9} \epsilon_{1H\ast}^3 + \order{\epsilon_{1H\ast}^4} \,,\nonumber \\
	\epsilon_{2Z\ast}^{(\textrm{s})} &=-2\epsilon_{1H\ast} + 7 \epsilon_{1H\ast}^2 - \frac{25}{3} \epsilon_{1H\ast}^3 + \order{\epsilon_{1H\ast}^4}\,, \nonumber \\ 
	\epsilon_{3Z\ast}^{(\textrm{s})}&=-2\epsilon_{1H\ast} + \frac{22}{3}\epsilon_{1H\ast}^2 + \frac{37}{3} \epsilon_{1H\ast}^3 + \order{\epsilon_{1H\ast}^4}\,. 
\end{align}
For tensor perturbations, we find
\begin{align}
Z_{\textrm{t} \ast} &= \frac{1}{96 \pi G\epsilon_{1H\ast} } \left[ 1 + \frac{5}{6} \epsilon_{1H\ast} + \frac{2}{9} \epsilon_{1H\ast}^2     \right.  \nonumber \\
&\qquad \left.  - \frac{8}{27} \epsilon_{1H\ast}^3 + \frac{2}{3} \epsilon_{1H\ast}^4 \right]  + \order{\epsilon_{1H\ast}^4}\,,
\end{align}
with Hubble-flow parameters for $Z_{\textrm{t}}$ given by
\begin{align}
\epsilon_{1Z\ast}^{(\textrm{t})}&=2\epsilon_{1H\ast} - 2\epsilon_{1H\ast}^2 + \frac{4}{3} \epsilon_{1H\ast}^3 + \order{\epsilon_{1H\ast}^4}\,, \nonumber \\
\epsilon_{2Z\ast}^{(\textrm{t})} &=-2\epsilon_{1H\ast} + \frac{7}{3}\epsilon_{1H\ast}^2 - \frac{14}{9} \epsilon_{1H\ast}^3 + \order{\epsilon_{1H\ast}^4} \,,\nonumber \\
\epsilon_{3Z\ast}^{(\textrm{t})}&=-2\epsilon_{1H\ast} + \frac{8}{3}\epsilon_{1H\ast}^2 - \frac{4}{3} \epsilon_{1H\ast}^3 + \order{\epsilon_{1H\ast}^4}\,. 
\end{align}
We can substitute the expressions found above into the the general formulas reported in Table~\ref{Tab:GenericTheory-Features} and \ref{Tab:GenericTheory-Amplitude}, to find the N3LO expressions of the power-law quantities for $R+\alpha R^2$ inflation. Then, equation \eqref{eq:eps-to-Ns} can be used to truncate the results in terms of the number of e-foldings until the end of inflation, $N_\ast$. In this way, a N3LO computation gives us a self-consistent and reliable truncation of the expressions, as long as it is taken up to order $N_\ast^{-3}$. Furthermore, given the phenomenological success of  $R+\alpha R^2$ inflation in accounting for current cosmological observations of primordial perturbations, it is useful to comment on the precision of its predictions. For this goal, let us consider a fiducial value of $N_\ast = 55$, and use it to compare different truncations allowed by the N3LO calculation. The numerical results are shown below in Table \ref{Tab:Starobinsky-values}.

\begin{table}[h]
	\caption{Values of power-law quantities for Starobinsky inflation in the geometric frame with a fiducial number of e-foldings $N_\ast =55$. In the N3LO calculations, we can trust the truncations up to order $N_\ast^{-3}$, according to the Table \ref{Tab:Starobinsky-Results}. We report the explicit numerical values for different truncations, illustrating the improvement from NLO to N3LO.}
	\label{Tab:Starobinsky-values}
	%	\rowcolors{2}{gray!25}{white}
	\begin{ruledtabular}
		\begin{tabular}{c c c c}
			Quantity &$\order{N_\ast^{-1}}$ & $\order{N_\ast^{-2}}$ & $\order{N_\ast^{-3}}$  \\
            \hline \noalign{\vskip 1pt}
			 $n_{\mathrm{s}}$ &$0.9636$ & $0.9642$ & $0.9642$  \\
    		\hline \noalign{\vskip 1pt}
            $r$ & $0$ & $3.967\times 10^{-3}$ &$3.694\times 10^{-3}$ \\
            \hline \noalign{\vskip 1pt}
            $n_{\mathrm{t}}$ & $0$ & $-4.959\times 10^{-4}$ & $-4.964\times 10^{-4}$  \\
            \hline \noalign{\vskip 1pt}
            $r+8n_{\mathrm{t}}$ & $0$ & $0$ & $-2.776\times 10^{-4}$\\
            \hline \noalign{\vskip 1pt}
            $\alpha_{\mathrm{s}}$& $0$ & $-6.612\times 10^{-4}$ & $-6.468\times 10^{-4}$ \\
            \hline \noalign{\vskip 1pt}
            $\alpha_{\mathrm{t}}$ & $0$  & $-1.803\times 10^{-5}$ & $-1.803\times 10^{-5}$  \\
            \hline \noalign{\vskip 1pt}
            $\beta_{\mathrm{s}}$ & $0$ & $0$ & $-2.404\times 10^{-5}$  
		\end{tabular}
	\end{ruledtabular}
\end{table}

Note that the order $\order{N^{-3}_\ast}$ correction to the tensor-to-scalar ratio is non-negligible and results in a decrease by  $7\%$ with respect to the value at order $\order{N^{-2}_\ast}$, see also Fig.~\ref{fig:rnsplane}. The standard $\order{N_\ast^{-2}}$ result is $r\approx 12/N_\ast^2$ 
 \cite{DeFelice2010} and one might expect that the correction has simply an extra $1/N_\ast$ factor. This is not the case as, in fact, the detailed calculation (Table~\ref{Tab:Starobinsky-Results}) shows that the correction comes with a large coefficient and also a logarithmic correction which cannot be neglected for $N_\ast= 55$.

\begin{figure}[t]
	\centering
	\includegraphics[width=\linewidth]{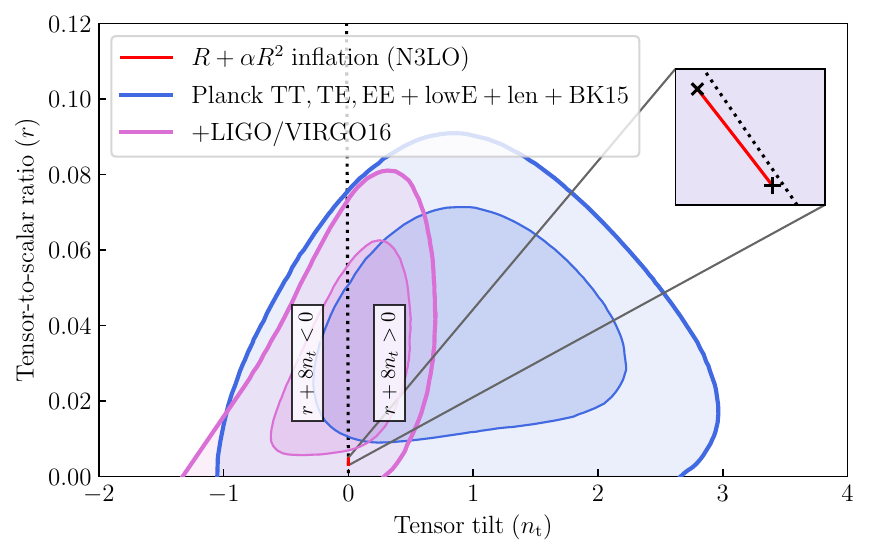}
	\caption{Marginalized joint 68\% and 95\% C.L. regions for $n_t$ and $r$ at $0.01$ Mpc$^{-1}$ as reported by the Planck Collaboration, assuming a power spectrum of the form $r\,\mathcal{A}_{\mathrm{s}} (k/k_\ast)^{n_{\mathrm{t}}}$ \cite{Akrami2020}. The purple region of LIGO/VIRGO is associated to the constraints on the stochastic gravitational-wave background, i.e., $\Omega_{\textrm{GW}}$ \cite{LSC2016}. The dotted black line illustrates the exact consistency relation $r = -8 n_t$. The symbol $\cross$ ($+$) indicates $N_\ast = 50\, (60)$ for Starobinsky inflation.}
	\label{fig:rntPlane}
\end{figure}

\begin{figure}[t]
	\centering
	\includegraphics[width=\linewidth]{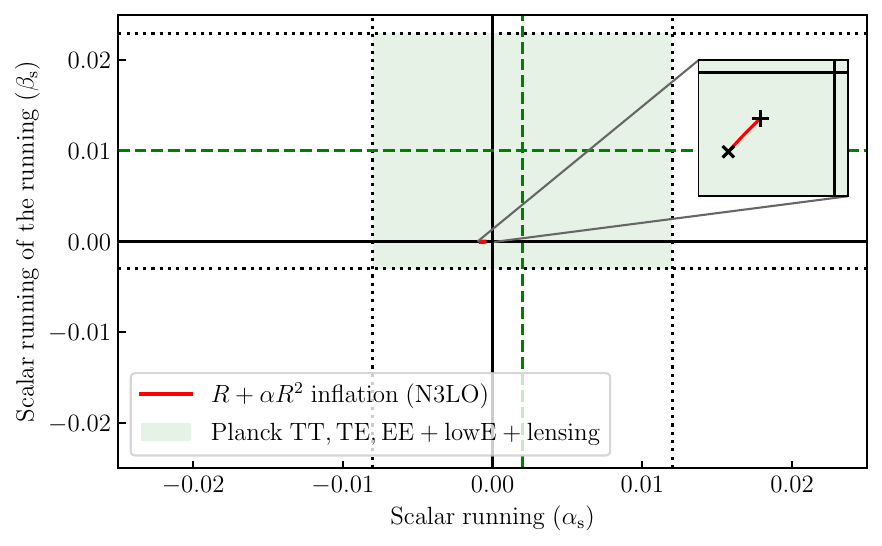}
	\caption{Schematic visualization of the constraints on the running and the running of the running of the scalar tilt. The solid black lines indicate $\alpha_{\textrm{s}} = \beta_{\textrm{s}}= 0$, and the dashed green lines the values $\alpha_{\textrm{s}} = 0.002 \pm 0.010$ and $\beta_{\textrm{s}} = 0.010\pm 0.013$, as reported by the Planck Collaboration, both at 68\% C.L.   \cite{Akrami2020}.  The $\cross$ ($+$) indicate $N_\ast = 50\, (60)$ for Starobinsky inflation. }
	\label{fig:alphabetaPlane}
\end{figure}

Our calculation also allows us to identify the order of magnitude of violation of the single-field consistency condition, generally stated as $r=-8n_\textrm{t}$  at LO \cite{Lidsey:1995np}. The formalism developed in this work provides a precise prediction of the amount of deviation from this condition for $R+\alpha R^2$ inflation, $\delta \equiv r + 8n_{\textrm{t}}  = - 48/{N_\ast^3} + \order{N_\ast^{-4}}$.
The values of $\delta$ and $n_{\mathrm{t}}$ up to order $\order{N^{-3}_\ast}$  are reported in Table \ref{Tab:Starobinsky-values}, and can be compared to the constraints imposed by Planck and LIGO/VIRGO on $r$ and $n_{\mathrm{t}}$, as shown in Fig.~\ref{fig:rntPlane}. Moreover, we find that $R+\alpha R^2$ inflation predicts a value of the running and running of the running for the scalar power spectrum, also reported in Table \ref{Tab:Starobinsky-values}. These values can also be compared with current constraints reported by Planck, as illustrated in Fig.~\ref{fig:alphabetaPlane}. Note also that the predicted value of the running $\alpha_{\textrm{s}}$ is negative and consistent with the $68\%$ C.L. interval $\alpha_{\textrm{s}} = (-6.75\pm 2.05)\times 10^{-4}$ recently obtained in  \cite{Martin:2024nlo} using the posterior probability distribution marginalized over nearly $300$ models of single-field inflation.

Furthermore, since the amplitude of curvature perturbation is constrained to be $\ln(10^{10} \mathcal{A}_{\textrm{s}}) = 3.044\pm 0.014 $, the corresponding value of the coupling constant $\alpha$ is $\alpha = 2.663\times 10^{10}\, G \hbar \simeq (2.7 \times 10^{-30} \, \textrm{m})^2$ for $N_\ast = 55$. It is interesting to remark also that if in the near future an amplitude $\mathcal{A}_\mathrm{t} \sim G\hbar/\alpha$ of tensor modes is observed, it will provide evidence for the quantization of gravity \cite{Krauss2014}. The geometric framework discussed here highlights how the observed amplitude $\mathcal{A}_\mathrm{s} \sim (G\hbar/\alpha )N_\ast^2$ of scalar perturbations via CMB temperature anisotropy already provides a probe of (perturbative) quantum gravity, as implied the Planck area $\ell_P^2=G\hbar$ in this expression.

\begin{table*}[t]
	\caption{Curvature and tensor perturbations for $R+\alpha R^2$ inflation, up to N3LO. }
	\label{Tab:Starobinsky-Results}
	\begin{ruledtabular}
		\begin{tabular}{p{2in} p{3.25in} }
			Quantity & Coefficients\footnote{Recall that $C\simeq -0.72963715$, $D_0 \simeq -0.63530380$, $D_1\simeq 0.25952645$, and $D_2 \simeq -0.056414205$. } \\
			\hline
			$\begin{aligned}
				\mathcal{A}_{\textrm{s}} &= \frac{G \hbar\, N_\ast^2}{18 \pi  \alpha}  \Bigg[ 1+ \frac{ p_{1}^{(\textrm{s})}    }{N_\ast}  - \frac{\ln( 2N_\ast)}{6N_\ast} \\ 
				& \quad  
				+\frac{p_{2}^{(\textrm{s})}}{N_\ast^2}  - \, p_{2L}^{(\textrm{s})} \frac{\ln(2N_\ast)}{N_\ast^2} +  \frac{\ln(2N_\ast)^2}{144N_\ast^2}  
				\\
				& \quad- p_{3L}^{(\textrm{s})} \frac{\ln(2N_\ast)}{N_\ast^3}  + p_{3L^2}^{(\textrm{s})}  \frac{\ln(2N_\ast)^2}{N_\ast^3} + \frac{\ln(2N_\ast)^3}{864N_\ast^3} \\
				& \quad+ \frac{p_{3}^{(\textrm{s})}}{N_\ast^3}  + \order{N_\ast^{-4}} \Bigg] 
			\end{aligned}$
			&$ \begin{aligned}
				p_1^{(\textrm{s})} &= 1/2 - 2 C - 2 D_0\simeq 3.2298819  \nonumber \\
				p_2^{(\textrm{s})}   &= -(283/48) + C^2 - D_0/2 + 3 D_0^2 \nonumber \\
				&\quad + C (-(4/3) + 2 D_0) - 4 D_1 + (
				7 \pi^2)/12 \nonumber \\
				&\simeq 2.7841172 \nonumber \\
				p_{2L}^{(\textrm{s})}  &=  (1 - 4 C - 12 D_0 + 96 D_2)/24 \simeq 0.25526794 \nonumber \\
				p_{3L}^{(\textrm{s})}  &= D_0^2 - D_1 + 1/12 D_0 (-1 + 4 C - 144 D_2) \nonumber \\
				&\quad + D_2 - 4 C D_2 \simeq -0.29960351 \nonumber \\
				p_{3L^2}^{(\textrm{s})}  &= -(D_0/24) + D_2 \simeq  -0.029943214 \nonumber \\
				p_{3}^{(\textrm{s})}  & = -(553/432) + (5 C^2)/12 + D_0^2/2 - 4 D_0^3 \nonumber \\
				&\quad - D_1 + 12 D_0 D_1 + 
				C [-(127/72) - 2 D_0^2 \nonumber \\
				&\quad + 4 D_1] + (25 \pi^2)/48 - 6 \zeta(3)  \nonumber \\
				&\simeq -3.0222532. 
			\end{aligned}$
			\\
			\hline
			$\vspace{0.2em}\begin{aligned}~\\[-3ex]  n_\textrm{s} &= 1 - \frac{2}{N_\ast} + \frac{\theta_2^{(\textrm{s})}}{N_\ast^2} - \frac{\ln(2N_\ast)}{6N_\ast^2} \nonumber \\
				&\quad - \theta_{3L}^{(\textrm{s})}\frac{\ln(2N_\ast )  }{N_\ast^3}  - \frac{\ln(2N_\ast )^2  }{72N_\ast^3} \nonumber \\
				&\quad + \frac{\theta_{3}^{(\textrm{s})}}{N_\ast^3} + \order{N_\ast^{-4}}
			\end{aligned}$ 
			& $\begin{aligned}
				\theta_{2}^{(\textrm{s})} &= -(1/3) - 2 C - 2 D_0 \simeq 2.3965486 \nonumber\\
				\theta_{3L}^{(\textrm{s})} &=1/18 + C/3 + 4 D_2 \simeq -0.41331365 \nonumber \\
				\theta_{3}^{(\textrm{s})} &= -(241/18) - 2 C^2 - (2 D_0)/3 - C (3 + 8 D_0)/2 \nonumber \\
				&\quad - 4 D_1 + (7 \pi^2)/6 \simeq -4.3133704. 
			\end{aligned}$
			\\
			\hline
			$\vspace{0.2em}\begin{aligned} ~\\[-3ex] 
				\alpha_{\textrm{s}}= 	- \frac{2}{N_\ast^2} - \frac{\ln(2N_\ast)}{3N_\ast^3} + \frac{\alpha_{3}^{(\textrm{s})}}{N_\ast^3} + \order{N_\ast^{-4}}
			\end{aligned}$ 
			& $\alpha_3^{(\textrm{s})} = -(3/2) - 4 C - 4 D_0 \simeq 3.9597638$
			\\
			\hline 
			$\vspace{0.2em}\begin{aligned} ~\\[-3ex] & 
				\beta_{\textrm{s}}= - \frac{4}{N_\ast^3} + \order{N_\ast^{-4}}
			\end{aligned}$\\
			\hline\hline	
			$\vspace{0.2em}\begin{aligned}~\\[-3ex] 
				\mathcal{A}_{\textrm{t}} &=  \frac{2 G \hbar}{3 \pi \alpha} \Bigg[  1 - \frac{3}{2N_\ast}  \\
				&\quad + \frac{p_2^{(\textrm{t})}}{N_\ast^2}  - \frac{\ln(2N_\ast)}{8N_\ast^2}   \nonumber \\
				&\quad -p_{3L}^{(\textrm{t})}  \frac{\ln(2N_\ast)}{N_\ast^3}- \frac{\ln(2N_\ast)^2}{96N_\ast^3}  +\frac{ p_3^{(\textrm{t})} }{N_\ast^3}  + \order{N_\ast^{-4}}\Bigg] 
			\end{aligned}$
			&$ \begin{aligned}
				p_2^{(\textrm{t})} &= -(1+24C+24D_0)/16 \simeq 1.9849114 \nonumber \\
				p_{3L}^{(\textrm{t})}  &= 1/96 + C/4 + 3 D_2 \simeq -0.34123524	 \nonumber \\
				p_{3}^{(\textrm{t})} &=-[85 +72 C^2 + 6 D_0 +36 C (1 + 4 D_0) \nonumber \\
				&\qquad + 144 D_1 - 6 \pi^2]/48  \simeq -2.8782506.
			\end{aligned}$
			\\
			\hline
			$\begin{aligned}  ~\\[-2ex] n_\textrm{t} &= -\frac{3}{2N_\ast^2} - \frac{\ln(2N_\ast)}{4N_\ast^3} + \frac{\theta_3^{(\textrm{t}) } }{N_\ast^3} + \order{N_\ast^{-4}} 
			\end{aligned}$ 
			& $\begin{aligned}
				\theta_3^{(\textrm{t}) } =-3 (C+D_0+1) \simeq 1.0948228
			\end{aligned}$
			\\
			\hline
			$\begin{aligned} ~\\[-3ex]
				\alpha_{\textrm{t}}= 	- \frac{3}{N_\ast^3} + \order{N_\ast^{-4}}
			\end{aligned}$ 
			& 
			\\
			\hline 
			$\begin{aligned} ~\\[-3ex]& 
				\hspace{-0.3em}	\beta_{\textrm{t}}= 0+ \order{N_\ast^{-4}}
			\end{aligned}$ \\ 
			\hline 
			\hline
			$\vspace{0.3em} \begin{aligned} ~\\[-3ex]
				r\equiv \dfrac{\mathcal{A}_{\textrm{t}}}{\mathcal{A}_\textrm{s}} = \dfrac{12}{N_\ast^2} + \dfrac{2\ln(2N_\ast)}{N_\ast^3} - r_3 \dfrac{24}{N_\ast^3} + \order{N_\ast^{-4}} 
			\end{aligned}$ 
			& $r_3 = 1-C-D_0 \simeq 2.3649409 $
			\\
			\hline 
			$\begin{aligned} ~\\[-1em]& 
				\hspace{-0.3em}\delta \equiv r+8n_{\textrm{t}} =  - \frac{48}{N_\ast^3} + \order{N_\ast^{-4}}
			\end{aligned}$
		\end{tabular}
	\end{ruledtabular}
\end{table*}

We note that the results presented in Table~\ref{Tab:Starobinsky-Results} are expressed in terms of the number of e-foldings $N_\ast$ computed in the geometric (or Jordan) frame. Alternatively one can express the power-law quantities directly in terms of the scalar tilt $n_\mathrm{s}$, which is one of the most accurately measured cosmological parameters, $n_\mathrm{s}-1= -0.0351\pm0.0042$ at $68\%$ CL \cite{Akrami2020}. Introducing a truncation in the parameter $|n_\mathrm{s}-1|\ll 1$, we find that the tensor-to-scalar ratio $r$, the tensor tilt $n_\mathrm{t}$, and the running of the scalar tilt $\alpha_\mathrm{s}$ are
\begin{align}
r\;=&+3\, (n_\mathrm{s}-1)^2+\tfrac{7}{2}(n_\mathrm{s}-1)^3\,+\,\order{(n_\mathrm{s}-1)^4},\label{eq:r-ns}\\[.6em]
n_\mathrm{t}=&-\tfrac{3}{8}(n_\mathrm{s}-1)^2+\tfrac{5}{16} (n_\mathrm{s}-1)^3 +\order{(n_\mathrm{s}-1)^4},\label{eq:nt-ns}\\[.6em]
\alpha_\mathrm{s}=&-\tfrac{1}{2}(n_\mathrm{s}-1)^2+\tfrac{5}{48} (n_\mathrm{s}-1)^3 +\order{(n_\mathrm{s}-1)^4}.\label{eq:alpha-ns}
\end{align}
These expressions are directly formulated in terms of the observed parameter $|n_\mathrm{s}-1|\ll 1$. Note that the equivalence between the Jordan and Einstein frames requires the identification of a mapping between pivot scales in the two frames, or of the number of e-foldings as discussed, for instance, in \cite{Karam2017}. On the other hand,  the expression \eqref{eq:r-ns}, for instance, gives the deparametrized curve in the $r-n_\mathrm{s}$ plane which is independent of the number of e-foldings from a given pivot scale. As a result, these expressions are independent of the pivot scale and provide a concrete illustration of how both frames lead to the same observational constraints. Therefore, the results on the decrease in $r$ at N3LO discussed in Fig.~\ref{fig:rnsplane}, the violation of the consistency relation $r+8n_{\textrm{t}}< 0$ (Fig.~\ref{fig:rntPlane}), and the negative value of the running of the scalar tilt $\alpha_\mathrm{s}$ are robust predictions of Starobinsky inflation, regardless of the frame one is working with.

\section{Discussion}
\label{sec:discussion}

\begin{table}[t]
	\caption{Summary of results.}
	\label{Tab:Final_Summary}
	%	\rowcolors{2}{gray!25}{white}
	\begin{ruledtabular}
		\begin{tabular}{l l}
			Result& Where to find it \\
			\hline
			Generic $Z_\psi$, $c_\psi$& $\vspace{0.2em}\begin{aligned} ~\\[-3ex] &\textrm{Power spectrum}: \textrm{Tables \ref{Tab:p0s},\ref{Tab:p1s},\ref{Tab:p2s},\ref{Tab:p3s} } \\ & \theta_\ast^{(\psi)}, \alpha_\ast^{(\psi)},  \beta_\ast^{(\psi)}: \, \textrm{Table \ref{Tab:GenericTheory-Features}} 
			\end{aligned}$ \\
            \hline 
			$R+\alpha R^2$& $\vspace{0.2em}\begin{aligned} ~\\[-3ex]& 
				\mathcal{A}_{\textrm{s,t}}, n_{\textrm{s,t}}, \alpha_{\textrm{s,t}}, \beta_{\textrm{s,t}}  : \;  \textrm{Table \ref{Tab:Starobinsky-Results}}   \end{aligned}$\\
			\hline
			Scalar field (App.~\ref{App:Power-law-single-field})& $\vspace{0.2em}\begin{aligned}  ~\\[-3ex]& 
				\textrm{Scalar} \; \mathcal{A}_{\textrm{s}}, n_{\textrm{s}}, \alpha_{\textrm{s}}, \beta_{\textrm{s}}  : \;  \textrm{Table \ref{Tab:SingleField-ScalarFeatures}}
				\\ & 	\textrm{Tensor} \; \mathcal{A}_{\textrm{t}}, n_{\textrm{t}}, \alpha_{\textrm{t}}, \beta_{\textrm{t}}  : \;  \textrm{Table \ref{Tab:SingleField-TensorFeatures}}
			\end{aligned}$ 
		\end{tabular}
	\end{ruledtabular}
\end{table}

In this paper, we derived N3LO expressions for the primordial power spectrum in a broad class of effective theories of inflation with an action for perturbations of the form \eqref{eq:Quadratic-Action-psi}.~We adopted the Green's function method \cite{Stewart2001,Auclair2022} to compute the late-time behavior of the mode functions of the quasi-Bunch-Davies initial state at N3LO, assuming a sufficiently long quasi-de Sitter inflationary phase $(N\gg N_\ast)$. Our main results are summarized in Table~\ref{Tab:Final_Summary}.

Current measurements of primordial observables already probe the amplitude and tilt of scalar modes and provide contraints on the amplitude and tilt of tensor modes \cite{Akrami2020}. The next generation of CMB experiments, such as CORE \cite{CORECollaboration2016}, CMB-S4 \cite{S4Collaboration2020}, LiteBIRD\cite{LiteBIRDCollaboration2022}, and PICO \cite{NASAPICO2019}, or surveys such as the Simons Observatory \cite{SOC2018} or EUCLID \cite{EuclidCollaboration2021}, are expected to measure N2LO corrections and put stronger constraints on N3LO terms, under the assumption of single-field inflation. In this work, we introduced a framework that covers up to N3LO all effective models parametrized by the two functions $Z(t)$ and $c(t)$, treated as independent here. As illustrated in Table~\ref{Tab:Models}, many effective theories fit within the framework developed in this paper. In the case of $R+\alpha R^2$ Starobinsky inflation, we computed the N3LO corrections, expressing them explicitly in terms of one single free parameter---the number of inflationary e-foldings $N_\ast$ from the exit of the pivot mode $k_\ast$ until the end of inflation. The explicit expressions are reported in Table~\ref{Tab:Starobinsky-Results} in terms of $N_\ast$ and in \eqref{eq:r-ns}--\eqref{eq:alpha-ns} in terms of $n_\mathrm{s}$. In particular, we predict a  negative running $\alpha_\mathrm{s}=-\frac{1}{2}(n_\mathrm{s}-1)^2+\ldots$ of the scalar tilt. We expect these results to be useful to further test this model with even more precise CMB observations in the future, as illustrated in Figs.~\ref{fig:rntPlane} and \ref{fig:alphabetaPlane}. 

\medskip

The fact that the primordial power spectrum probes physics at a scale that is only $\sim 5$ orders of magnitude away from the Planck length $\ell_P$ is remarkable. This is a regime that lies at the interface of effective field theory and quantum gravity. While, on one hand, it is important to identify top-down derivations of the cosmological regime of quantum gravity theories such as \cite{Agullo2013,Fernandez-Mendez:2012poe,Bianchi:2010zs,Gielen:2013kla,Ashtekar:2020gec}, on the other hand, working at this interface where one parametrizes quantum gravity effects into an effective field theory can allow us to put observational constraints and identify features of quantum geometry in the CMB sky \cite{Ashtekar2021}. In particular, it would be interesting to develop a similar N3LO framework for functions $Z_{\mathrm{t}}(t,k)$ and $c_{\mathrm{t}}(t,k)$ with a Fourier mode dependence, such as the ones that appear in models with a parity-violating coupling to the Chern-Simons density \cite{Alexander:2009tp}. In fact, extracting precise predictions for effective theories such as \cite{Bianchi:2024mrt} and \cite{Daniel:2024lev} could allow us to distinguish quantum gravity theories with observations of primordial parity violation.

\medskip

\begin{acknowledgments}
We thank Monica Rincon-Ramirez, Miguel Fernandez, Lucas Hackl, Paul John Balderston, Samarth Khandelwal, Erick Mui\~no, Daniel Paraizo, Parampreet Singh, Nelson Yokomizo, Ivan Agullo, Sergey Ketov and Abhay Ashtekar for useful discussions, and an anonymous referee for comments which led to improvements of this manuscript. M.G. is supported by the \href{https://anid.cl}{Agencia Nacional de Investigación y Desarrollo} (ANID) and \href{http://www.fulbright.cl/}{ Fulbright Chile} through the Fulbright Foreign Student Program and ANID BECAS/Doctorado BIO Fulbright-ANID 56190016.~M.G. also acknowledges  support from the \href{https://blaumannfoundation.org}{Blaumann Foundation} for participating at the Loops'24 conference where this work was first presented. E.B. acknowledges support from the National Science Foundation, Grant No. PHY-2207851. This work was made possible through the support of the ID 62312 grant from the John Templeton Foundation, as part of the project \href{https://www.templeton.org/grant/the-quantum-information-structure-of-spacetime-qiss-second-phase}{``The Quantum Information Structure of Spacetime'' (QISS)}. The opinions expressed in this work are those of the authors and do not necessarily reflect the views of the John Templeton Foundation.
\end{acknowledgments}

\appendix

\begin{widetext}

\section{Generalized conformal time with speed of sound}
\label{app:conformaltime}
In conformal time $\eta$, the FLRW metric takes the form
\begin{equation}
	\bar{g}_{\mu\nu} \dd{x}^\mu \dd{x}^\nu = a(t)^2 \big( - \dd{\eta}^2 + \delta_{ij} \dd{x}^i \dd{x}^j\big),
\end{equation}
which corresponds to the following relation to the cosmic time $t$:
\begin{equation}
	\dv{\eta}{t} = \frac{1}{a(t)}.
\end{equation}
In de Sitter space we have the exact relation $\eta_{\textrm{dS}} = -1/(a H_0)$. Here we consider the case of quasi-de Sitter with, in addition, a speed of sound $c_\psi \neq 1$. Then, we will use a generalized conformal time $\tau$, such that $x = -k_\ast \tau$, and which solves \eqref{eq:xdot}. Hence, the goal is to write 
\begin{equation}
	\dv{\tau}{t} =   \frac{c_\psi(t)}{a(t)} = - \dv{}{t} \qty(\frac{c_\psi(t)}{a(t)H(t)}) + \; \textrm{corrections},
\end{equation}
in an order-by-order expansion. At zero order, we can start with the ansatz
\begin{equation}
	\tau_{\textrm{ansatz}}^{(0)} = - \frac{c_\psi(t)}{a(t ) H(t)} \quad \to \quad  \frac{c_\psi(t)}{a(t)} - \dv{}{t} \tau_{\textrm{ansatz}}^{(0)}  = 0 + \order{\epsilon} \quad \to \quad 	\tau^{(0)} = - \frac{c_\psi(t)}{a(t ) H(t)} \,.
\end{equation}
For the next order, we consider the most general ansatz of order one,
\begin{equation}
	\tau_{\textrm{ansatz}}^{(1)} = - \frac{c_\psi(t)}{a(t ) H(t)}  \qty(1+ b_1 \epsilon_{1H}(t) + b_2 \epsilon_{1c}(t) )\; \to \; \frac{c_\psi(t)}{a(t)} - \dv{}{t} \tau_{\textrm{ansatz}}^{(1)}  = -\frac{  c_\psi(t) ((b_1-1) \epsilon_{1H}(t)+(b_2+1) \epsilon_{1c}(t))}{a(t)} + \order{\epsilon^2},
\end{equation}
which vanishes for $b_1 = 1$ and $b_2 = -1$. Hence, 
\begin{equation}
	\tau^{(1)} = - \frac{c_\psi(t)}{a(t ) H(t)}  \qty(1+ \epsilon_{1H}(t) - b_2 \epsilon_{1c}(t) ).
\end{equation}
Similarly, at the next order we have
\begin{align}
	\tau_{\textrm{ansatz}}^{(2)} &= -\frac{c(t)}{a(t) H(t)} \Big(1 + \epsilon_{1H}(t) - \epsilon_{1c}(t) + c_{11} \epsilon_{1H}(t)^2 + c_{22}\, \epsilon_{2H}(t)^2 + c_{12}\, \epsilon_{1H}(t) \epsilon_{2H}(t) + b_{11} \,\epsilon_{1c}(t)^2 + b_{22}\, \epsilon_{2c}(t)^2 \nonumber \\
	&\qquad+ b_{12}\, \epsilon_{1c}(t) \epsilon_{2c}(t)
	 + bc_{1}\, \epsilon_{1c}(t) \epsilon_{1H}(t) + bc_{2} \,\epsilon_{2c}(t) \epsilon_{1H}(t) + bc_{3}\, \epsilon_{1c}(t) \epsilon_{2H}(t) + bc_{4} \epsilon_{2c}(t) \epsilon_{2H}(t) \Big)\,.
\end{align}
After replacing the ansatz, we find that 
\begin{equation}
	\frac{c_\psi(t)}{a(t)} - \dv{}{t} \tau_{\textrm{ansatz}}^{(2)} =0 + \order{\epsilon^3} 
\end{equation}
 for $c_{11} \to 1, c_{12} \to -1, c_{22} \to 0, b_{11} \to 1, b_{22} \to 0, bc_4 \to 0, c_{22} \to 0, bc_1 \to -2, b_{12 }\to 1, bc_3 \to 0, bc_2 \to 0, c_{12} \to -1 $. Then, up to second order,
\begin{align}
		\tau^{(2)} &=  - \frac{ c_\psi(t)}{a(t)H(t)} \Big[  1 + \epsilon_{1H}(t)  - \epsilon_{1c}(t)  + \epsilon_{1H}(t)^2 - \epsilon_{1H}(t) \epsilon_{2H}(t)  - 2 \epsilon_{1c}(t) \epsilon_{1H}(t)  + \epsilon_{1c}(t)\epsilon_{2c}(t) + \epsilon_{1c}(t)^2    \Big]	\,.
\end{align}
The same procedure can be extended order-by-order. In particular, for the next order we need to assume an ansatz with all the possible combinations of third-order quantities. Repeating the same process, we find that the conformal time up to third order is given by
\begin{align}
\tau^{(3)}&\equiv \frac{\tilde{c}_\psi(t)}{a(t)H(t)} \nonumber\\
&=  -\frac{ c_\psi(t)}{a(t)H(t)} \Big[  1 + \epsilon_{1H}(t)  - \epsilon_{1c}(t)  + \epsilon_{1H}(t)^2 - \epsilon_{1H}(t) \epsilon_{2H}(t)  - 2 \epsilon_{1c}(t) \epsilon_{1H}(t)  + \epsilon_{1c}(t)\epsilon_{2c}(t) + \epsilon_{1c}(t)^2   \nonumber \\
	&\quad 
	+\epsilon_{1H}(t)^3 + \epsilon_{1H}(t) \epsilon_{2H}(t) \epsilon_{3H}(t)  - 
	3 \epsilon_{1H}(t)^2 \epsilon_{2H}(t) + \epsilon_{1H}(t) \epsilon_{2H}(t)^2 - \epsilon_{1c}(t) \epsilon_{2c}(t)^2 + 
	3 \epsilon_{1c}(t) \epsilon_{1H}(t) \epsilon_{2H}(t)  \nonumber \\
	&\quad - 
	3 \epsilon_{1c}(t)^2 \epsilon_{2c}(t)  +  
	3 \epsilon_{1c}(t) \epsilon_{1H}(t) \epsilon_{2c}(t)  - 
	\epsilon_{1c}(t) \epsilon_{2c}(t) \epsilon_{3c}(t) 	-\epsilon_{1c}(t)^3 + 3 \epsilon_{1c}(t)^2 \epsilon_{1H}(t) - 
	3 \epsilon_{1c}(t) \epsilon_{1H}(t)^2  \Big]\,. \label{eq:tau-full}
\end{align}
Note that by setting $c_\psi(t)=1$ and $\epsilon_{1c}(t) = 0$, $\epsilon_{2c}(t)=0$, and $\epsilon_{3c}(t)=0$, we recover an expression for the standard conformal time $\eta$ in a quasi-de Sitter background:
\begin{align}
\eta^{(3)} &= -\frac{1}{a(t) H(t)} \bigg( 1 + \epsilon_{1H}(t) + \epsilon_{1H}^2(t) 
- \epsilon_{1H}(t) \epsilon_{2H}(t) \nonumber \\
&\qquad\qquad\qquad - 3 \epsilon_{1H}^2(t) \epsilon_{2H}(t) 
+ \epsilon_{1H}(t) \epsilon_{2H}^2(t) + \epsilon_{1H}(t) \epsilon_{2H}(t) \epsilon_{3H}(t)+ \epsilon_{1H}^3(t)  \bigg). \label{eq:eta-qdS-full}
\end{align}
The generalized conformal time $\tau$ can also be expressed as $\tau(t) =\hat{c}_\psi (t) \, \eta(t)$, where
\begin{align}
\hat{c}_\psi (t) &\equiv c_\psi(t)\, \Big\{ 1 - \epsilon_{1c\ast}(t) + \epsilon_{1c\ast}(t)^2 - \epsilon_{1c\ast}(t)^3 
- \epsilon_{1c\ast}(t) \epsilon_{1H\ast}(t) 
+ 2 \epsilon_{1c\ast}(t)^2 \epsilon_{1H\ast}(t) 
- \epsilon_{1c\ast}(t) \epsilon_{1H\ast}(t)^2 
+ \epsilon_{1c\ast}(t) \epsilon_{2c\ast}(t) \nonumber\\ 
&\quad - 3 \epsilon_{1c\ast}(t)^2 \epsilon_{2c\ast}(t) 
+ 2 \epsilon_{1c\ast}(t) \epsilon_{1H\ast}(t) \epsilon_{2c\ast}(t) 
- \epsilon_{1c\ast}(t) \epsilon_{2c\ast}(t)^2 
+ 2 \epsilon_{1c\ast}(t) \epsilon_{1H\ast}(t) \epsilon_{2H\ast}(t) 
- \epsilon_{1c\ast}(t) \epsilon_{2c\ast}(t) \epsilon_{3c\ast}(t)
  \Big\}.
\end{align}

\vfill
\section{Comparing two power spectra at different pivot scales}
\label{app:comparing-pivot-scales}
To illustrate the procedure, let us consider without loss of generality two different SVT modes, $A$ and $B$, such that $\tau^{(A)} = \hat{c}^{(A)}\, \eta$ and $\tau^{(B)}=\hat{c}^{(B)}\, \eta$, with $\hat{c}^{(A)} \neq \hat{c}^{(B)}$. Different speeds of sound imply that we have two different pivot scales, $k_\ast \tau_\ast^{(A)} = -1$ and $k_\diamond \tau_\diamond^{(B)} = -1$, so one SVT mode will have a power $\mathcal{P}_0^{(A)}(k)$ expanded around $k_\ast$ and the other will have $\mathcal{P}_0^{(B)}(k)$ expanded around $k_\diamond$. More explicitly, we would have 
\begin{align}
	\mathcal{P}_0^{(A)}(k) &=  \frac{\hbar H_\ast^2}{4\pi^2 c_\ast^3 Z_\ast} \Bigg[ 1 + p_{0\ast}^{(A)} + p_{1\ast}^{(A)}  \ln(\frac{k}{k_\ast}) +\, p_{2\ast}^{(A)}  \ln(\frac{k}{k_\ast})^2 + p_{3\ast}^{(A)}  \ln(\frac{k}{k_\ast})^3 \Bigg]\,, \\
 \mathcal{P}_0^{(B)}(k) &=  \frac{\hbar H_\diamond^2}{4\pi^2 c_\diamond^3 Z_\diamond} \Bigg[ 1 + p_{0\diamond}^{(B)} + p_{1\diamond}^{(B)} \ln(\frac{k}{k_\diamond}) +\, p_{2\diamond}^{(B)} \ln(\frac{k}{k_\diamond})^2 + p_{3\diamond}^{(B)} \ln(\frac{k}{k_\diamond})^3 \Bigg]\,. \label{eq:app-PowerB} 
\end{align}
Let us consider the standard conformal time $\eta$, as defined in \eqref{eq:eta-qdS-full}. We can implicitly assume that $\eta_\diamond= \eta_\ast$, i.e., replacing the coefficients $\rho^{(B)}_\diamond$ by $\rho^{(B)}_\ast$, while the change of pivot gets encoded in the running of the scale. To find this running, note first that
\begin{align}
\frac{\tau^{(A)}}{\tau^{(B)}} &= \frac{\hat{c}^{(A)}(t)}{\hat{c}^{(B)}(t)}\nonumber\\
&=\frac{c^{(A)}(t)}{c^{(B)}(t)} \Big\{
1 +  \big( -\epsilon_{1c}^{(A)}(t) + \epsilon_{1c}^{(B)}(t) \big) 
+  \big( \epsilon_{1c}^{(A)}(t)^2 - \epsilon_{1c}^{(A)}(t) \epsilon_{1c}^{(B)}(t) 
- \epsilon_{1c}^{(A)}(t) \epsilon_{1H\ast}(t) + \epsilon_{1c}^{(B)}(t) \epsilon_{1H\ast}(t) \nonumber \\
&\quad + \epsilon_{1c}^{(A)}(t) \epsilon_{2c}^{(A)}(t) - \epsilon_{1c}^{(B)}(t) \epsilon_{2c}^{(B)}(t) \big)
+  \Big( -\epsilon_{1c}^{(A)}(t)^3 + \epsilon_{1c}^{(A)}(t)^2 \epsilon_{1c}^{(B)}(t) 
+ 2 \epsilon_{1c}^{(A)}(t)^2 \epsilon_{1H\ast}(t) \nonumber \\
&\quad - 2 \epsilon_{1c}^{(A)}(t) \epsilon_{1c}^{(B)}(t) \epsilon_{1H\ast}(t) 
- \epsilon_{1c}^{(A)}(t) \epsilon_{1H\ast}(t)^2 + \epsilon_{1c}^{(B)}(t) \epsilon_{1H\ast}(t)^2 
- 3 \epsilon_{1c}^{(A)}(t)^2 \epsilon_{2c}^{(A)}(t) 
+ \epsilon_{1c}^{(A)}(t) \epsilon_{1c}^{(B)}(t) \epsilon_{2c}^{(A)}(t) \nonumber \\
&\quad + 2 \epsilon_{1c}^{(A)}(t) \epsilon_{1H\ast}(t) \epsilon_{2c}^{(A)}(t) 
- \epsilon_{1c}^{(A)}(t) \epsilon_{2c}^{(A)}(t)^2 
+ \epsilon_{1c}^{(A)}(t) \epsilon_{1c}^{(B)}(t) \epsilon_{2c}^{(B)}(t) 
+ \epsilon_{1c}^{(B)}(t)^2 \epsilon_{2c}^{(B)}(t) 
- 2 \epsilon_{1c}^{(B)}(t) \epsilon_{1H\ast}(t) \epsilon_{2c}^{(B)}(t) \nonumber \\
&\quad + \epsilon_{1c}^{(B)}(t) \epsilon_{2c}^{(B)}(t)^2 
+ 2 \epsilon_{1c}^{(A)}(t) \epsilon_{1H\ast}(t) \epsilon_{2H\ast}(t) 
- 2 \epsilon_{1c}^{(B)}(t) \epsilon_{1H\ast}(t) \epsilon_{2H\ast}(t) 
- \epsilon_{1c}^{(A)}(t) \epsilon_{2c}^{(A)}(t) \epsilon_{3c}^{(A)}(t) \nonumber \\
&\quad+ \epsilon_{1c}^{(B)}(t) \epsilon_{2c}^{(B)}(t) \epsilon_{3c}^{(B)}(t) \Big) \Big\} .
\end{align}
Then, since $k_\diamond/k_\ast=\tau^{(A)}_\diamond/\tau^{(B)}_\diamond$, one can compute the following expression order-by-order,
\begin{equation}
\ln(k_\diamond) = \ln(k_\ast) + \ln(\frac{\tau^{(A)}_\ast}{\tau^{(B)}_\ast}) \quad \to \quad \ln(\frac{k}{k_\diamond}) = \ln(\frac{k}{k_\ast}) - \ln(\frac{\hat{c}^{(A)}_\ast}{\hat{c}^{(B)}_\ast}).
\end{equation}
Finally, by replacing the last expression into \eqref{eq:app-PowerB}, we will find the expression for the power spectrum $\mathcal{P}_0^{(B)}(k)$ now fully expanded around the pivot scale $k_\ast$, which now can be consistently compared with $\mathcal{P}_0^{(A)}(k)$, as both are expanded around the same pivot scale, i.e., in powers of $\ln(k/k_\ast)$. 
\vfill
\section{Finite Expression}
\label{app:finite}
We report the N3LO expression of $	p_{\circledast}$:
\begin{align*}
	p_{\circledast}&=1 + (2 + 3 C) \epsilon_{1c\circledast} - (8 + 3 C + \frac{9 C^2}{2} + \frac{9 \pi^2}{8}) \epsilon_{1c\circledast}^2 - 2 (1 + C) \epsilon_{1H\circledast} + (10 - 5 C - 6 C^2 - \frac{3 \pi^2}{2}) \epsilon_{1c\circledast} \epsilon_{1H\circledast} \\
	&+ \frac{1}{2} (-6 + 4 C + 4 C^2 + \pi^2) \epsilon_{1H\circledast}^2 + C \epsilon_{1Z\circledast} - (6 - C + 3 C^2 + \frac{3 \pi^2}{4}) \epsilon_{1c\circledast} \epsilon_{1Z\circledast} \\
	&- \frac{1}{2} (-8 + 2 C + 4 C^2 + \pi^2) \epsilon_{1H\circledast} \epsilon_{1Z\circledast} + \frac{1}{8} (-8 + 4 C^2 + \pi^2) \epsilon_{1Z\circledast}^2 - (2 + 2 C + \frac{3 C^2}{2} - \frac{\pi^2}{8}) \epsilon_{1c\circledast} \epsilon_{2c\circledast} \\
	&- \frac{1}{8} (-96 + 36 C^2 + 36 C^3 + 13 \pi^2 + 15 C (-8 + \pi^2)) \epsilon_{1c\circledast}^2 \epsilon_{2c\circledast} + (-8 + 3 C^2 + 3 C^3 + \frac{13 \pi^2}{12} + \frac{5}{4} C (-8 + \pi^2)) \epsilon_{1c\circledast} \epsilon_{1H\circledast} \epsilon_{2c\circledast} \\
	&- \frac{1}{24} (-96 + 36 C^2 + 36 C^3 + 13 \pi^2 + 15 C (-8 + \pi^2)) \epsilon_{1c\circledast} \epsilon_{1Z\circledast} \epsilon_{2c\circledast} + (2 + 2 C + C^2 - \frac{\pi^2}{12}) \epsilon_{1H\circledast} \epsilon_{2H\circledast} \\
	&+ \left(-12 + \frac{9 C^2}{2} + 3 C^3 + \frac{13 \pi^2}{8} + C (-9 + \frac{5 \pi^2}{4})\right) \epsilon_{1c\circledast} \epsilon_{1H\circledast} \epsilon_{2H\circledast} + \left(8 - 3 C^2 - 2 C^3 - \frac{13 \pi^2}{12} + C \left(6 - \frac{5 \pi^2}{6}\right)\right) \epsilon_{1H\circledast}^2 \epsilon_{2H\circledast} \\
	&+ \left(-4 + \frac{3 C^2}{2} + C^3 + \frac{13 \pi^2}{24} + C \left(-3 + \frac{5 \pi^2}{12}\right)\right) \epsilon_{1H\circledast} \epsilon_{1Z\circledast} \epsilon_{2H\circledast} + \frac{1}{24} (-12 C^2 + \pi^2) \epsilon_{1Z\circledast} \epsilon_{2Z\circledast} \\
	&- \frac{1}{8} C (-48 + 12 C^2 + 5 \pi^2) \epsilon_{1c\circledast} \epsilon_{1Z\circledast} \epsilon_{2Z\circledast} + C \left(-4 + C^2 + \frac{5 \pi^2}{12}\right) \epsilon_{1H\circledast} \epsilon_{1Z\circledast} \epsilon_{2Z\circledast} \\
	&- \frac{1}{24} C (-48 + 12 C^2 + 5 \pi^2) \epsilon_{1Z\circledast}^2 \epsilon_{2Z\circledast} + \epsilon_{1c\circledast} \epsilon_{1H\circledast} \epsilon_{1Z\circledast} (24 - 6 C^3 - \frac{9}{2} C (-8 + \pi^2) - 21 \zeta(3)) \\
	&+ \frac{1}{3} \epsilon_{1H\circledast}^3 (16 - 4 C^3 - 3 C (-8 + \pi^2) - 14 \zeta(3)) - \frac{1}{24} \epsilon_{1Z\circledast} \epsilon_{2Z\circledast}^2 (16 - 4 C^3 + C \pi^2 - 8 \zeta(3)) \\
	&- \frac{1}{24} \epsilon_{1Z\circledast} \epsilon_{2Z\circledast} \epsilon_{3Z\circledast} (16 - 4 C^3 + C \pi^2 - 8 \zeta(3)) + \epsilon_{1H\circledast} \epsilon_{1Z\circledast}^2 (4 - C^3 - \frac{3}{4} C (-8 + \pi^2) - \frac{7 \zeta(3)}{2}) \\
	&+ \epsilon_{1c\circledast} \epsilon_{2c\circledast}^2 (C^2 + \frac{C^3}{2} - \frac{\pi^2}{12} + C (2 - \frac{\pi^2}{8}) + \zeta(3)) + \epsilon_{1c\circledast} \epsilon_{2c\circledast} \epsilon_{3c\circledast} (C^2 + \frac{C^3}{2} - \frac{\pi^2}{12} + C (2 - \frac{\pi^2}{8}) + \zeta(3)) \\
	&+ \epsilon_{1H\circledast}^2 \epsilon_{1Z\circledast} (-8 + 2 C^3 + \frac{3}{2} C (-8 + \pi^2) + 7 \zeta(3)) + \frac{1}{12} \epsilon_{1H\circledast} \epsilon_{2H\circledast}^2 (-12 C^2 - 4 C^3 + \pi^2 + C (-24 + \pi^2) - 8 (1 + \zeta(3))) \\
	&+ \frac{1}{12} \epsilon_{1H\circledast} \epsilon_{2H\circledast} \epsilon_{3H\circledast} (-12 C^2 - 4 C^3 + \pi^2 + C (-24 + \pi^2) - 8 (1 + \zeta(3))) + \frac{9}{8} \epsilon_{1c\circledast}^3 (4 C^3 + 3 C (-8 + \pi^2) + 2 (-8 + 7 \zeta(3))) \\
	&- \frac{9}{4} \epsilon_{1c\circledast}^2 \epsilon_{1H\circledast} (4 C^3 + 3 C (-8 + \pi^2) + 2 (-8 + 7 \zeta(3))) + \frac{9}{8} \epsilon_{1c\circledast}^2 \epsilon_{1Z\circledast} (4 C^3 + 3 C (-8 + \pi^2) + 2 (-8 + 7 \zeta(3))) \\
	&+ \frac{3}{8} \epsilon_{1c\circledast} \epsilon_{1Z\circledast}^2 (4 C^3 + 3 C (-8 + \pi^2) + 2 (-8 + 7 \zeta(3))) + \frac{1}{24} \epsilon_{1Z\circledast}^3 (4 C^3 + 3 C (-8 + \pi^2) + 2 (-8 + 7 \zeta(3))) \\
	&+ \epsilon_{1c\circledast} \epsilon_{1H\circledast}^2 (6 C^3 + \frac{9}{2} C (-8 + \pi^2) + 3 (-8 + 7 \zeta(3)))
\end{align*}

\vfill
\newpage

\section{Amplitude of power spectrum for a generic theory at N3LO evaluated at $k_\ast$}
\label{App:Power-law-generic}

We report the NLO, N2LO, and N3LO corrections to the amplitude of the power spectrum for any SVT mode in a theory with generic $Z_\psi$ and $c_\psi$.

\begin{table*}[ht]
	\caption{Amplitude of the power spectrum for a theory with generic $Z_\psi$ and $c_\psi$, up to N3LO.}
	\label{Tab:GenericTheory-Amplitude}
	\begin{ruledtabular}
		\begin{tabular}{p{0.2in} p{6.2in}}
			 Order & Expression \\
			\hline
			  $\begin{aligned} ~\\[-3ex]
				\textrm{LO+NLO}:
				\hspace{0.1cm}  \mathcal{A}_\ast^{(\psi)} &= \frac{\hbar H_\ast^2 }{4 \pi^2 c_\ast^3  Z_{\ast} } \Big[ 1 - 2 (1 + C) \epsilon_{1H\ast}  + C \epsilon_{1Z\ast} +  (2 + 3 C) \epsilon_{1c\ast}  \\ 
				\textrm{N2LO}: \hspace{2.6em} &+  \frac{1}{2} (-6 + 4 C + 4 C^2 + \pi^2) \epsilon_{1H\ast}^2 + (2 + 2 C + C^2 - \frac{\pi^2}{12}) \epsilon_{1H\ast} \epsilon_{2H\ast} - \frac{1}{2} (-8 + 2 C + 4 C^2 + \pi^2) \epsilon_{1H\ast} \epsilon_{1Z\ast} \\
					& + \frac{1}{8} (-8 + 4 C^2 + \pi^2) \epsilon_{1Z\ast}^2+ \frac{1}{24} (-12 C^2 + \pi^2) \epsilon_{1Z\ast} \epsilon_{2Z\ast} + (-6 + C + 3 C^2 + \frac{3 \pi^2}{4}) \epsilon_{1c\ast} \epsilon_{1Z\ast} \\
					&+  \frac{1}{8} (-64 + 24 C + 36 C^2 + 9 \pi^2) \epsilon_{1c\ast}^2 + (10 - 5 C - 6 C^2 - \frac{3 \pi^2}{2}) \epsilon_{1c\ast} \epsilon_{1H\ast}  + \frac{1}{8} (-16 - 16 C - 12 C^2 + \pi^2) \epsilon_{2c\ast} \epsilon_{1c\ast}  \\
				\textrm{N3LO}: \hspace{2.6em}  & 
			+ \frac{1}{3} \epsilon_{1H\ast}^3 (16 - 4 C^3 - 3 C (-8 + \pi^2) - 14 \zeta(3)) + (8 - \frac{3 C^2}{2} - 2 C^3 - \frac{13 \pi^2}{12} + C (6 - \frac{5 \pi^2}{6})) \epsilon_{1H\ast}^2 \epsilon_{2H\ast} \\
			& + \frac{1}{12} \epsilon_{1H\ast} \epsilon_{2H\ast}^2 (-12 C^2 - 4 C^3 + \pi^2 + C (-24 + \pi^2) - 8 (1 + \zeta(3)))  \nonumber \\
			&	+ \frac{1}{12} \epsilon_{1H\ast} \epsilon_{2H\ast} \epsilon_{3H\ast} (-12 C^2 - 4 C^3 + \pi^2 + C (-24 + \pi^2) - 8 (1 + \zeta(3)))  \nonumber \\
			&  + (-4 + \frac{3 C^2}{2} + C^3 + \frac{13 \pi^2}{24} + C (-3 + \frac{5 \pi^2}{12})) \epsilon_{1H\ast} \epsilon_{1Z\ast} \epsilon_{2H\ast}  - \frac{1}{24} C (-48 + 12 C^2 + 5 \pi^2) \epsilon_{1Z\ast}^2 \epsilon_{2Z\ast}  \nonumber \\
			& - \frac{1}{8} C (-48 + 12 C^2 + 5 \pi^2) \epsilon_{1c\ast} \epsilon_{1Z\ast} \epsilon_{2Z\ast} + C (-4 + C^2 + \frac{5 \pi^2}{12}) \epsilon_{1H\ast} \epsilon_{1Z\ast} \epsilon_{2Z\ast}  \\
						&+ \frac{1}{24} \epsilon_{1Z\ast} \epsilon_{2Z\ast}^2 (4 C^3 - C \pi^2 + 8 (-2 + \zeta(3)))  + \epsilon_{1H\ast} \epsilon_{1Z\ast}^2 (4 - C^3 - \frac{3}{4} C (-8 + \pi^2) - \frac{7 \zeta(3)}{2})\\
						& + \epsilon_{1H\ast}^2 \epsilon_{1Z\ast} (-8 + 2 C^3 + \frac{3}{2} C (-8 + \pi^2) + 7 \zeta(3))  + \frac{1}{24} \epsilon_{1Z\ast} \epsilon_{2Z\ast} \epsilon_{3Z\ast} (4 C^3 - C \pi^2 + 8 (-2 + \zeta(3)))    \\
						&  + \frac{1}{24} \epsilon_{1Z\ast}^3 (4 C^3 + 3 C (-8 + \pi^2) + 2 (-8 + 7 \zeta(3)))   + \frac{9}{8} \epsilon_{1c\ast}^3 (4 C^3 + 3 C (-8 + \pi^2) + 2 (-8 + 7 \zeta(3)))    \\
						&+ \epsilon_{1c\ast} \epsilon_{2c\ast}^2 (C^2 + \frac{C^3}{2} - \frac{\pi^2}{12} + C (2 - \frac{\pi^2}{8}) + \zeta(3))   + \frac{1}{8} (96 - 36 C^2 - 36 C^3 - 13 \pi^2 - 15 C (-8 + \pi^2)) \epsilon_{1c\ast}^2 \epsilon_{2c\ast} \\
								& + \epsilon_{1c\ast} \epsilon_{1H\ast} \epsilon_{1Z\ast} (24 - 6 C^3 - \frac{9}{2} C (-8 + \pi^2) - 21 \zeta(3))   + \epsilon_{1c\ast} \epsilon_{2c\ast} \epsilon_{3c\ast} (C^2 + \frac{C^3}{2} - \frac{\pi^2}{12} + C (2 - \frac{\pi^2}{8}) + \zeta(3)) \nonumber \\
				&+ (-8 + 3 C^2 + 3 C^3 + \frac{13 \pi^2}{12} + \frac{5}{4} C (-8 + \pi^2)) \epsilon_{1c\ast} \epsilon_{1H\ast} \epsilon_{2c\ast} \nonumber \\
				& + \frac{1}{24} (96 - 36 C^2 - 36 C^3 - 13 \pi^2 - 15 C (-8 + \pi^2)) \epsilon_{1c\ast} \epsilon_{1Z\ast} \epsilon_{2c\ast} \\
				& + (-12 + \frac{9 C^2}{2} + 3 C^3 + \frac{13 \pi^2}{8} + C (-9 + \frac{5 \pi^2}{4})) \epsilon_{1c\ast} \epsilon_{1H\ast} \epsilon_{2H\ast} \\
				& - \frac{9}{4} \epsilon_{1c\ast}^2 \epsilon_{1H\ast} (4 C^3 + 3 C (-8 + \pi^2) + 2 (-8 + 7 \zeta(3)))  + \frac{9}{8} \epsilon_{1c\ast}^2 \epsilon_{1Z\ast} (4 C^3 + 3 C (-8 + \pi^2) + 2 (-8 + 7 \zeta(3)))\\
				& + \frac{3}{8} \epsilon_{1c\ast} \epsilon_{1Z\ast}^2 (4 C^3 + 3 C (-8 + \pi^2) + 2 (-8 + 7 \zeta(3))) + \epsilon_{1c\ast} \epsilon_{1H\ast}^2 (6 C^3 + \frac{9}{2} C (-8 + \pi^2) + 3 (-8 + 7 \zeta(3)))	 \Big] 
			\end{aligned}$
		\end{tabular}
	\end{ruledtabular}
\end{table*}

\newpage

\section{Power-law quantities for single field inflation at N3LO}
\label{App:Power-law-single-field}
In general, we can compute power-law quantities, i.e., the amplitude $\mathcal{A}_\ast$ at the pivot mode $k_\ast$, together with its log-derivatives: the spectral tilt $\theta_\ast$, the running of the tilt $\alpha_\ast$, and the running-of-the-running of the tilt $\beta_\ast$, which are defined as 
\begin{align}
	\mathcal{A}_\ast &\equiv \mathcal{P}_0(k_\ast)\, , \label{eq:power-law-quantities-amplitude} \\
 \theta_\ast &\equiv \eval{ k \dv{}{k} \ln(\mathcal{P}_0(k)) }_{k=k_\ast} \, , \label{eq:power-law-quantities-tilt}\\
		\alpha_\ast &\equiv \eval{ k \dv{}{k} \qty[ k \dv{}{k} \ln(\mathcal{P}_0(k))] }_{k=k_\ast}\, , \label{eq:power-law-quantities-run}\\
  \beta_\ast &\equiv \eval{k \dv{}{k} \qty{ k \dv{}{k} \qty[ k \dv{}{k} \ln(\mathcal{P}_0(k))] } }_{k=k_\ast}\, .\label{eq:power-law-quantities-runrun}
\end{align}
Note that the power-law quantities above can be straightforwardly obtained from an expansion of $\ln(\mathcal{P}_0(k))$ up to N3LO, since
\begin{equation}
\ln(\mathcal{P}_0(k)) = \ln(\mathcal{A}_\ast) + \theta_\ast \ln(\frac{k}{k_\ast}) + \frac{\alpha_\ast}{2!} \ln(\frac{k}{k_\ast})^2 + \frac{\beta_\ast}{3!} \ln(\frac{k}{k_\ast})^3 + \order{\textrm{N4LO}} \, .
\end{equation}
Below, we report these quantities for minimally coupled single field inflation.
\begin{table*}[ht]
	\caption{Power-law quantities of curvature perturbations for a minimally coupled single field up to N3LO.}
	\label{Tab:SingleField-ScalarFeatures}
	\begin{ruledtabular}
		\begin{tabular}{p{0.7in} p{0.67in} p{6in}}
			Quantity &  Order & Expression \\
			\hline

			$  \hspace{0.5cm}  \mathcal{A}_{\textrm{s}}$ &  $		\vspace{0.3em} \begin{aligned}  ~\\[-3ex] 
				\textrm{LO+NLO}:
				\hspace{0.1cm}  & \frac{G\hbar \, H_\ast^2}{\pi \epsilon_{1H\ast} } \Big[ 1 - 2 (1 + C) \epsilon_{1H\ast} + C \epsilon_{2H\ast}   \\ 
				\textrm{N2LO}: \hspace{0.1cm} &  + \frac{1}{2} (-6 + 4 C + 4 C^2 + \pi^2) \epsilon_{1H\ast}^2 + \qty(6 + C - C^2 - \frac{7 \pi^2}{12}) \epsilon_{1H\ast} \epsilon_{2H\ast}  \nonumber\\
				& + \frac{1}{8} (-8 + 4 C^2 + \pi^2) \epsilon_{2H\ast}^2 + \frac{1}{24} (-12 C^2 + \pi^2) \epsilon_{2H\ast} \epsilon_{3H\ast} \nonumber  \\
				\textrm{N3LO}: \hspace{0.1cm}  & + \frac{1}{24} \epsilon_{2H\ast} \epsilon_{3H\ast}^2 (4 C^3 - C \pi^2 + 8 (-2 + \zeta(3))) + \frac{1}{24} \epsilon_{2H\ast} \epsilon_{3H\ast} \epsilon_{4H\ast} (4 C^3 - C \pi^2 + 8 (-2 + \zeta(3))) \nonumber \\
				& + \frac{1}{3} \epsilon_{1H\ast}^3 (16 - 4 C^3 - 3 C (-8 + \pi^2) - 14 \zeta(3)) + \epsilon_{1H\ast}^2 \epsilon_{2H\ast} (-3 C^2 - \frac{13 \pi^2}{12} + \frac{2}{3} C (-9 + \pi^2) + 7 \zeta(3))  \nonumber\\
				& + \frac{1}{12} \epsilon_{1H\ast} \epsilon_{2H\ast} \epsilon_{3H\ast} (-12 C^2 + 8 C^3 + \pi^2 + 6 C (-12 + \pi^2) - 8 (1 + \zeta(3)))  \nonumber\\
				&  + \frac{1}{24} \epsilon_{2H\ast}^3 (4 C^3 + 3 C (-8 + \pi^2) + 2 (-8 + 7 \zeta(3)))  - \frac{1}{24} C (-48 + 12 C^2 + 5 \pi^2) \epsilon_{2H\ast}^2 \epsilon_{3H\ast} \nonumber \\
				& + \frac{1}{24} \epsilon_{1H\ast} \epsilon_{2H\ast}^2 (12 C^2 - 8 C^3 + 15 \pi^2 - 6 C (-4 + \pi^2) - 4 (4 + 25 \zeta(3))) \Big] 
			\end{aligned}$
			\\
			\hline
			$\hspace{0.5cm} n_\textrm{s}$  & $ \vspace{0.2em} \begin{aligned}   ~\\[-3ex]  \textrm{LO+NLO}:
				\hspace{0.2cm}  &
				1 -2 \epsilon_{1H\ast} + \epsilon_{2H\ast} \\ 
				\textrm{N2LO}: \hspace{0.2cm}  &
				-2 \epsilon_{1H\ast}^2 + (3 + 2C) \epsilon_{1H\ast} \epsilon_{2H\ast} - C \epsilon_{2H\ast} \epsilon_{3H\ast} \\
				\textrm{N3LO}: \hspace{0.2cm}&  -2 \epsilon_{1H\ast}^3 + \left(15 + 6C - \pi^2\right) \epsilon_{1H\ast}^2 \epsilon_{2H\ast} + \frac{1}{12} \left(-84 - 36C - 12C^2 + 7\pi^2\right) \epsilon_{1H\ast} \epsilon_{2H\ast}^2 \nonumber \\
				&+ \frac{1}{12} \left(-72 - 48C - 12C^2 + 7\pi^2\right) \epsilon_{1H\ast} \epsilon_{2H\ast} \epsilon_{3H\ast} + \frac{1}{4} \left(8 - \pi^2\right) \epsilon_{2H\ast}^2 \epsilon_{3H\ast} \nonumber \\
				& + \frac{1}{24} \left(12C^2 - \pi^2\right) \epsilon_{2H\ast} \epsilon_{3H\ast}^2+\frac{1}{2} C^2 \epsilon_{2H\ast} \epsilon_{3H\ast} \epsilon_{4H\ast} - \frac{1}{24} \pi^2 \epsilon_{2H\ast} \epsilon_{3H\ast} \epsilon_{4H\ast} 
			\end{aligned}$ \\
			\hline
			$\hspace{0.5cm} \alpha_{\textrm{s}}$ & $\vspace{0.2em}\begin{aligned}  ~\\[-3ex] 
				\hspace{0.55cm} 	\textrm{N2LO}: & \hspace{0.2cm} 2 \epsilon_{1H\ast} \epsilon_{2H\ast} - \epsilon_{2H\ast} \epsilon_{3H\ast} 
				\\   \textrm{N3LO}: & \hspace{0.2cm} 6 \epsilon_{1H\ast}^2 \epsilon_{2H\ast} + (-3 - 2C) \epsilon_{1H\ast} \epsilon_{2H\ast}^2 - 2(2 + C) \epsilon_{1H\ast} \epsilon_{2H\ast} \epsilon_{3H\ast} + C \epsilon_{2H\ast} \epsilon_{3H\ast}^2 + C \epsilon_{2H\ast} \epsilon_{3H\ast} \epsilon_{4H\ast}  \end{aligned}$ \\
			\hline 
			$\hspace{0.5cm} \beta_{\textrm{s}}$ & $\vspace{0.2em} \begin{aligned}  ~\\[-3ex]  & 
				\hspace{0.45cm}  \textrm{N3LO}: -2 \epsilon_{1H\ast} \epsilon_{2H\ast}^2 - 2 \epsilon_{1H\ast} \epsilon_{2H\ast} \epsilon_{3H\ast} + \epsilon_{2H\ast} \epsilon_{3H\ast}^2 + \epsilon_{2H\ast} \epsilon_{3H\ast} \epsilon_{4H\ast}  \end{aligned}$
		\end{tabular}
	\end{ruledtabular}
\end{table*}

\begin{table*}[h]
	\caption{\label{Tab:SingleField-TensorFeatures}Power-law quantities of tensor perturbations for a minimally coupled single field up to N3LO. Notice that a deviation from the exact consistency relation, denoted by $\delta $, is already expected at N2LO.}
	\begin{ruledtabular}
		\begin{tabular}{p{0.7in} p{0.67in} p{6in}}
			Quantity & Order & Expression \\
			\hline
				$\hspace{0.5cm} \mathcal{A}_{\textrm{t}} $ & $\vspace{0.2em}\begin{aligned} ~\\[-3ex] 
				\textrm{LO+NLO}:
				\hspace{0.1cm}  &\frac{16  G \hbar \, H_\ast^2}{\pi} \Big[ 1 - 2 (1 + C) \epsilon_{1H\ast} \\
				\textrm{N2LO}: \hspace{0.1cm}   & + \frac{1}{2} \left(-6 + 4 C + 4 C^2 + \pi^2\right) \epsilon_{1H\ast}^2 + (2 + 2 C + C^2 - \frac{\pi^2}{12}) \epsilon_{1H\ast} \epsilon_{2H\ast}  \nonumber \\
				\textrm{N3LO}: \hspace{0.1cm}  & + \frac{1}{3} \epsilon_{1H\ast}^3 \left(16 - 4 C^3 - 3 C (-8 + \pi^2) - 14 \zeta(3)\right) \nonumber  \\
				& + \left(8 - 3 C^2 - 2 C^3 - \frac{13 \pi^2}{12} + C \left(6 - \frac{5 \pi^2}{6}\right)\right) \epsilon_{1H\ast}^2 \epsilon_{2H\ast} \\
				& + \frac{1}{12} \epsilon_{1H\ast} \epsilon_{2H\ast}^2 \left(-12 C^2 - 4 C^3 + \pi^2 + C (-24 + \pi^2) - 8 (1 + \zeta(3))\right) \nonumber \\
				& + \frac{1}{12} \epsilon_{1H\ast} \epsilon_{2H\ast} \epsilon_{3H\ast} \left(-12 C^2 - 4 C^3 + \pi^2 + C (-24 + \pi^2) - 8 (1 + \zeta(3))\right)\Big]
			\end{aligned}$
			\\
			\hline
			$\hspace{0.5cm} n_\textrm{t}$ & $\vspace{0.2em}\begin{aligned} ~\\[-3ex]  \hspace{0.8cm} \textrm{NLO}: 
				& \hspace{0.1cm}  -2 \epsilon_{1H\ast}  \\ 				 \textrm{N2LO}:
				& \hspace{0.1cm}   -2 \epsilon_{1H\ast}^2 + 2 (1 + C) \epsilon_{1H\ast} \epsilon_{2H\ast} \\
				\textrm{N3LO}: & \hspace{0.1cm}   -2 \epsilon_{1H\ast}^3 + \left(14 + 6 C - \pi^2\right) \epsilon_{1H\ast}^2 \epsilon_{2H\ast} + \frac{1}{12} \left(-24 - 24 C - 12 C^2 + \pi^2\right) \epsilon_{1H\ast} \epsilon_{2H\ast}^2 \nonumber   \\
				& \hspace{0.1cm} + \frac{1}{12} \left(-24 - 24 C - 12 C^2 + \pi^2\right) \epsilon_{1H\ast} \epsilon_{2H\ast} \epsilon_{3H\ast} 
			\end{aligned}$ \\
			\hline
			$\hspace{0.5cm}\alpha_{\textrm{t}}$ & $\vspace{0.2em}\begin{aligned} ~\\[-3ex]  \hspace{0.6cm}	\textrm{N2LO}: &\hspace{0.2cm}   2 \epsilon_{1H\ast} \epsilon_{2H\ast} 
				\\ 	\textrm{N3LO}: &  \hspace{0.2cm}6 \epsilon_{1H\ast}^2 \epsilon_{2H\ast} - 2 (1 + C) \epsilon_{1H\ast} \epsilon_{2H\ast}^2 - 2 (1 + C) \epsilon_{1H\ast} \epsilon_{2H\ast} \epsilon_{3H\ast} \end{aligned}$ \\
			\hline 
				$\hspace{0.5cm} \beta_{\textrm{t}}$ & $\vspace{0.2em}\begin{aligned} ~\\[-3ex] \hspace{0.6cm}	\textrm{N3LO}:  &\hspace{0.1cm}	-2 \epsilon_{1H\ast} \epsilon_{2H\ast}^2  - 2 \epsilon_{1H\ast} \epsilon_{2H\ast} \epsilon_{3H\ast}  \end{aligned}$\\
							\hline 
							\hline
			$\hspace{0.2cm} r\equiv \dfrac{\mathcal{A}_{\textrm{t}}}{\mathcal{A}_\textrm{s}}$ & $\vspace{0.2em}\begin{aligned} ~\\[-3ex] \hspace{0.8cm} \textrm{NLO}: 
	& \hspace{0.1cm} 16 \epsilon_{1H\ast}  \\ 				 \textrm{N2LO}:
	& \hspace{0.1cm}  - 16 C \epsilon_{1H\ast} \epsilon_{2H\ast}  \\
	\textrm{N3LO}: & + 8 (-8 - 2 C + \pi^2) \epsilon_{1H\ast}^2 \epsilon_{2H\ast} + 2 (8 + 4 C^2 - \pi^2) \epsilon_{1H\ast} \epsilon_{2H\ast}^2 - \frac{2}{3} (-12 C^2 + \pi^2) \epsilon_{1H\ast} \epsilon_{2H\ast} \epsilon_{3H\ast}
\end{aligned}$  \\
\hline 
	$\delta \equiv r+8n_{\textrm{t}}$ & $\begin{aligned} ~\\[-3ex] \hspace{0.65cm}  \textrm{N2LO}: 
	& \hspace{0.1cm}  -16 \epsilon_{1H\ast}^2 + 16 \epsilon_{1H\ast} \epsilon_{2H\ast}  \\
	\textrm{N3LO}: & -16 \epsilon_{1H\ast}^3 + 16 (3 + 2 C) \epsilon_{1H\ast}^2 \epsilon_{2H\ast} - \frac{4}{3} (12 C + \pi^2) \epsilon_{1H\ast} \epsilon_{2H\ast}^2 - 16 (1 + C) \epsilon_{1H\ast} \epsilon_{2H\ast} \epsilon_{3H\ast}
\end{aligned}$ 
		\end{tabular}
	\end{ruledtabular}
\end{table*}

\vfill
\newpage

\section{Deriving the expansion of $H(t)$ in terms of $\epsilon_{1H}(t)$}
\label{app:derivation-H(eps)}
Let us recall the modified Friedmann equation for the model $R+\alpha R^2$, reported in \eqref{eq:Friedmann_Modified}, which can be rewritten as
\begin{equation}
\label{eq:Friedmann-appendix}
1-36\, \alpha\, H(t)^2 \, \epsilon_{1H}(t) + 18 \, \alpha \, H(t)^2 \, \epsilon_{1H}(t)^2 - 12\, \alpha\, H(t)\, \dot{\epsilon}_{1H}(t) = 0\,.
\end{equation}
From the above expression one can also solve for $\dot{\epsilon}_{1H}(t)$,
\begin{equation}
\label{eq:dot-eps1}
\dot{\epsilon}_{1H}(t) = \dv{\epsilon_{1H}(t)}{t} =  \frac{1-36\, \alpha\, H(t)^2 \epsilon_{1H}(t) + 18 \, \alpha H(t)^2 \epsilon_{1H}(t)^2 }{12\, \alpha \, H(t)}\,.
\end{equation}
If we neglect contributions of order $\order{\epsilon^2}$, the Friedmann equation \eqref{eq:Friedmann-appendix} is solved by $H(t) \sim \frac{1}{\sqrt{36 \,\alpha\, \epsilon_{1H}(t)}}$. Hence, one would like to determine an expansion of $H(t)$ order-by-order in $\epsilon_{1H}(t)$, of the form
\begin{equation}
\label{eq:H(t)-ansatz}
H(t)^{(\mathrm{anz})} = \frac{1}{\sqrt{36 \,\alpha\, \epsilon_{1H}(t)}} \Big[ 1 + a_1 \,\epsilon_{1H}(t) + a_2 \, \epsilon_{1H}(t)^2 + a_3 \, \epsilon_{1H}(t)^3 + a_4 \, \epsilon_{1H}(t)^4 + \cdots \Big] 
\end{equation}
This also defines an ansatz $\epsilon_{1H}(t)^{(\mathrm{anz})} = - \dot{H}^{(\mathrm{anz})}/H^{(\mathrm{anz})^2}$. Using \eqref{eq:H(t)-ansatz} as an ansatz, we can impose the condition, $\epsilon_{1H} (t)^{(\mathrm{anz})} = \epsilon_{1H}(t) +\order{\epsilon^5}$. From this self-consistency condition, we find the coefficients $a_1$, $a_2$, $a_3$, and $a_4$, which are finally reported in \eqref{eq:H-to-eps}. This expression also allows us to expand $\epsilon_{2H}(t)$, $\epsilon_{3H}(t)$, etc. in terms of $\epsilon_{1H}(t)$, as reported in \eqref{eq:Epsilon_Expansion_234}. It can be checked that the resulting Hubble rate is a solution of \eqref{eq:Friedmann-appendix}, up to N3LO corrections.

\vfill
\newpage

\end{widetext}

\newpage

\providecommand{\href}[2]{#2}\begingroup\raggedright\endgroup

%\bibliographystyle{JHEP}
%%\bibliographystyle{apsrev4-2}
%\bibliography{RefP1}

%spphys
\end{document}